\documentclass[aps,pra,twocolumn,groupedaddress]{revtex4-1}

\pdfoutput=1

\usepackage{eurosym}
\usepackage{amsfonts}
\usepackage{amsmath}
\usepackage{amssymb,epsf}
\usepackage{color}
\usepackage{epstopdf}
\usepackage{graphicx}
\usepackage{caption}
\usepackage{subfig}
\usepackage{hyperref}
\usepackage{mathtools}
\usepackage{lipsum}
\usepackage{floatrow}
\usepackage{multirow}
\usepackage{bigstrut}
\definecolor{aogreen}{rgb}{0.0, 0.5, 0.0}

\def\ketm#1{  \left\vert  #1   \right\rangle   }

\def\bram#1{  \left\langle  #1   \right\vert   }

\begin{document}

\title{Toward simulation of topological phenomenas with one-, two- and three-dimensional quantum walks}

\author{S. Panahiyan $^{1,2,3}$}
\email{email address: shahram.panahiyan@uni-jena.de}
\author{S. Fritzsche $^{1,2,3}$ }
\email{email address: s.fritzsche@gsi.de}
\affiliation{$^1$Helmholtz-Institut Jena, Fr\"{o}belstieg 3, D-07743 Jena, Germany  \\
	$^2$GSI Helmholtzzentrum f\"{u}r Schwerionenforschung, D-64291 Darmstadt, Germany \\
	$^3$Theoretisch-Physikalisches Institut, Friedrich-Schiller-University Jena, D-07743 Jena, Germany}

\date{\today}

\begin{abstract}
We study the simulation of the topological phases in three subsequent dimensions with quantum walks. We are mainly focused on the completion of a table for the protocols of the quantum walk that could simulate different family of the topological phases in one, two dimensions and take the first initiatives to build necessary protocols for three-dimensional cases. We also highlight the possible boundary states that can be observed for each protocol in different dimensions and extract the conditions for their emergences or absences. To further enrich the simulation of the topological phenomenas, we include step-dependent coins in the evolution operators of the quantum walks. Consequently, this leads to step-dependency of the simulated topological phenomenas and their properties which in turn introduce dynamicality as a feature to simulated topological phases and boundary states. This dynamicality provides the step-number of the quantum walk as a mean to control and engineer the number of topological phases and boundary states, their populations, types and even occurrences.
\end{abstract}

\maketitle
\section{Introduction} \label{Intro}

The emergences of topological phenomenas such as the integer Hall effect \cite{Thouless}, fractional charges \cite{Qi2008} and topological insulators \cite{Hasan,Qi,Fidkowski,Koenig} are among the recent significant hallmarks in condensed matter physics. The topological phases and boundary states are states of the matter that are characterized by presence or absence of a geometrical quantity known as topological invariant. This quantity parameterizes the global structure of ground-state wave function of matters and remains fixed through continuous modification \cite{Thouless}. The topological phases (gapped energy bands) are characterized by well-defined topological invariants while boundary state (gapless energy bands) by ill-defined ones. The classification of topological phases is done based on symmetries that they posses and their dimensionality \cite{Schnyder2008,Ryu2010,Chiu,Zhang2018}. On the other hand, the type of boundary state is determined by how energy bands close their gap, i.e. linearly, nonlinearly or else \cite{Schnyder,Jin}.

One of the methods to simulate topological phases and boundary states is through quantum walks \cite{KitagawaExp,Cardano,Wang2018,Cardano2017,Barkhofen,Flurin,Zhan,Xiao,Nitsche,Wang,Errico,Xu}. The quantum walks are among the pronounced protocols of the quantum information and computation which are universal computational primitives \cite{Lovett} and simulators of other quantum systems and phenomenas \cite{Mohseni,Vakulchyk} such as topological phases \cite{Kitagawa}. In fact, all types of the topological phases in one and two dimensions are simulable by one- and two-dimensional quantum walks \cite{Kitagawa,Kitagawa2012,Asboth,Obuse,Chen}. The simulated topological phenomenas with the quantum walks can be used to extract topological invariants \cite{Ramasesh}, suppress the limitations on strongly-driven systems, check the robustness of the edge states \cite{KitagawaExp} and investigation of topological phase transitions \cite{Rakovszky,Mera,Panahiyan2019}. So far, the simulation of the topological phases and boundary states in three dimensions with quantum walks have not been addressed in literature.

Following our earlier work \cite{Panahiyan2020}, one of our main goals in this paper includes completion of a table for protocols of the quantum walks to simulate of all types of the topological phases in one and two dimensions. In addition, we take the first initiatives to introduce novel protocols for three-dimensional cases as well. We also study the boundary states, their types and conditions for their presences in each protocol. In order to further enrich the simulated topological phenomenas, we include step-dependent coins in the protocols of the quantum walks \cite{Panahiyan2018,Dhar,Katayama}. Due to the step-dependent coins, simulated topological phases and boundary states become dynamical with step-number being dynamical factor. By adjusting the step-number, we are able to control or engineer the type of topological phases and boundary states, their populations and their absences or presences for the protocols of the quantum walks in different dimensions. In addition, by setting step-number to $1$, the results for protocols with step-dependent coins reduce to those for the protocols with step-independent coins. Finally, the quantum walks with step-dependent coins can simulate all types of boundary states \cite{Panahiyan2020}.       

The structure of the paper is as follows. First, we present theoretical basis that map the protocol of the quantum walk and its properties to Hamiltonian, energy bands and other properties governing topological phenomena. Next, we investigate simulation of topological phases by one-dimensional quantum walk with two protocols each having only chiral symmetry or particle-hole symmetry. Then, we study two classes of two-dimensional quantum walks that simulate topological phases with zero and non-zero Chern number. Finally, we take the first steps to build protocols of three-dimensional quantum walks for simulation of topological phenomena that are observed in three dimensions. The paper is concluded with some closing remarks and an appendix.

\section{Quantum walk and effective Hamiltonian} \label{QWEH}

The generation of the quantum walk is the result of a successive application of a protocol, $\widehat{U}$, which manipulates both internal and external degrees of freedoms of a walker. The protocol of the quantum walk includes number of coin and shift operators. The coin operators modify the internal degrees of freedom of the walker and create superpositions of them. The shift operators change the external degrees of freedom of the walker conditioned to its internal degrees of freedom. 

Stroboscopically, the protocol of the quantum walk realizes a periodic Floquet evolution which can be mapped to a (dimensionless) effective Floquet Hamiltonian

\begin{eqnarray}
\widehat{H}& = &i \ln\widehat{U}= E \boldsymbol n \cdot \boldsymbol \sigma,  \label{Hamiltonian}
\end{eqnarray}
where $E$ is the (quasi)energy dispersion defined up to $2\pi$, $\boldsymbol \sigma$ are Pauli matrices and $\boldsymbol n$ defines the eigenstates of the energy. We can obtain the energy dispersion through  

\begin{eqnarray}
E& = &i \ln \lambda,  \label{energy}
\end{eqnarray}
in which $\lambda$ is the eigenvalue of $\widehat{U}$. In addition to energy, we can also investigate the topological phase and boundary states through group velocity given by  

\begin{eqnarray}
V(k_{i})& = & \frac{d E}{dk_{i}}, \label{groupv}
\end{eqnarray}
where $k_{i}$ corresponds to $k_{x}$, $k_{y}$ and $k_{z}$ which are (quasi)momenta. It is possible to decompose the protocol of the quantum walk into 

\begin{eqnarray}
\widehat{U}& = & d_{0} \sigma_{0}- i \boldsymbol d (k_{i}) \cdot \boldsymbol \sigma, 
\end{eqnarray}
where, one can find $\boldsymbol n$ using $\boldsymbol d(k_{i})$ through 

\begin{eqnarray}
\boldsymbol n& = & \frac{\boldsymbol d(k_{i})}{\sqrt{d_{x}^2+d_{y}^2+d_{z}^2}}.  \label{n} 
\end{eqnarray} 

Different classes of topological phases are determined by two factors of dimensionality of the system and the symmetries that the Hamiltonian possesses. In this paper, we focus on three distinctive symmetries of particle-hole (PHS), tieme reversal (TRS) and chiral (CHS). The existences of these PHS, CHS and TRS are conditioned to the satisfaction of the following relations, respectively \cite{Kitagawa} 

\begin{eqnarray}
\widehat{\mathcal{P}}\widehat{H}(k)\widehat{\mathcal{P}}^{-1} &=& -\widehat{H}(k), \label{PHS}
\\
\widehat{\Gamma}^{-1}\widehat{H}(k)\widehat{\Gamma} &=& -\widehat{H}(k),   \label{CHS}
\\
\widehat{\mathcal{T}}\widehat{H}(k)\widehat{\mathcal{T}}^{-1} &=& \widehat{H}(k),  \label{TRS}
\end{eqnarray}
where $\widehat{\mathcal{P}}$ and $\widehat{\mathcal{T}}$ are antiunitary operators while $\widehat{\Gamma}$ is an unitary one. If the protocol of the quantum walk is real, then we will have

\begin{eqnarray}
\widehat{U}^{\ast}=\widehat{U} \Rightarrow \widehat{H}^{\ast}(k) = -\widehat{H}(k), 
\end{eqnarray}
which indicates that we can define $\widehat{\mathcal{P}}\equiv \widehat{K}$ where $\widehat{K}$ is the complex conjugation operator that satisfies the condition for presence of PHS \eqref{PHS}. Generally speaking, if the coin operators (rotation matrices) are performed around an axis that lies in $xy$ plane, then the Hamiltonian possesses PHS \cite{Kitagawa}. Therefore, the recipe for breaking the PHS is choosing a rotation axis that contains a nonzero $z$ component. 

The unitary operator of the CHS is also Hermitian. Therefore, we have an additional condition to Eq. \eqref{CHS} which is $\widehat{\Gamma}^2=I$. We can build up the CHS operator as

\begin{eqnarray}
\widehat{\Gamma} & = &\boldsymbol A \cdot \boldsymbol\sigma, \label{gamma}
\end{eqnarray}
in which $\boldsymbol A$ is a vector labeling a point on the Bloch sphere and perpendicular to $\boldsymbol n$. For one-dimensional quantum walk, presence of CHS confines $\boldsymbol n$ on a great circle on Bloch sphere and winding number, number of times $\boldsymbol n(k)$ winds around origin as $k$ transverses through the first Brillouin zone, becomes topological invariant. 

The presence of PHS and CHS guarantees existence of TRS where $\widehat{\mathcal{T}}\equiv \widehat{\Gamma}\widehat{\mathcal{P}}$. Another signature of the TRS is through studying the energy. One of the requirements imposed by the presence of TRS is validation of $E(k)=E(-k)$. In the case of violation of this condition, the TRS would be absent for the considered Hamiltonian.

The quantum walk protocols that we will consider gives us two bands of energy. If these two bands of the energy are gapped, then the Hamiltonian and the energy bands describe a topological phase with specific topological invariant. In contrast, if the gap between the energy band closes, gapless energy bands, we have an boundary state. The boundary states are located between two topological phases and they mark phase transition points. 

There are three types of boundary states which characterize how energy bands close their gaps \cite{Chiu}. If the energy bands close their gaps linearly, the resulting boundary state is Dirac cone. In contrast, Fermi arc, another type of boundary state, are formed when energy bands close up their gap nonlinearly. Finally, flat bands boundary states are gapless energy bands which remain constant for variations of $k_{i}$. 

In this paper, we focus on a single walker with two internal degrees of freedom. Therefore, the coin Hilbert space ($\mathcal{H}_{C}$) is spanned by $\{ \ketm{0},\: \ketm{1} \}$. As for external degrees of freedom, we investigate three cases of one-, two- and three-dimensional position spaces. Consequently, the position Hilbert spaces of one- ($\mathcal{H}_{x}$), two- ($\mathcal{H}_{x} \otimes \mathcal{H}_{y}$) and three- ($\mathcal{H}_{x} \otimes \mathcal{H}_{y} \otimes \mathcal{H}_{z}$) dimensional position spaces are spanned by $\{ \ketm{x}_{P}: x\in \mathbb{Z}\}$, $\{ \ketm{x,y}_{P}: x,y \in \mathbb{Z}\}$ and $\{ \ketm{x,y,z}_{P}: x,y,z \in \mathbb{Z}\}$, respectively. The Hilbert space of the walk is given by tensor product of subspaces of coin and position spaces. 

The protocols for quantum walk can be divided into two groups: a) Simple-step protocol which contains only one coin and only one shift operator. b) Split-step protocol where there are at least two different coin and two different shift operators \cite{Kitagawa}. 

For the sake of brevity, we rename the following terms in our formulas 

\begin{align*}
\cos(\frac{T j}{2})&= \kappa_{j}   &  \sin(\frac{T j}{2})&= \lambda_{j} 
\end{align*}
where $j$ could be $\alpha$, $\beta$, $\gamma$, $\lambda$ and $\zeta$, and $T$ corresponds to step number of the quantum walk.

\section{one-dimensional quantum walk} \label{One dimensions}

In this section, we consider one-dimensional quantum walk with two different split-step protocols. One of these protocols admits only PHS \ref{OPHS} whereas in the other protocol, both TRS and PHS are absent and only CHS is present \ref{OCHS}. 

\subsection{Quantum walk with only PHS} \label{OPHS}

In order to have PHS, we consider the protocol of the quantum walk to be real and the rotation axes of the coin operators lie in the $xy$ plane. Such protocol is given by 

\begin{eqnarray}
\widehat{U} & = & \widehat{S}_{\uparrow}(x) \widehat{C}_{y}(\alpha) \widehat{S}_{\downarrow}(x) \widehat{C}_{y}(\beta)\widehat{S}_{\uparrow \downarrow}(x)  \label{protocol1},
\end{eqnarray}
where $\widehat{C}_{y}(\beta)$ and $\widehat{C}_{y}(\alpha)$ are rotation matrices (coin operators) around $y$ axis, and $\widehat{S}_{\uparrow}(x)$, $\widehat{S}_{\downarrow}(x)$ and $\widehat{S}_{\uparrow \downarrow}(x)$ are shift operators. A single step of the walk chronologically includes displacement of the walker by $\widehat{S}_{\uparrow \downarrow}(x)$, rotation of its internal states by $\widehat{C}_{y}(\beta)$, a displacement of the walker with $\widehat{S}_{\downarrow}(x)$, another rotation of the internal states with $\widehat{C}_{y}(\alpha)$ and final displacement of the walker by $\widehat{S}_{\uparrow} (x)$. Since the the protocol of the quantum walk is real, we find $\widehat{\mathcal{P}}\equiv \widehat{K}$ which satisfies \eqref{PHS}. 

The coin operators are 

\begin{eqnarray}
\widehat{C}_{\alpha} & = & e^{-\frac{i T \alpha}{2}\sigma_{y}},  \label{coin1}
\\
\widehat{C}_{\beta} & = & e^{-\frac{i T \beta}{2}\sigma_{y}},  \label{coin2}
\end{eqnarray}
in which $\alpha$ and $\beta$ are rotation angles belonging to $[-\pi,\pi]$ and $T$ characterizes step-dependency of the coins. The shift operators are

\begin{equation}
\widehat{S}_{\uparrow \downarrow} (x)= \ketm{\uparrow} \bram{\uparrow} \otimes \sum_{x} \ketm{x+1} \bram{x}
\ketm{\downarrow} \bram{\downarrow} \otimes \sum_{x} \ketm{x-1} \bram{x}\, , \label{shift1}
\end{equation} 

\begin{equation}
\widehat{S}_{\downarrow}(x)=\sum_{x}[\ketm{\uparrow} \bram{\uparrow} \otimes \ketm{x} \bram{x} + \ketm{\downarrow} \bram{\downarrow} \otimes \ketm{x-1} \bram{x}\,]  \label{shift2},
\end{equation}

\begin{equation}
\widehat{S}_{\uparrow}(x)=\sum_{x}[\ketm{\uparrow} \bram{\uparrow} \otimes \ketm{x+1} \bram{x} + \ketm{\downarrow} \bram{\downarrow} \otimes \ketm{x} \bram{x}\,]  \label{shift3}.
\end{equation}

One can use the Discrete Fourier Transformation, $\ketm{k_{x}}=\sum_{x}e^{-\frac{i k_{x} x}{2}}\ketm{x}$, and rewrites shift operators as $\widehat{S}_{\uparrow \downarrow}(x)= e^{ik_{x} \sigma_{z}}$, $\widehat{S}_{\downarrow}(x)=e^{\frac{i k_{x}}{2}(\sigma_{z}+1)}$ and $\widehat{S}_{\uparrow}(x)=e^{\frac{i k_{x}}{2}(\sigma_{z}-1)}$. The presence of the $\widehat{S}_{\uparrow \downarrow}(x)$ breaks down the CHS and the protocol of the quantum walk and corresponding Hamiltonian have only PHS. 

It is a matter of calculation to find the energy as  

\begin{eqnarray}
E & = & \pm\cos^{-1}(\rho), \label{energy1}
\end{eqnarray}
where $\rho= \kappa_{\alpha}\kappa_{\beta} [\cos^2(k_{x})- \sin^2(k_{x})]-\cos(k_{x})\lambda_{\alpha}\lambda_{\beta}$. The $\pm$ indicates two bands of the energy. The energy spans $[-\pi,\pi]$ and the energy bands close their gap at $E=0$ and $\pm \pi$. $k_{x}$ traverses the first Brillioun zone. 

Based on obtained energy bands \eqref{energy2}, we can highlight the following behaviors for energy bands ($c$ is integer):

I) If $\beta=\pm (2c+1) \pi/T$, then energy bands become $E= \pm\cos^{-1}(\pm \cos(k_{x})\lambda_{\alpha})$. In case of $\alpha=\pm (2c+1) \pi/T$, energy bands would be linear function of the $k_{x}$, hence boundary states are only Dirac cones. On the other hand, for $\alpha=\pm 2c\pi/T$, energy bands are independent of $k_{x}$ and constant, $E=\pm \pi/2$, which indicates that energy bands are flat bands.   

II) If $\beta=\pm 2c\pi/T$, then energy bands reduce to $E= \pm\cos^{-1}(\kappa_{\alpha} [\cos^2(k_{x})- \sin^2(k_{x})])$. Identically, for $\alpha=\pm 2c\pi/T$, the energy bands close up their gap only at $k_{x}=0$, $\pm \pi/2$ and $\pm \pi$. In such case, the boundary states again would be Dirac cone. 

Next, we find $\boldsymbol n$ and group velocity as

\begin{equation}
n_{x}(k_{x})  =  -\lambda_{\alpha} \kappa_{\beta} \sin (k_{x}), \notag
\end{equation}
\begin{equation}
n_{y}(k_{x})  =  \lambda_{\alpha} \kappa_{\beta} \cos (k_{x}) + \kappa_{\alpha} \lambda_{\beta}, \notag
\end{equation}
\begin{equation}
n_{z}(k_{x})  =\sin(k_{x})\lambda_{\alpha}\lambda_{\beta}- 2 \kappa_{\alpha} \kappa_{\beta} \cos(k_{x}) \sin (k_{x}), 
\end{equation}  
\begin{equation}
V(k_{x})= \pm \frac{\sin(k_{x})\lambda_{\alpha}\lambda_{\beta}-4\kappa_{\alpha}\kappa_{\beta}\cos(k_{x})\sin(k_{x})}{ \sqrt{1-\rho^2}}. \label{groupv1}
\end{equation}

The obtained $\boldsymbol n$ and group velocity are ill-defined where the energy bands close their gap. Therefore, one of the characterizations of the boundary states are ill-defined group velocity and $\boldsymbol n$.

\subsection{Quantum walk with only CHS} \label{OCHS}

In this section, we investigate topological phases with only CHS. Previously, we discussed that in order to break the PHS, one can change the rotation axis of the coin operators to the one that has nonzero $z$ component. To do so, we perform the rotations around the axis $\nu=\frac{1}{\sqrt{2}}(0,1,1)$ which was introduced by Kitagawa \textit{et al} in Ref. \cite{Kitagawa}. The protocol of the quantum walk with only CHS will be

\begin{eqnarray}
\widehat{U} & = & \widehat{S}_{\uparrow}(x) \widehat{C}_{\nu}(\alpha) \widehat{S}_{\downarrow}(x) \widehat{C}_{\nu}(\beta)  \label{protocol2},
\end{eqnarray}
where $\widehat{C}_{\nu}(\beta)=e^{-i \frac{T \beta}{2} \boldsymbol \nu . \sigma }$ and $\widehat{C}_{\nu}(\alpha)$ are step-dependent rotation matrices (coin operators) around $ \boldsymbol \nu=(0,\frac{1}{\sqrt{2}},\frac{1}{\sqrt{2}})$ axis, and $\widehat{S}_{\uparrow}(x)$ and $\widehat{S}_{\downarrow}(x)$ are shift operators. A Single step of the quantum  walk includes rotoatin of internal states by $\widehat{C}_{\nu}(\beta)$, displacement of walker with $\widehat{S}_{\downarrow}(x)$, another rotation of the internal states with $\widehat{C}_{\nu}(\alpha)$ and final displacement with $\widehat{S}_{\uparrow}(x)$. Using the Discrete Fourier Transformation, we rewrite $\widehat{S}_{\uparrow}(x)$ and $\widehat{S}_{\downarrow}(x)$ in terms of moment, $\widehat{S}_{\downarrow}(x)=e^{\frac{i k_{x}}{2}(\sigma_{z}+1)}$ and $\widehat{S}_{\uparrow}(x)=e^{\frac{i k_{x}}{2}(\sigma_{z}-1)}$. Therefore, we can find the eigenvalues of the $\widehat{U}$ and by using Eq. \eqref{energy}, energy as 

\begin{eqnarray}
E & = & \pm\cos^{-1}(\rho), \label{energy2}
\end{eqnarray}
where $\rho= -\frac{1}{2}\lambda_{\alpha}\lambda_{\beta} [1+\cos(k_{x})]+ \kappa_{\alpha}\kappa_{\beta} \cos(k_{x})+\frac{1}{\sqrt{2}} \sin(\frac{T \alpha}{2}+\frac{T \beta}{2})\sin(k_{x})$. According to obtained energy, we can highlight two specific cases:

I) By setting $\beta=\pm (2c+1) \pi/T$, the energy bands reduce $E= \pm \cos^{-1}(\frac{1}{2}\lambda_{\beta} [1-\cos(k_{x})]+\frac{1}{\sqrt{2}} \sin(\frac{T \alpha}{2} \pm \frac{(2c+1)\pi}{T})\sin(k_{x}))$. If $\alpha=\pm (2c+1) \pi/T$, then energy bands would be nonlinear function of the $k_{x}$. Therefore, the boundary state are Fermi arc type.  

II) For $\alpha=\beta=\pm 2c\pi/T$ the energy bands become $E= \pm k_{x}$ which indicates that the energy bands close up their gap linearly at $k_{x}=0$ and $\pm \pi$ and boundary states are Dirac cones.

Using obtained energy \eqref{energy2} and the protocol of quantum walk \eqref{protocol2}, we can find $\boldsymbol d$ (hence $\boldsymbol n$) and group velocity as

\begin{eqnarray}
d_{x}  =  \frac{\lambda_{\beta} }{2}[\sqrt{2} \kappa_{\alpha}\sin(k_{x})+\lambda_{\alpha}(1-\cos(k_{x}))], \notag
\end{eqnarray}
\begin{eqnarray}
d_{y}  = \frac{\lambda_{\alpha}\kappa_{\beta}}{\sqrt{2}}+\frac{\lambda_{\beta}}{2}[\lambda_{\alpha}\sin(k_{x})+ \sqrt{2}\kappa_{\alpha}\cos(k_{x})], \notag
\end{eqnarray}
\begin{equation}
d_{z}  = \frac{1}{2}[\lambda_{\alpha}\lambda_{\beta}\sin(k_{x})-2\kappa_{\alpha} \kappa_{\beta}]+\frac{1}{\sqrt{2}}\sin(\frac{T (\alpha+\beta)}{2})\cos(k_{x}) 
\end{equation}  
\begin{eqnarray}
V(k_{x})& = & \pm (-n_{z}(k_{x})). \label{groupv2}
\end{eqnarray}

First of all, both $\boldsymbol n$ and group velocity are ill-defined where boundary states are formed. Therefore, well- or ill-definition of $\boldsymbol n$ and group velocity can be used as a mean to find boundary states. In addition, since $V(k_{x})= |n_{z}(k_{x})|$, the group velocity spans $[-1,1]$ which is in contrast of what we observed for PHS where group velocity traverses $[-2,2]$.

Finally, one can build up the chiral symmetry operator using $\Gamma= \boldsymbol A \cdot \boldsymbol \sigma$ in which $\boldsymbol A =(\kappa_{\beta},-\lambda_{\beta}/\sqrt{2},\lambda_{\beta}/\sqrt{2})$ is a vector perpendicular to $\boldsymbol n$ for variation of $k_{x}$.

\subsection{Discussion on results of one-dimensional quantum walks}

The presence of the step-dependent coin in protocols of the quantum walk provide us with step-dependent simulated topological phenomena. The major consequence of this step-dependency is a dynamical nature for topological phases, boundary states, their places and numbers. Therefore, from one step to another one, we observe considerably different simulated topological phenomena. In return, we can use the step number as a mean to control or engineer the properties of the topological phenomena. 

\begin{figure*}[htb]
	\centering
	{\begin{tabular}[b]{cc}%
		\sidesubfloat[]{\includegraphics[width=1\linewidth]{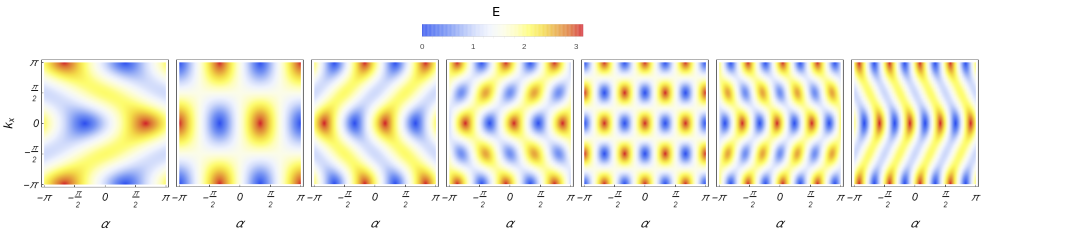}}\\[0.0001cm]
		\sidesubfloat[]{\includegraphics[width=1\linewidth]{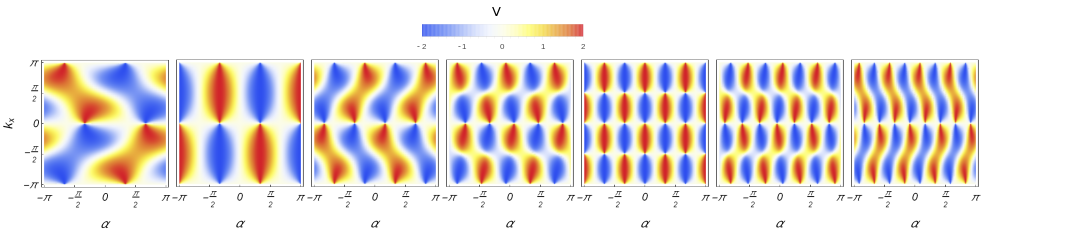}}\\[0.0001cm]	
	\end{tabular}}				
	\caption{Quantum walk with only PHS (one-dimensional): Modification of energy (a) and group velocity (b) (both positive branches) as functions of momentum and rotation angle $\alpha$ ($\beta=\pi/3$) for subsequent steps of $T=2,...,8$ from left to right, respectively. In (a), each step contains only one type of boundary states, Dirac cone or Fermi arc, and there is no flat band boundary states. As step number increases, the size of topological phases decreases. Therefore, population of topological phases and boundary states (topological phase transitions) increase step-dependently. In (b), we observe that the characteristic behavior of the group velocity around gapless point of each boundary state is different enabling us to recognize the type of boundary states. On the other hand, for both of these boundary states, the sign of the group velocity swaps from positive to negative or vise versa at gapless point of energy bands.} \label{Fig1}
\end{figure*}

\begin{figure*}[htb]
	\centering
	{\begin{tabular}[b]{cc}%
		\sidesubfloat[]{\includegraphics[width=1\linewidth]{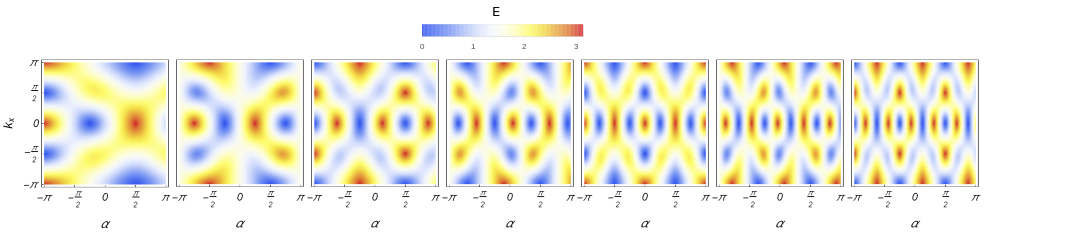}}\\[0.0001cm]
		\sidesubfloat[]{\includegraphics[width=1\linewidth]{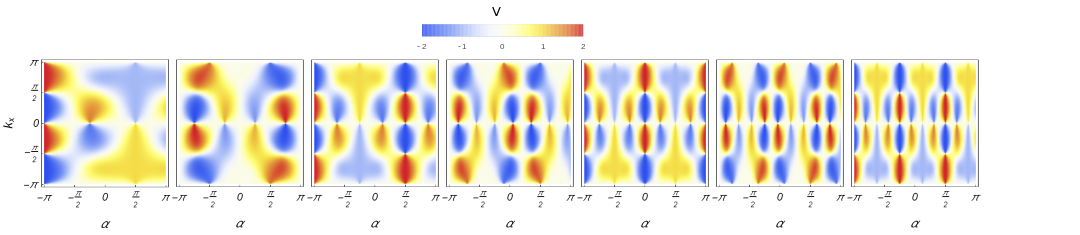}}\\[0.0001cm]	
	\end{tabular}}				
	\caption{Quantum walk with only PHS (one-dimensional): Modification of energy (a) and group velocity (b) (both positive branches) as functions of momentum and rotation angle $\alpha$ ($\beta=(\alpha+\pi)/3$) for subsequent steps of $T=2,...,8$ from left to right, respectively. In (a), we observe the simulation of the Dirac cone and Fermi arc boundary states together in each step while flat band boundary states are absent. There are two types of Dirac cone boundary states where the first type, gapless energy bands are observed at $k_{x}=0$ and $\pm \pi$ while for type two we have additionally $k_{x}=\pm \pi/2$. In (b), for type one Dirac cone, the group velocity is fixed at $\pm 1$ around gapless point while it is fixed at $\pm 2$ for type two Dirac cones. At each gapless point, the sign of group velocity swaps to its opposite.} \label{Fig2}
\end{figure*}

The topological phases are located between two boundary states which mark the phase transition points. The increment in step number results into shrinking the size of topological phases. In other words, the population of boundary states, phase transitions and topological phases step-dependently increases. This is valid irrespective of the protocol used for the quantum walk (see Figs. \ref{Fig1}, \ref{Fig2}, \ref{Fig4} and \ref{Fig5}). 

\begin{figure*}[htb]
	\centering
	{\begin{tabular}[b]{cc}%
		\subfloat[]{\includegraphics[width=0.4\linewidth]{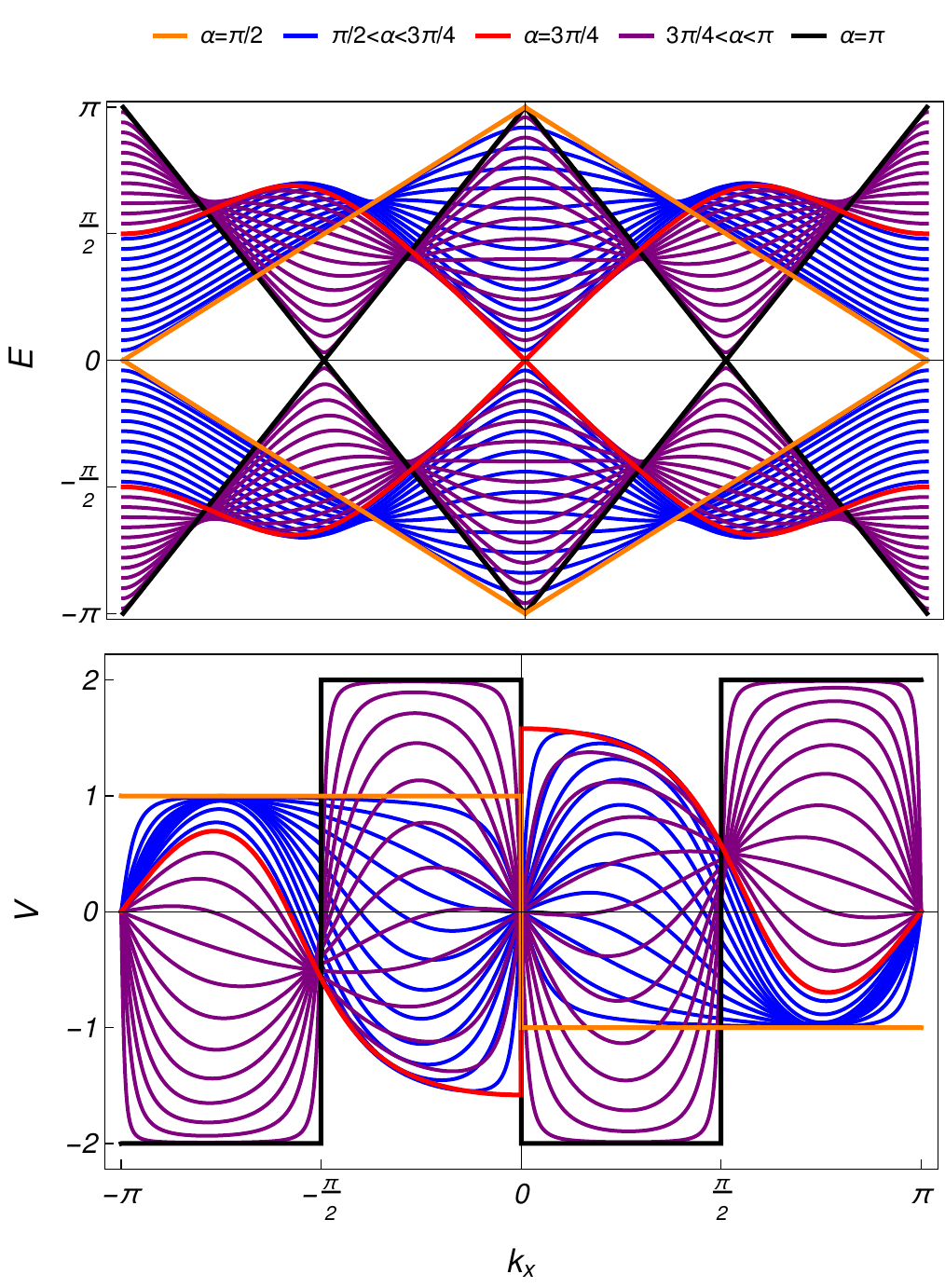}}
		\subfloat[]{\includegraphics[width=0.4\linewidth]{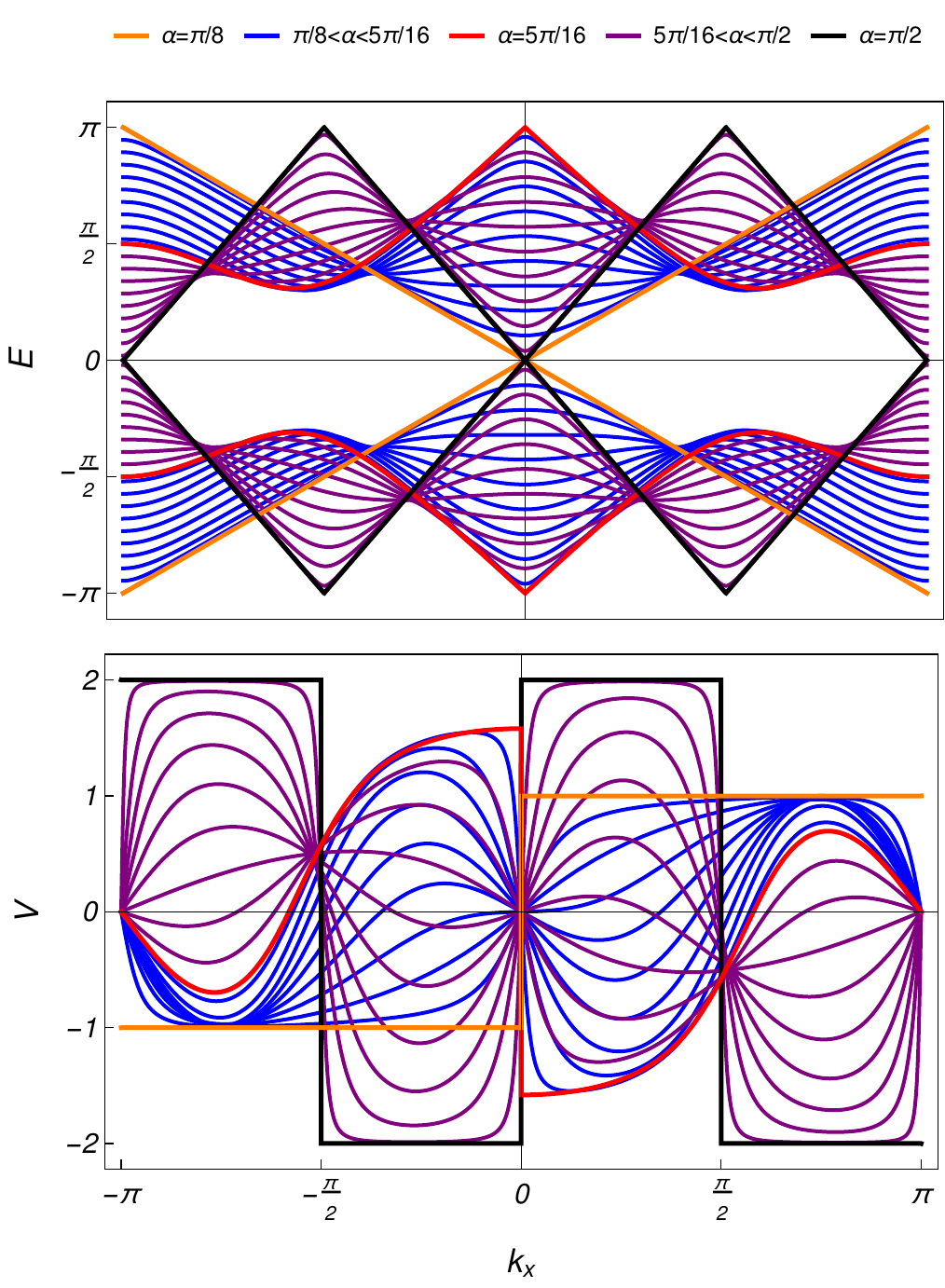}}				
	\end{tabular}}				
	\caption{Quantum walk with only PHS (one-dimensional): Energy (up panels) and group velocity (down panels) (only positive branch) for $6th$ (a) and $8th$ (b) steps with $\beta=(\alpha+\pi)/3$. As we scan the rotation angle $\alpha$ through limited range, the protocol starts with type one (two) Dirac cone boundary state followed by series of topological phases (blue lines), confronted Fermi arc boundary state. After the Fermi arc another set of topological phases are emerged (purple lines) and finally we reach Type two (one) Dirac cone boundary state. Comparing (a) with (b), we see that the place of topological phases and boundary states change step-dependently. in down panels of (a) and (b), we observe that around type one Dirac cone boundary state, the group velocity approximates to $\approx \pm 1$ while for around the type two Dirac, it approximates to $\approx \pm 2$.}	\label{Fig3}
\end{figure*}

\begin{figure*}[htb]
	\centering
	{\begin{tabular}[b]{cc}%
		\sidesubfloat[]{\includegraphics[width=0.95\linewidth]{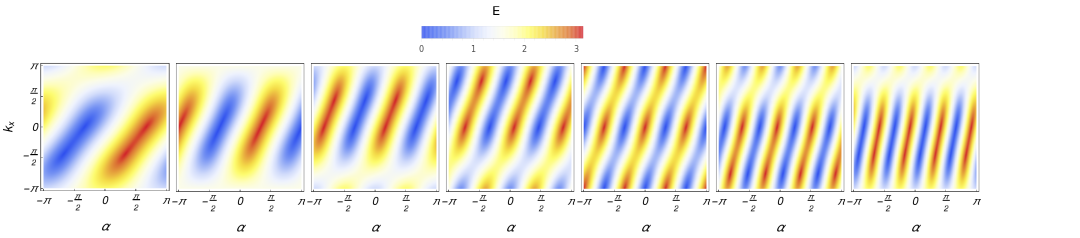}}\\[0.0001cm]
		\sidesubfloat[]{\includegraphics[width=0.95\linewidth]{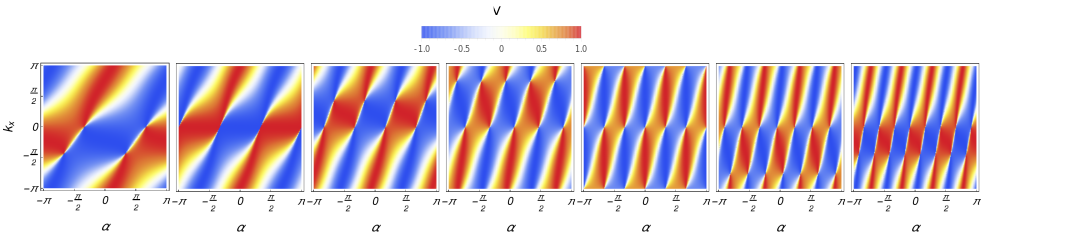}}\\[0.0001cm]	
	\end{tabular}}				
	\caption{Quantum walk with only CHS (one-dimensional): Modification of energy (a) and group velocity (b) (both positive branches) as functions of momentum and rotation angle $\alpha$ ($\beta=\pi/3$) for subsequent steps of $T=2,...,8$ from left to right, respectively. In (a), we observe that for early stages of walk, only Fermi arc boundary states are observable and increment in step number results into simulation of both Dirac cone and Fermi arc boundary states together. Only type one Dirac cones are present for boundary states. Contrary to PHS protocol, the boundary states formed for arbitrary values of the $k_{x}$. In (b), we observed that group velocity for Fermi arcs nonlinearly is modified as we scan the first Brillouin zone for $k_{x}$ while it is linearly modified for Dirac cones. The swamping of the group velocity's sign to the opposite happens at gapless points of energy bands.} \label{Fig4}
\end{figure*}

\begin{figure*}[htb]
	\centering
	{\begin{tabular}[b]{cc}%
		\sidesubfloat[]{\includegraphics[width=1\linewidth]{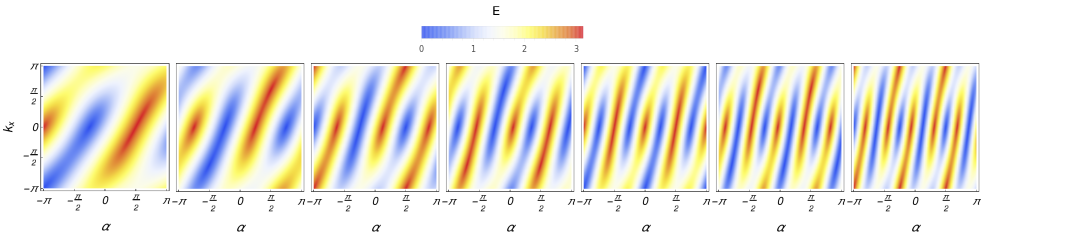}}\\[0.0001cm]
		\sidesubfloat[]{\includegraphics[width=1\linewidth]{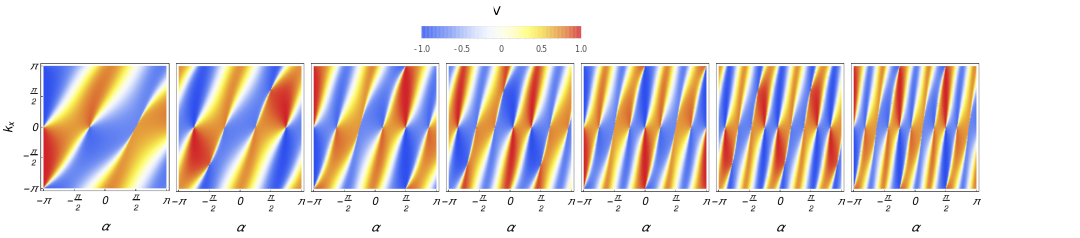}}\\[0.0001cm]	
	\end{tabular}}				
	\caption{Quantum walk with only CHS (one-dimensional): Modification of energy (a) and group velocity (b) (both positive branches) as functions of momentum and rotation angle $\alpha$ ($\beta=(\alpha+\pi)/3$) for subsequent steps of $T=2,...,8$ from left to right, respectively. In (a), we notice that rotation angles being linearly related to each other guarantees simulation of the Fermi arc and type one Dirac cone boundary states in each step. In (b), we observe that group velocity spans $[-1,1]$. This is due to $V(k_{x})= |n_{z}(k_{x})|$ with highest (lowest) value of group velocity around the gapless energy bands.} \label{Fig5}
\end{figure*}

In the quantum walk with PHS, the boundary states are only Dirac cone and Fermi arc types and the flat band boundary states are absent. If rotation angles are independent of each other, every step includes only one type of boundary states (see Fig. \ref{Fig1}). In contrast, if rotation angles are linear function of each other, in each step, we observe two types of boundary states together (see Fig. \ref{Fig2}) and possible emergences of a characteristic behavior. This characteristic behavior points to a structure for simulated topological phenomena which contains two Dirac cones as exterior (boundary) of the structure and two Fermi arcs with an additional Dirac cone located between them (see $6th$ and $8th$ steps in Fig. \ref{Fig2}). For the Exterior Dirac cones (type one Dirac cone), the energy gap closes only at $k_{x}=0$ and $\pm \pi$ whereas for the interior Dirac cone (type two Dirac cone), the gapless energy bands happen  $k_{x}=0$, $\pm \pi/2$ and $\pm \pi$ (see Fig. \ref{Fig3}). Therefore, we have two types of Dirac cones. 

The group velocity shows specific behaviors around each boundary state and topological phases (see Fig. \ref{Fig3}). In case of the topological phases, the group velocity is smooth function of $k_{x}$. In contrast, at the closing gap, it becomes ill-defined. If the boundary state is Fermi arc type, around the gapless point, the group velocity is a nonlinear function of $k_{x}$ and a sudden swap in its sign happens as we pass through gapless point. In contrast, for Dirac cones, the group velocity is constant around gapless point. For type one Dirac cone boundary state, group velocity is fixed at $1$ around the gapless point and it swaps from $+1$ to $-1$ or the opposite. In case of the type two Dirac cone boundary states, the group velocity remains at $2$ around the gapless point and it swaps from $+2$ to $-2$ or vice versa at gapless point. Therefore, the topological phases around type one Dirac cone boundary state have group velocities about $\approx \pm 1$ while for the type two Dirac cone boundary states, their group velocities rise to about $\approx \pm 2$.  

\begin{figure*}[htb]
	\centering
	{\begin{tabular}[b]{cc}%
		\subfloat[]{\includegraphics[width=0.4\linewidth]{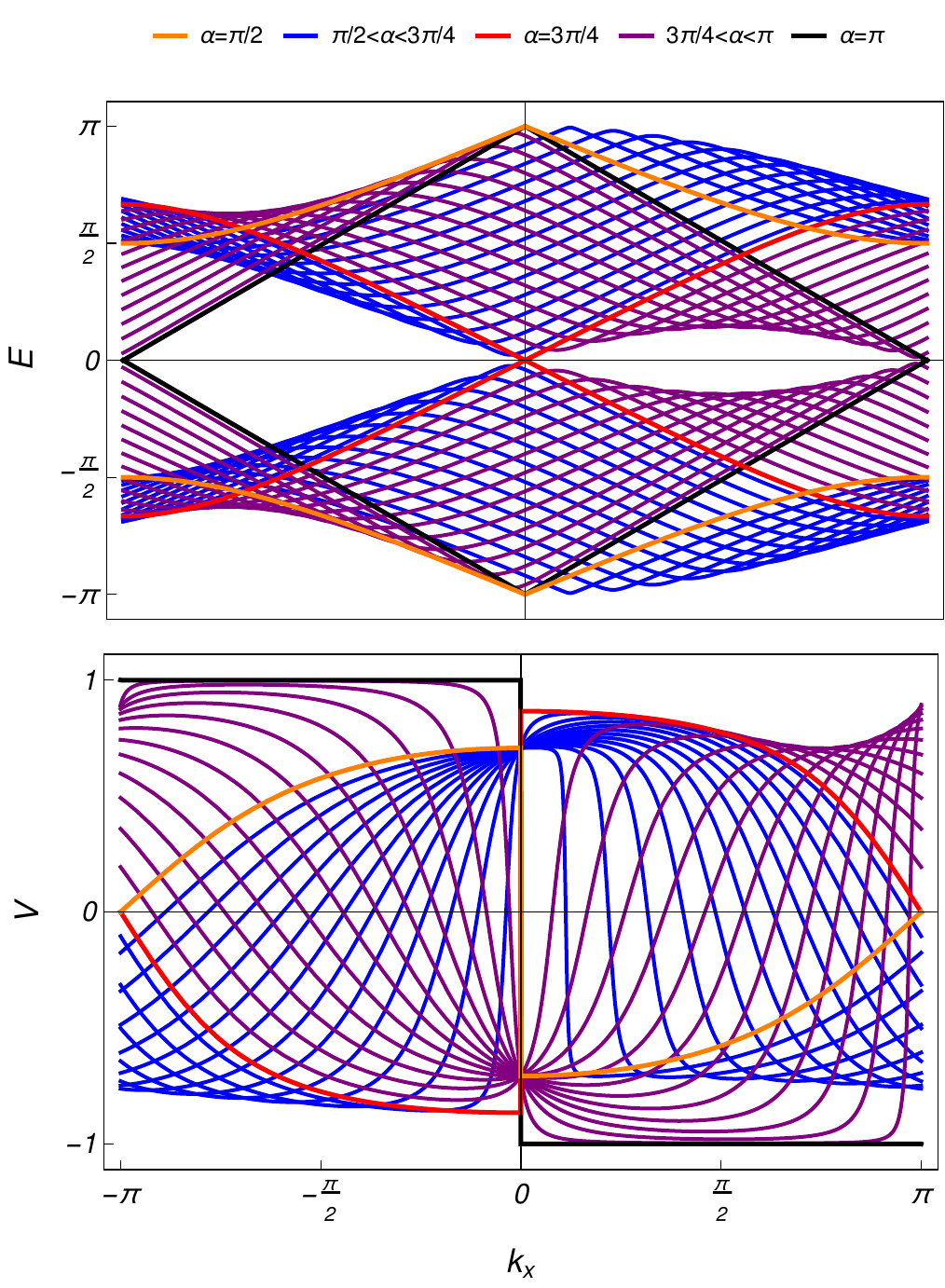}}
		\subfloat[]{\includegraphics[width=0.4\linewidth]{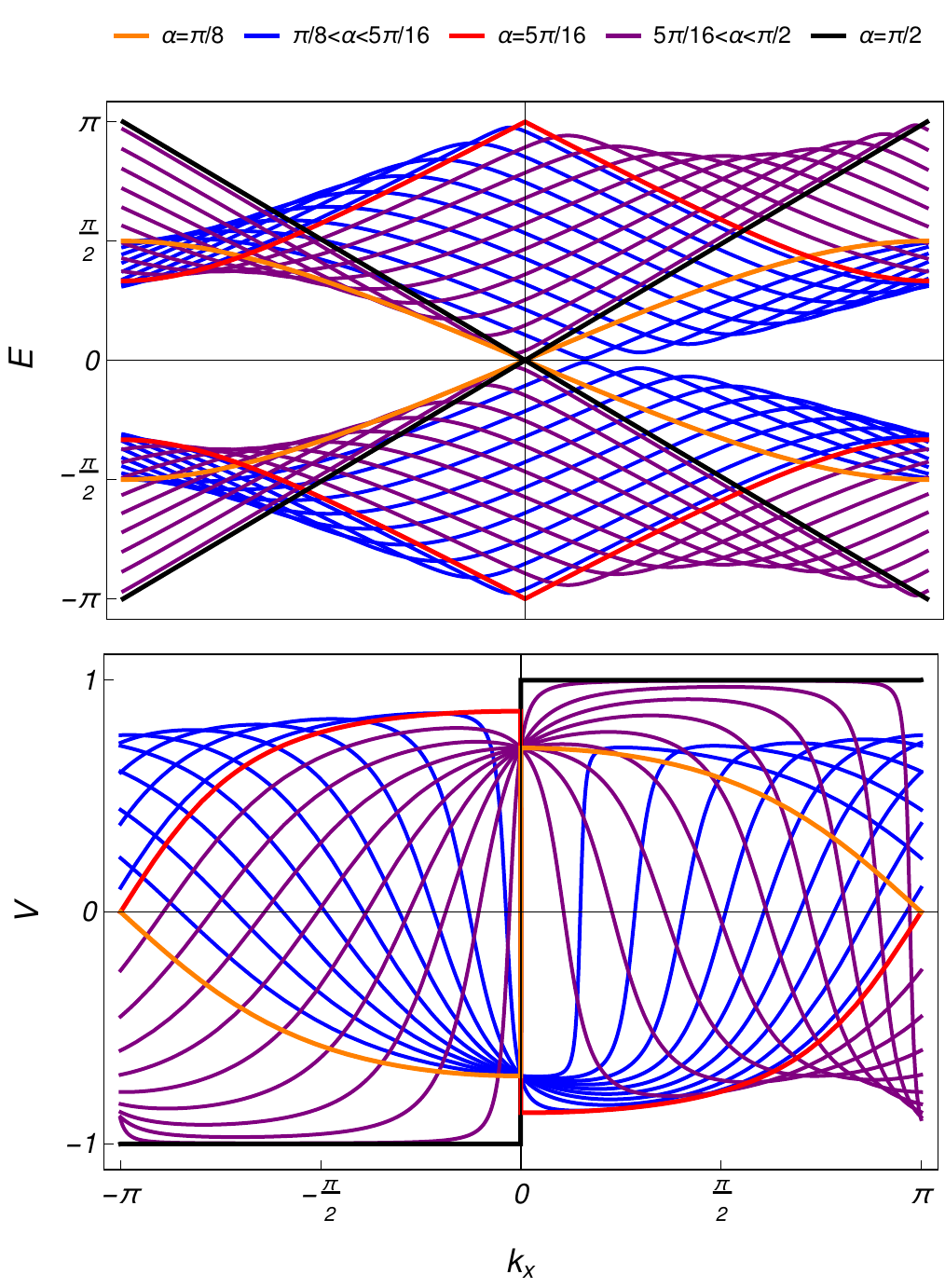}}\\
		\subfloat[]{\includegraphics[width=0.8\linewidth]{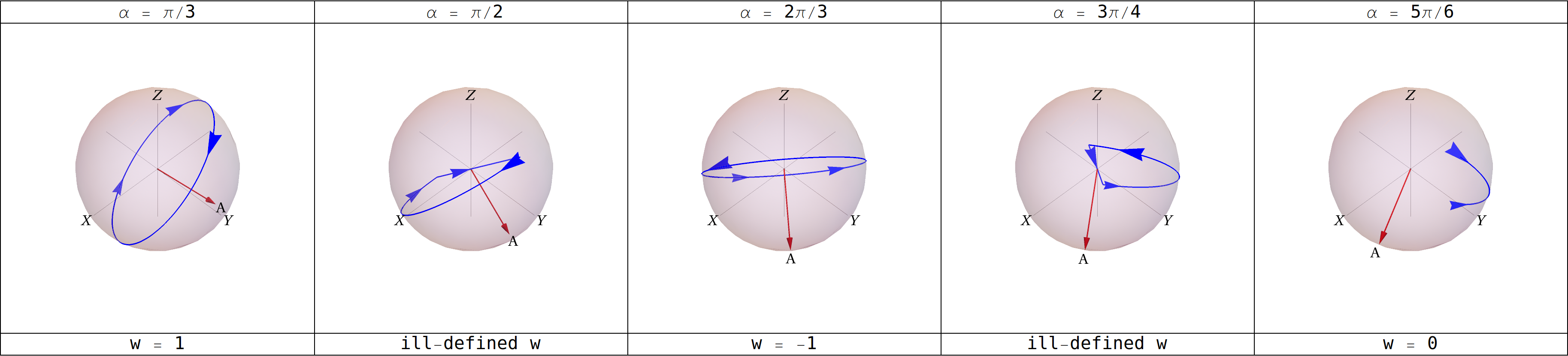}}				
	\end{tabular}}				
	\caption{Quantum walk with only CHS (one-dimensional): Energy (up panels) and group velocity (down panels) (only positive branche) for $6th$ (a) and $8th$ (b) steps with $\beta=(\alpha+\pi)/3$. As we scan rotation angle $\alpha$ through limited range (up panels of (a) and (b), we start with a Fermi arc boundary state which is followed by topological phases (blue lines) until we reach another Fermi arc boundary state. By further increasing the rotation angle, we first come across topological phases and finally meet a Type one Dirac cone boundary state. In down panels of (a) and (b), we observe that highest (lowest) value of group velocity for topological phases is $\approx 1$. In (c), we have modification of $\boldsymbol n$ (blue curves) as $k_{x}$ traverses the first Brillouin zone in a plane orthogonal to $\boldsymbol A$ for $6th$ step with $\beta=(\alpha+\pi)/3$. The closed loops around the origin correspond to nontrivial topological phases in which the clockwise winding direction has $w=-1$ while counterclockwise has $w=1$. If the loop is not closed and does not cross the origin, it indicates a trivial phase with $w=0$. The boundary states are those loops with $\boldsymbol n$ passing the origin.}	\label{Fig6}
\end{figure*}

\begin{figure}[htb]
	\centering
	{\begin{tabular}[b]{cc}%
			\includegraphics[width=0.7\linewidth]{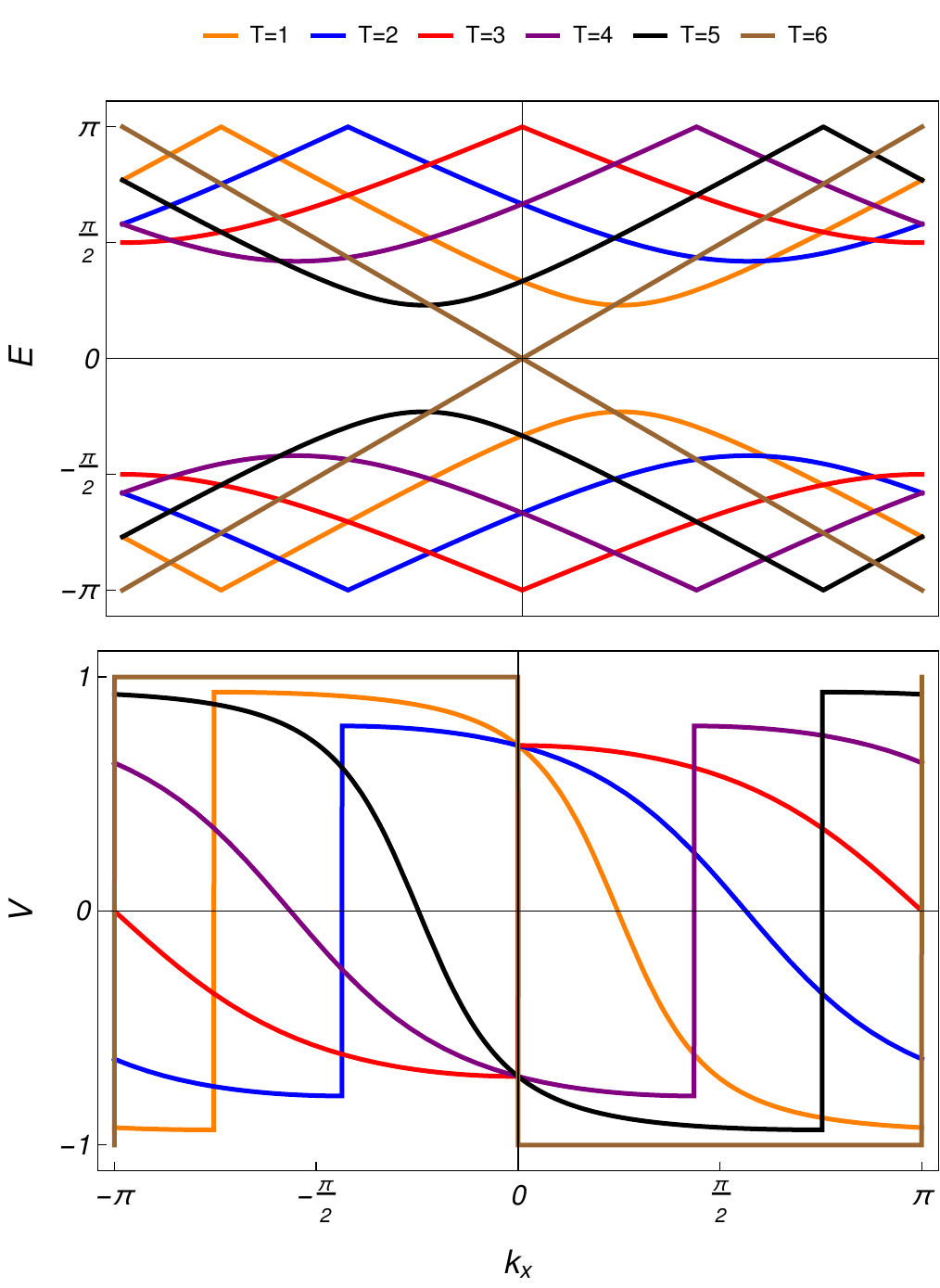}				
		\end{tabular}}				
		\caption{Quantum walk with only CHS (one-dimensional): Energy (up panel) and group velocity (down panel) (only positive branche) with $\alpha=\beta=\pi/3$ for six subsequent steps. The formations of Fermi arcs happen for arbitrary values of $k_{x}$ which step-dependently changes.}	\label{Fig7}
\end{figure}

In the quantum walk with CHS, rotation angles being linear function of each other guarantees the observation of both Dirac cone and Fermi arc boundary states in a single step. Whereas if rotation angles are independent of each other, it is possible to observe both Dirac cone and Fermi arc boundary state at a single step otherwise only Fermi arc boundary states (see Figs. \ref{Fig4} and \ref{Fig5}). The flat band boundary states are absent and Dirac cones are only type one. The characteristic behavior of the group velocity for different types of boundary states and around their gapless points are similar to those described for quantum walk with PHS. On the other hands, contrary to the quantum walk with PHS where the energy gap nonlinearly closes in only $k_{x}=0$ and $\pm \pi$, the Fermi arcs are observed for arbitrary values of $k_{x}$ in quantum walk with CHS (see Figs. \ref{Fig6} and \ref{Fig7}).  

In our previous studies in Refs. \cite{Panahiyan2019,Panahiyan2020}, we showed that simple- and split-step quantum walks with protocols 

\begin{eqnarray}
\widehat{U} & = & \widehat{S}_{\uparrow}(x) \widehat{S}_{\downarrow}(x) \widehat{C}_{y}(\beta),
\\
\widehat{U} & = & \widehat{S}_{\uparrow}(x) \widehat{C}_{y}(\alpha) \widehat{S}_{\downarrow}(x) \widehat{C}_{y}(\beta), \label{GGH}
\end{eqnarray}
have PHS, TRS and CHS with $\widehat{\mathcal{P}}^2=+1$, $\widehat{\mathcal{T}}^2=+1$ and $\widehat{\Gamma}^2=1$. Therefore, these two protocols simulate BDI family of topological phases with integer topological invariant, $\mathbb{Z}$. In the simple-step protocol only Dirac cone boundary states are observed while split-step protocol has all three types of Dirac cone, Fermi arc and flat band boundary states. Therefore, in one-dimensional quantum walk, only the split-step protocol in Eq. \eqref{GGH} can simulate all types of boundary state together. 

The protocol with only PHS \eqref{protocol1} has $\widehat{\mathcal{P}}^2=+1$. This indicates that D family of topological phases can be simulated by this protocol and topological invariant is a binary, $\mathbb{Z}_{2}$. On the other hand, since the TRS is absent for this protocol, we can use doubling procedure introduced in Ref. \cite{Kitagawa} to construct protocol with PHS, TRS and CHS where $\widehat{\mathcal{P}}^2=+1$, $\widehat{\mathcal{T}}^2=-1$ and $\widehat{\Gamma}^2=+1$. To do so, we consider the walker to have two additional flavor index, $A$ and $B$. The protocol of the quantum walk should be diagonal in the flavor index with  

\begin{eqnarray}
\widehat{U} = \begin{pmatrix}
\widehat{U}_{a} & 0 \\
0 & \widehat{U}_{a}^t,
\end{pmatrix} \label{DPHS1}
\end{eqnarray}
in which $\widehat{U}_{a}$ is given by Eq. \eqref{protocol1}, $\widehat{U}_{a}$ acts on flavor $A$ and $\widehat{U}_{a}^t$ acts on flavor $B$. The protocol given in Eq. \eqref{DPHS1} simulates DIII family of topological phases with topological invariant being $\mathbb{Z}_{2}$.

In case of protocol with only CHS \eqref{protocol2}, AIII family of topological phases can be simulated with integer valued winding number (w), $\mathbb{Z}$, being the topological invariant (see Fig. \ref{Fig6}). Since the TRS is absent, we can use the doubling procedure and build the following protocol

\begin{eqnarray}
\widehat{U} = \begin{pmatrix}
\widehat{U}_{b} & 0 \\
0 & \widehat{U}_{b}^t,
\end{pmatrix} \label{DPHS2}
\end{eqnarray} 
where $\widehat{U}_{b}$ is given by Eq. \eqref{protocol2}. The new protocol has PHS, TRS and CHS with $\widehat{\mathcal{P}}^2=-1$, $\widehat{\mathcal{T}}^2=-1$ and $\widehat{\Gamma}^2=+1$ which indicates that it can simulate CII family of the topological phases with topological invariant being integer, $\mathbb{Z}$. 

Using all of the protocols given here, we can simulate different classes of the topological phases that are available in one-dimensional systems (see Table \ref{table}). The main contributions of the step-dependent coins comparing to their step-independent counterparts are high level of control over the simulation of topological phases, boundary states and their numbers, and additionally simulation of different types of boundary states.


\section{Two-dimensional quantum walk} \label{Two dimensions}

In this section, we generalize our one-dimensional quantum walk to two-dimensional case and investigate two protocols each having different symmetry properties. 

\subsection{Split-step quantum walk with only PHS} 

The protocol of such quantum walk was proposed by Kitagawa in Ref. \cite{Kitagawa2012} as  

\begin{eqnarray}
\widehat{U} & = & \widehat{S}_{\uparrow \downarrow}(x) \widehat{C}_{y}(\beta) \widehat{S}_{\uparrow \downarrow}(y) \widehat{C}_{y}(\alpha) \widehat{S}_{\uparrow \downarrow} \widehat{C}_{y}(\beta) \label{protocol3},
\end{eqnarray}
which indicates that each step of the walk comprises of rotation of internal states by $\widehat{C}_{y}(\beta)$, displacement of the walker in both $x$ and $y$ positions by $ \widehat{S}_{\uparrow \downarrow}(x,y)$, another rotation of internal states by $\widehat{C}_{y}(\alpha)$, displacement of walker in $y$ position space by $\widehat{S}_{\uparrow \downarrow}(y)$, again rotation of internal states by $\widehat{C}_{y}(\beta)$ and final displacement in $x$ position space by $\widehat{S}_{\uparrow \downarrow}(x)$. The existence of $\widehat{S}_{\uparrow \downarrow}$ breaks the CHS, and since the matrix elements are real and rotation matices are done in $xy$ plane, the PHS is preserved for this protocol. 

The coin operators are given by Eqs. \eqref{coin1} and \eqref{coin2}. The shift operators are \cite{Kitagawa2012,Panahiyan2020} 

\begin{eqnarray}
\widehat{S}_{\uparrow \downarrow} =  && \ketm{\uparrow} \bram{\uparrow} \otimes \sum_{x,y} \ketm{x+1,y+1} \bram{x,y}+ \notag
\\
 && \ketm{\downarrow} \bram{\downarrow} \otimes \sum_{x,y} \ketm{x-1,y-1} \bram{x,y}\, \label{shift4}
\end{eqnarray}
\begin{eqnarray}
\widehat{S}_{\uparrow \downarrow} (x) = && \ketm{\uparrow} \bram{\uparrow} \otimes \sum_{x,y} \ketm{x+1,y} \bram{x,y} + \notag
\\
 && \ketm{\downarrow} \bram{\downarrow} \otimes \sum_{x,y} \ketm{x-1,y} \bram{x,y}\, , \label{shift5}
\end{eqnarray}
\begin{eqnarray}
\widehat{S}_{\uparrow \downarrow} (y) =  && \ketm{\uparrow} \bram{\uparrow} \otimes \sum_{x,y} \ketm{x,y+1} \bram{x,y} + \notag
\\
 && \ketm{\downarrow} \bram{\downarrow} \otimes \sum_{x,y} \ketm{x,y-1} \bram{x,y}\, , \label{shift6}
\end{eqnarray}
where by using Discrete Fourier Transformation, we can rewrite them as $\widehat{S}_{\uparrow \downarrow} (x,y)= e^{i(k_{x}+k_{y}) \sigma_{z}}$, $\widehat{S}_{\uparrow \downarrow} (x)= e^{ik_{x} \sigma_{z}}$ and $\widehat{S}_{\uparrow \downarrow} (y)= e^{ik_{y} \sigma_{z}}$ where $k_{x}$ and $k_{y}$ traverse the first Brillioun zone. It is a matter of calculation to find eigenvalues of Eq. \eqref{protocol3} which we use to find energy as

\begin{eqnarray}
E(k) & = & \pm\cos^{-1}(\rho), \label{energy3}
\end{eqnarray}
where $\rho= \kappa_{\alpha} [\cos(T \beta)\cos(k_{x})\cos(k_{x}+2k_{y})-\sin(k_{x})\sin(k_{x}+2k_{y})]-\lambda_{\alpha} \sin(T \beta) \cos^2(k_{x})$ and $\pm$ corresponds to existence of two bands of energy. The following could be highlighted about the obtained energy:

I) By setting $\beta=\pm (2c+1) \pi/2T$, the energy bands reduce $E= \pm \cos^{-1}(-\kappa_{\alpha} \sin(k_{x})\sin(k_{x}+2k_{y})-\lambda_{\alpha} \sin(T \beta) \cos^2(k_{x})$. If $\alpha=\pm (2c+1) \pi/T$, then energy bands become independent of $k_{y}$ and nonlinearly close their gaps. In addition, for $\alpha=\beta=\pm 2c\pi/T$, the energy bands would be also nonlinearly closing their gap provided $k_{x}=\pm (2c+1) \pi/2$ $k_{y}=\pm c\pi$. Therefore, for both cases, we have Fermi arc boundary states. 

II) For $\beta=\pm 2c\pi/T$ the energy bands reduce to  $E= \pm \cos^{-1}(\kappa_{\alpha} [\pm \cos(k_{x})\cos(k_{x}+2k_{y})-\sin(k_{x})\sin(k_{x}+2k_{y})])$ which for $\alpha=\pm 2c\pi/T$, and appropriate values of $k_{x}$ and $k_{y}$, energy bands would close their gap linearly, hence the boundary states are of the Dirac cone type. 

Next, we find $\boldsymbol d$ (hence $\boldsymbol n$) and components of group velocity as

\begin{eqnarray}
d_{x}  =  2 \lambda _{\beta } \sin \left(k_x\right) [\kappa _{\alpha } \kappa _{\beta } \cos \left(k_x+2 k_y\right)-\lambda _{\alpha } \lambda _{\beta } \cos \left(k_x\right)], \notag
\end{eqnarray}
\begin{eqnarray}
d_{y}  =  \lambda _{\alpha } \kappa _{\beta }^2-\lambda _{\alpha } \lambda _{\beta }^2 \cos \left(2 k_x\right)+2 \kappa _{\alpha } \kappa _{\beta } \lambda _{\beta } \cos \left(k_x\right) \cos \left(k_x+2 k_y\right), \notag
\end{eqnarray}
\begin{equation}
d_{z}  = \lambda _{\alpha } \kappa _{\beta } \lambda _{\beta } \sin \left(2 k_x\right)-\kappa _{\alpha } [\kappa _{\beta }^2 \sin \left(2 \left(k_x+k_y\right)\right)+\lambda _{\beta }^2 \sin \left(2 k_y\right)]
\end{equation}
\begin{widetext}
\begin{eqnarray}
V(k_{x})= && \pm (1-\rho^2)^{-\frac{1}{2}}\bigg(2\lambda_{\alpha}\sin(T \beta)\sin(k_{x})\cos(k_{x})-\cos(k_{x}+2k_{y})\sin(k_{x})\lambda_{\alpha}[1+\cos(k_{x}+2k_{y})]-
\\ &&  \notag
\kappa_{\alpha} \cos(k_{x})\sin(k_{x}+2k_{y})[1+\cos(T \beta)]\bigg), \notag
\end{eqnarray}
\begin{eqnarray}
V(k_{y})& = & \pm \frac{-2\kappa_{\alpha} \cos(T \beta)\cos(k_{x})\sin(k_{x}+2k_{y})-2\kappa_{\alpha} \sin(k_{x})\cos(k_{x}+2k_{y})}{ \sqrt{1-\rho^2}}. \label{groupv32}
\end{eqnarray}
\end{widetext}

Evidently, the group velocity becomes ill-defined at gapless points of energy bands. Therefore, we can recognize the gapless energy bands by ill-defined group velocity. In two-dimensional systems, $\boldsymbol n$ plays the role of a mapping between an area on a two-dimensional torus and area on Bloch sphere \cite{Kitagawa}. Therefore, the topological invariant is an integer-valued number which counts number of times that $\boldsymbol n$ surrounds the origin as $k_{x}$ and $k_{y}$ traverse the first Brillouin zone. Such topological invariant is known as Chern number. Instead of $\boldsymbol n$, one can also use $\boldsymbol d$ to obtain Chern number \cite{Verga}. We find that in case of boundary states, $\boldsymbol d$ passes the origin as momenta traverse the first Brillouin zone. This characteristic behavior is a mean to recognize boundary state for two-dimensional quantum walk.   

\subsection{Split-step quantum walk without PHS, TRS and CHS} 

In order to remove PHS, TRS and CHS for the protocol of a two-diemnsional quantum walk, we build the protocol as 

\begin{equation}
\widehat{U} = \widehat{S}_{\uparrow \downarrow}(y) \widehat{C}_{y}(\gamma)  \widehat{S}_{\uparrow \downarrow}(x) \widehat{C}_{\nu}(\alpha) \widehat{S}_{\uparrow \downarrow} \widehat{C}_{y} (\beta)  \label{protocol33},
\end{equation}
in which due to $\widehat{C}_{\nu}(\alpha)$, the PHS is broken while $\widehat{S}_{\uparrow \downarrow}$ also omits CHS for the considered protocol. Each step of the walk includes of rotation of internal states by $\widehat{C}_{y} (\beta)$, displacement of the walker in $xy$ position space by $\widehat{S}_{\uparrow \downarrow}$, rotation of internal states and movement of walker in $x$ position space by $ \widehat{C}_{\nu}(\alpha)$ and $\widehat{S}_{\uparrow \downarrow}(x)$, additional rotation of internal states by $\widehat{C}_{y}(\gamma)$ and final displacement of the walker in $y$ position space by $\widehat{S}_{\uparrow \downarrow}(y)$. The coin operator of $\widehat{C}_{y}(\gamma)$ is likewise $\widehat{C}_{y} (\beta)$

Using Discrete Fourier Transformation, we rewrite the shift operators as a function of momenta which span the first Brillouin zone. Consequently, we can find energy as

\begin{eqnarray}
E(k) & = & \pm\cos^{-1}(\rho), \label{energy33}
\end{eqnarray}
where $\pm$ corresponds to existence of two bands of energy and $\rho = \frac{\sqrt{2}\lambda _{\alpha }}{2}[ \kappa _{\beta } \kappa_{\gamma } \sin \left(2 k_x+2 k_y\right)-  \lambda _{\beta } \kappa _{\gamma }- \kappa _{\beta } \lambda _{\gamma } \cos\left(2k_y\right)- \lambda _{\beta } \lambda _{\gamma } \sin \left(2 k_x\right)]+ \kappa _{\alpha }[ \kappa _{\beta } \kappa _{\gamma } \cos \left(2 k_x+2 k_y\right)- \lambda _{\beta } \lambda _{\gamma } \cos \left(2 k_x\right)]$.

The energy bands span $[-\pi,\pi]$ and they close up its gap only at $E=0$, $\pm pi$. The following could be highlighted about the obtained energy:

I) By setting $\alpha=\beta=\gamma=\pm 2c\pi/T$, the energy bands become linear function of the momenta. Therefore, the energy bands close their gap linearly and the boundary states are Dirac cone type. 

II) For $\beta=\pm 2c\pi/T$ with $\alpha=\gamma=\pm (2c+1)\pi/T$, the energy bands become independent of $k_{y}$ and linear function of the $k_{x}$.

III) Finally, if $\gamma=\pm 2c\pi/T$ with $\alpha=\beta=\pm (2c+1)\pi/T$, the energy bands will be independent of momenta and therefore, we will have flat band energy bands with $E=\pm \pm/4$.

Next, we can calculate $\boldsymbol d$ and components of group velocity as 

\begin{widetext} 
\begin{eqnarray}
d_{x} = && \frac{\sqrt{2}\lambda _{\alpha }}{2}[\kappa _{\beta } \lambda _{\gamma } \cos \left(2 k_x\right)- \lambda _{\beta } \kappa _{\gamma } \cos \left(2 k_x+2 k_y\right)-\lambda _{\beta } \lambda_{\gamma } \sin \left(2 k_y\right)]+
\kappa _{\alpha }[\lambda_{\beta } \kappa _{\gamma } \sin \left(2 k_x+2 k_y\right)- \kappa _{\beta } \lambda _{\gamma } \sin \left(2 k_x\right)], \notag
\end{eqnarray}
\begin{eqnarray}
d_{y}  = && \frac{\sqrt{2} \lambda _{\alpha }}{2}[\kappa _{\beta } \kappa _{\gamma }+\kappa _{\beta } \lambda _{\gamma } \sin \left(2 k_x\right)+\lambda _{\beta } \kappa _{\gamma } \sin \left(2 k_x+2 k_y\right)- \lambda _{\beta } \lambda _{\gamma } \cos \left(2 k_y\right)]+\kappa _{\alpha }[\kappa _{\beta } \lambda_{\gamma } \cos \left(2 k_x\right)+\lambda _{\beta } \kappa _{\gamma } \cos \left(2 k_x+2k_y\right)], \notag
\end{eqnarray}
\begin{equation}
d_{z} = \frac{\sqrt{2} \lambda _{\alpha }}{2}[ \kappa _{\beta } \kappa _{\gamma } \cos \left(2 k_x+2 k_y\right)+ \kappa _{\beta }	\lambda _{\gamma } \sin \left(2 k_y\right)+ \lambda _{\beta } \lambda _{\gamma } \cos \left(2 k_x\right)]-\kappa _{\alpha } [ \lambda _{\beta } \lambda _{\gamma } \sin \left(2 k_x\right)+ \kappa _{\beta } \kappa _{\gamma } \sin \left(2 k_x+2 k_y\right)]
\end{equation}
\begin{eqnarray}
V(k_{x})= && \pm \frac{ 2 \kappa _{\alpha } \lambda _{\beta } \lambda _{\gamma } \sin \left(2 k_x\right)+ \sqrt{2} \lambda _{\alpha } \kappa _{\beta } \kappa _{\gamma } \cos \left(2 k_x+2 k_y\right)-2 \kappa _{\alpha }\kappa _{\beta } \kappa _{\gamma } \sin \left(2 k_x+2 k_y\right)-\sqrt{2} \lambda _{\alpha } \lambda _{\beta } \lambda _{\gamma } \cos \left(2 k_x\right)}{ \sqrt{1-\rho^2}}, 
\end{eqnarray}
\begin{eqnarray}
V(k_{y}) &=& \pm \frac{2 \sqrt{2} \lambda _{\alpha } \kappa _{\beta } \kappa _{\gamma } \cos \left(2 k_x+2 k_y\right)-4 \kappa _{\alpha } \kappa _{\beta } \kappa _{\gamma } \sin \left(2 k_x+2 k_y\right)+2 \sqrt{2} \lambda _{\alpha }\kappa _{\beta } \lambda _{\gamma } \sin \left(2 k_y\right)}{ \sqrt{1-\rho^2}}. \label{groupv332}
\end{eqnarray}
\end{widetext}

Both of the group velocity and $\boldsymbol n$ becomes ill-defined as energy bands close their gaps. Therefore, characteristically, we can find boundary states by ill-defined group velocity and $\boldsymbol n$ or if $\boldsymbol d$ passes origin as momenta traverse first Brillouin zone.

\subsection{Discussion on results of two-dimensional quantum walks}

\begin{figure*}[htb]
	\centering
	{\begin{tabular}[b]{cc}%
			\sidesubfloat[]{
				\includegraphics[width=0.3\linewidth]{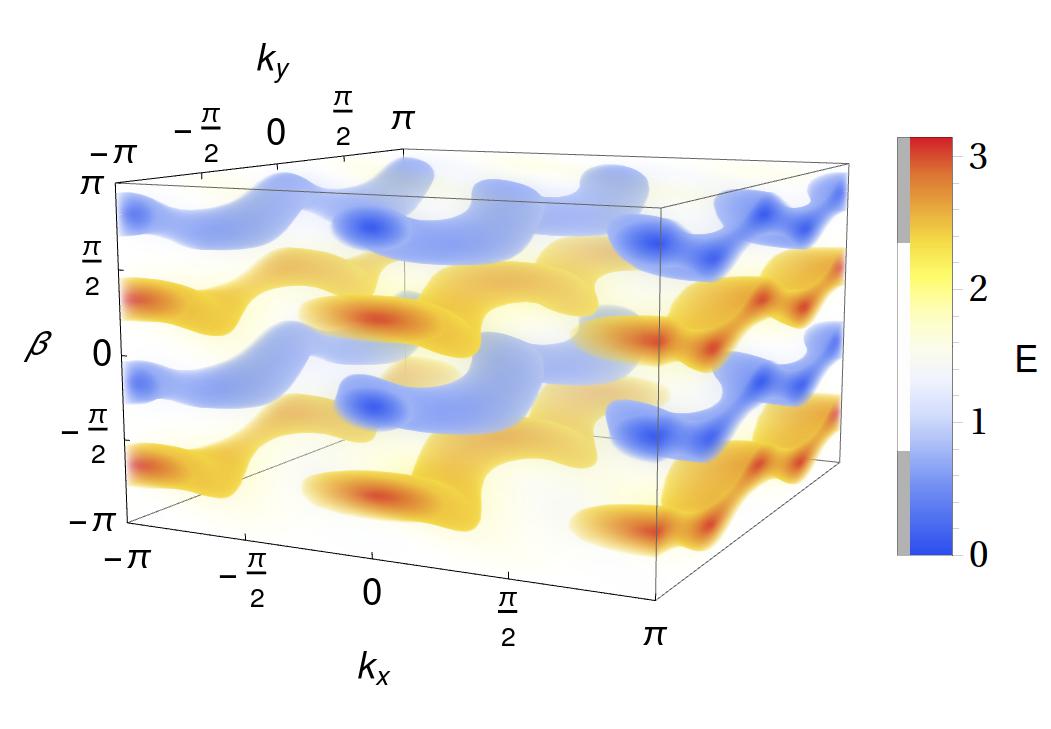}
				\includegraphics[width=0.3\linewidth]{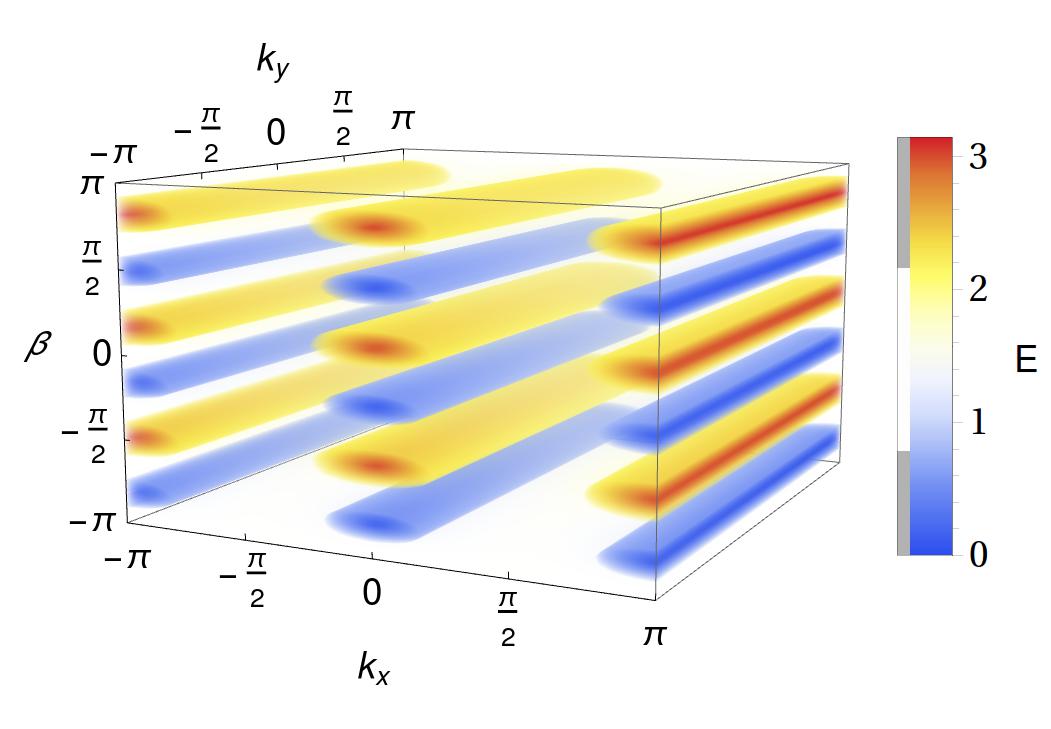}
				\includegraphics[width=0.3\linewidth]{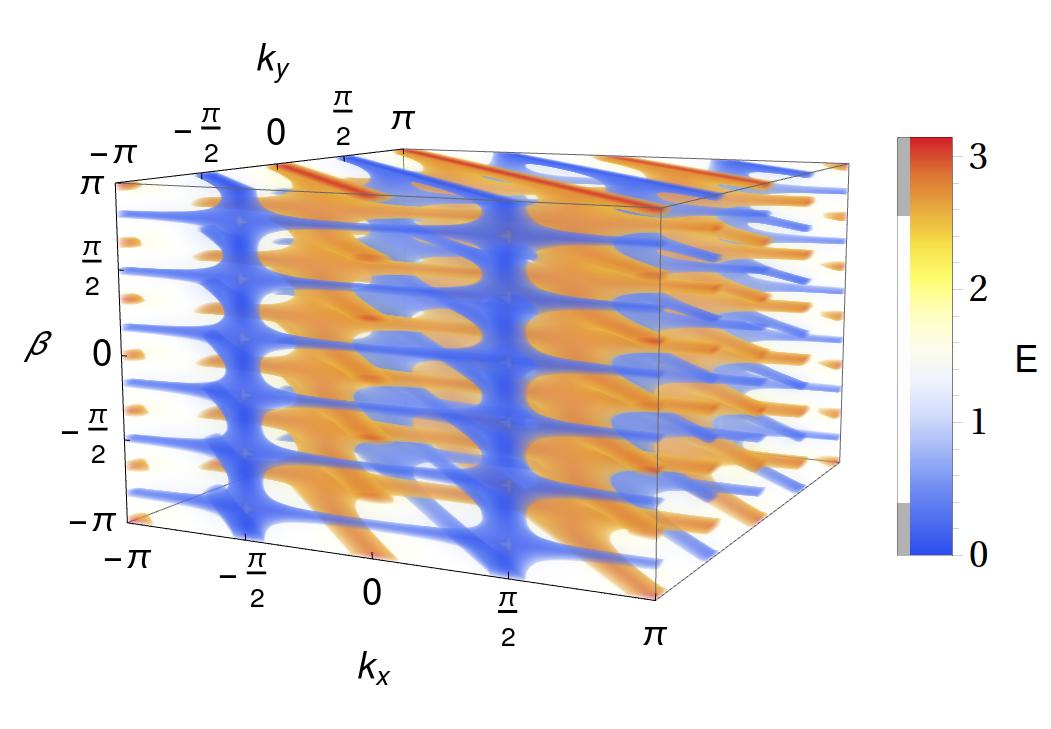}
			}\\			
			\sidesubfloat[]{
				\includegraphics[width=0.3\linewidth]{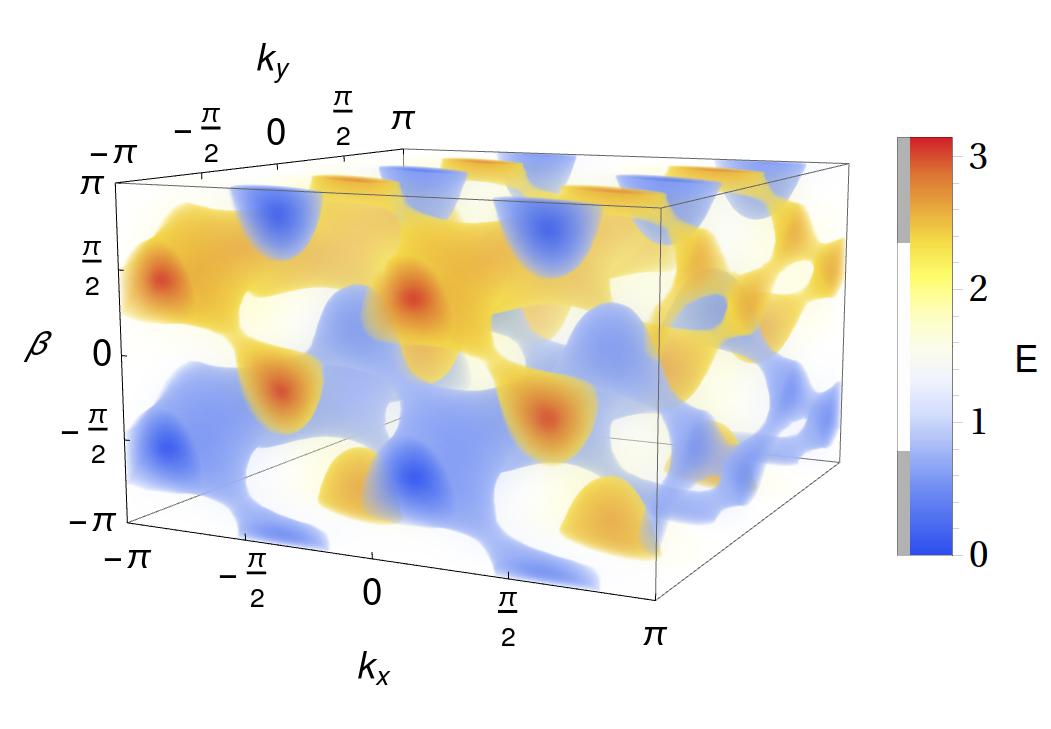}
				\includegraphics[width=0.3\linewidth]{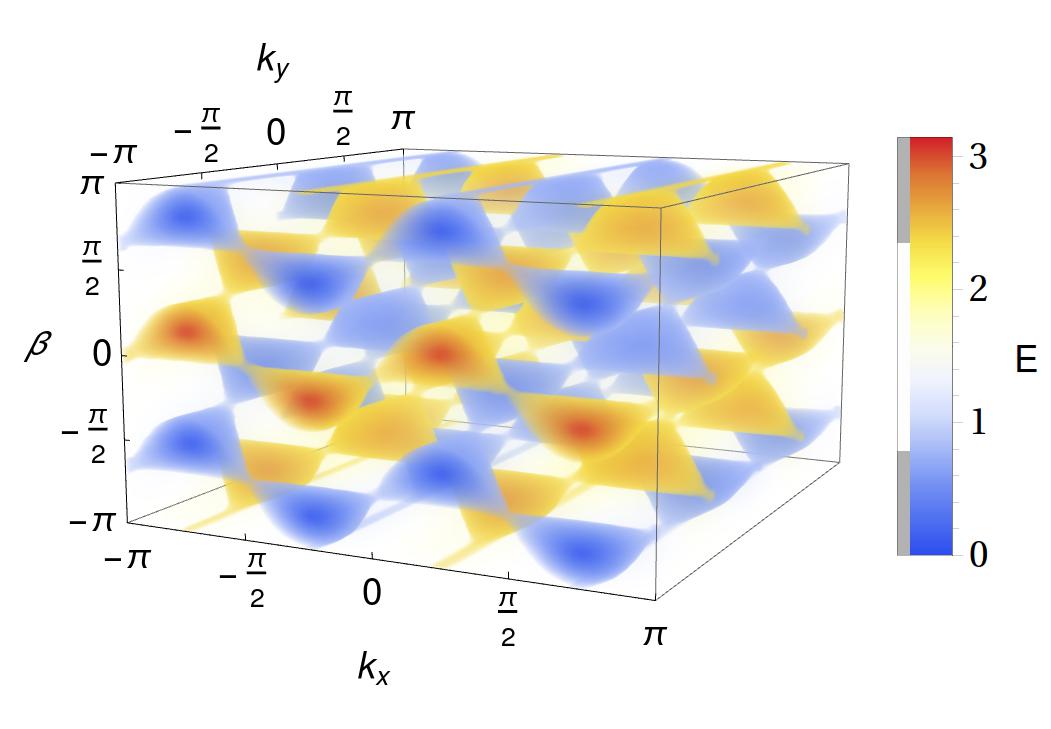}
				\includegraphics[width=0.3\linewidth]{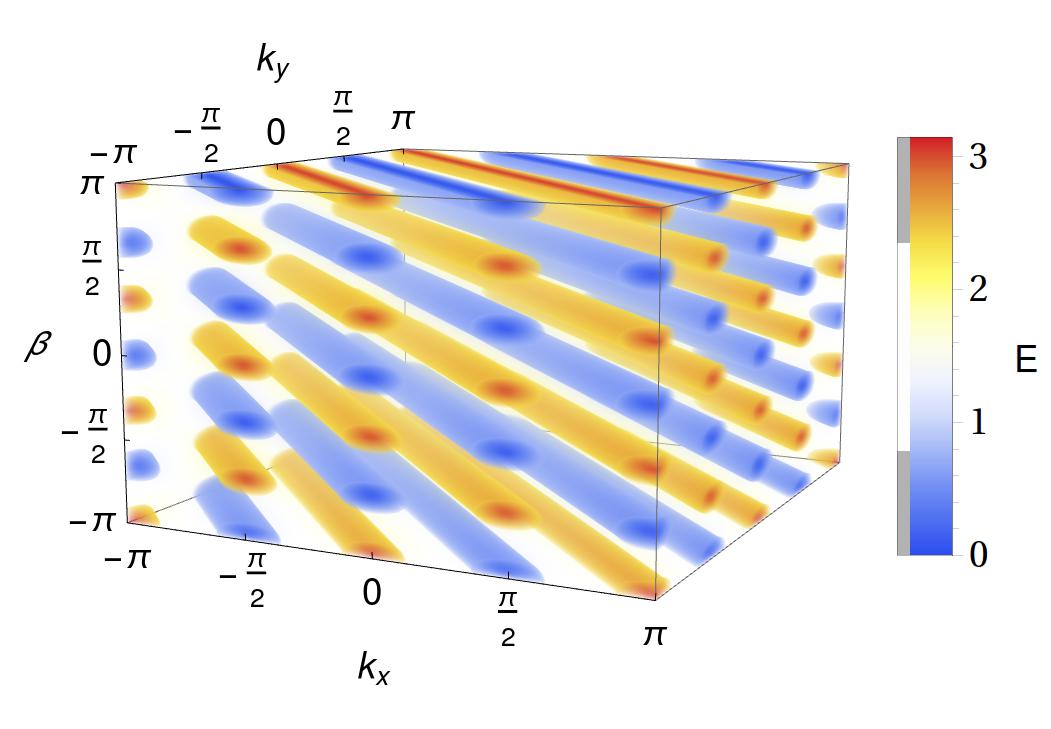}				
			}							
	\end{tabular}}				
	\caption{Quantum walk in two dimensions: Modification of energy (only positive branch) with $\alpha=\pi/3$ and $\gamma=\pi/4$ for subsequents steps of $T=2$, $3$ and $6$ from left to right. In (a), we have quantum walk with only PHS while in (b), the quantum walks without three symmetries are presented. We observe that structure of the simulated topological phenomena changes step-dependently and from one protocol to another one. The populations of topological phases and boundary states (topological phase transitions) increases as quantum walks proceed. In $3th$ step of the quantum walk with only PHS, we have type one Dirac cone dge states while for the $6th$ of the other protocol, we have type two Dirac cone boundary states. Usually, each step contains both Dirac cones and Fermi arcs with Characteristical flat band boundary states. } \label{Fig8}
\end{figure*}

The dynamicality of properties of the simulated topological phenomena that was reported for one-dimensional quantum walks is also present for the protocols of the two-dimensional cases. This indicates that topological phases and boundary states, their types and numbers are subject to step number of the quantum walk. As the quantum walks proceed, the population of the topological phases and boundary states increase (see Fig. \ref{Fig8}). Therefore, the step number can be used to engineer the structure of the simulated topological phenomena. For example, we can tune the parameters of the quantum walk so the simulated boundary states would be type one Dirac cone (see $3th$ step in (a) of Fig. \ref{Fig8}) or type two Dirac cone (see $6th$ step in (a) of Fig. \ref{Fig8}).

\begin{figure*}[htb]
	\centering
	{\begin{tabular}[b]{cc}%
		\sidesubfloat[]{
			\includegraphics[width=0.22\linewidth]{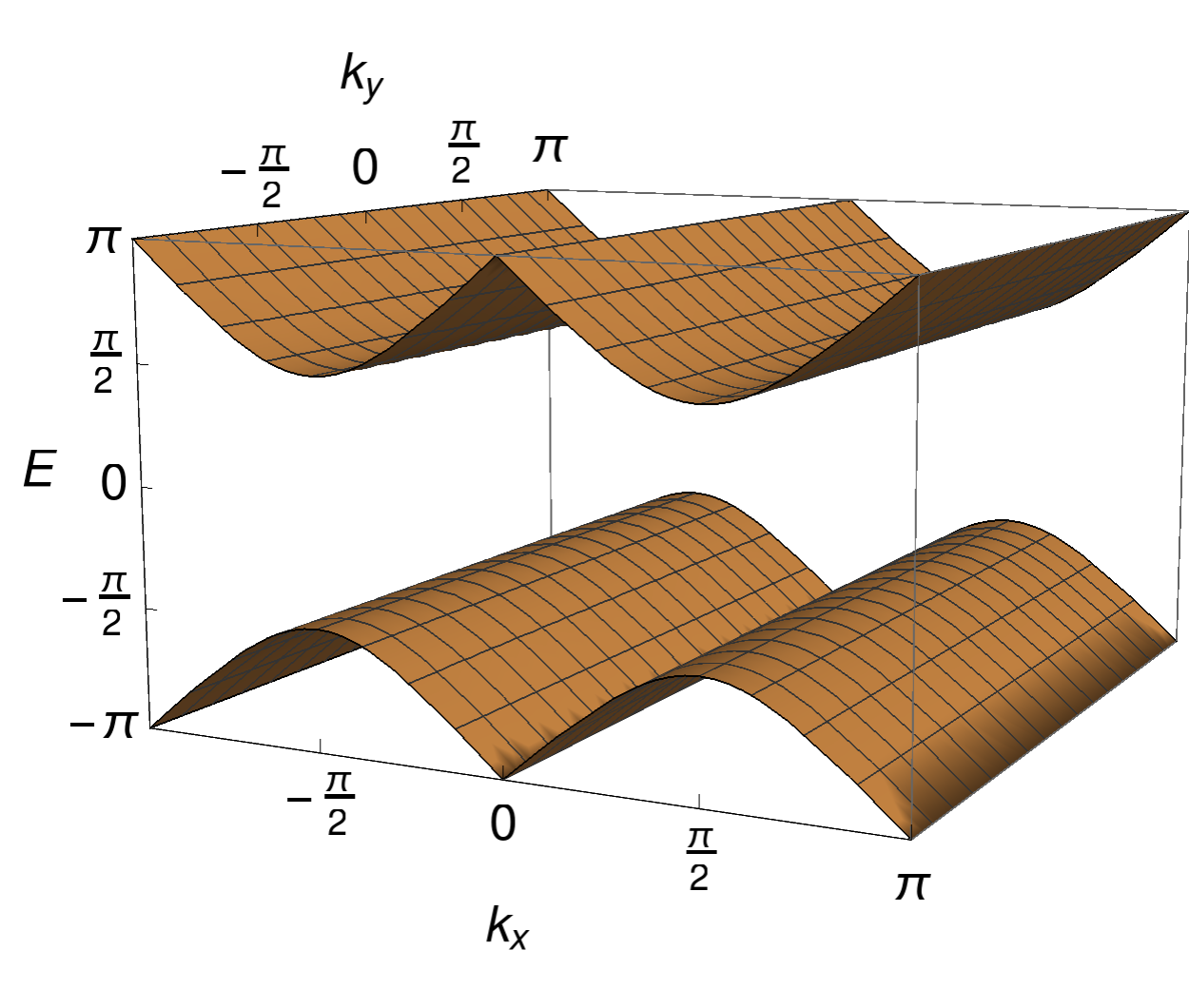}
			\includegraphics[width=0.22\linewidth]{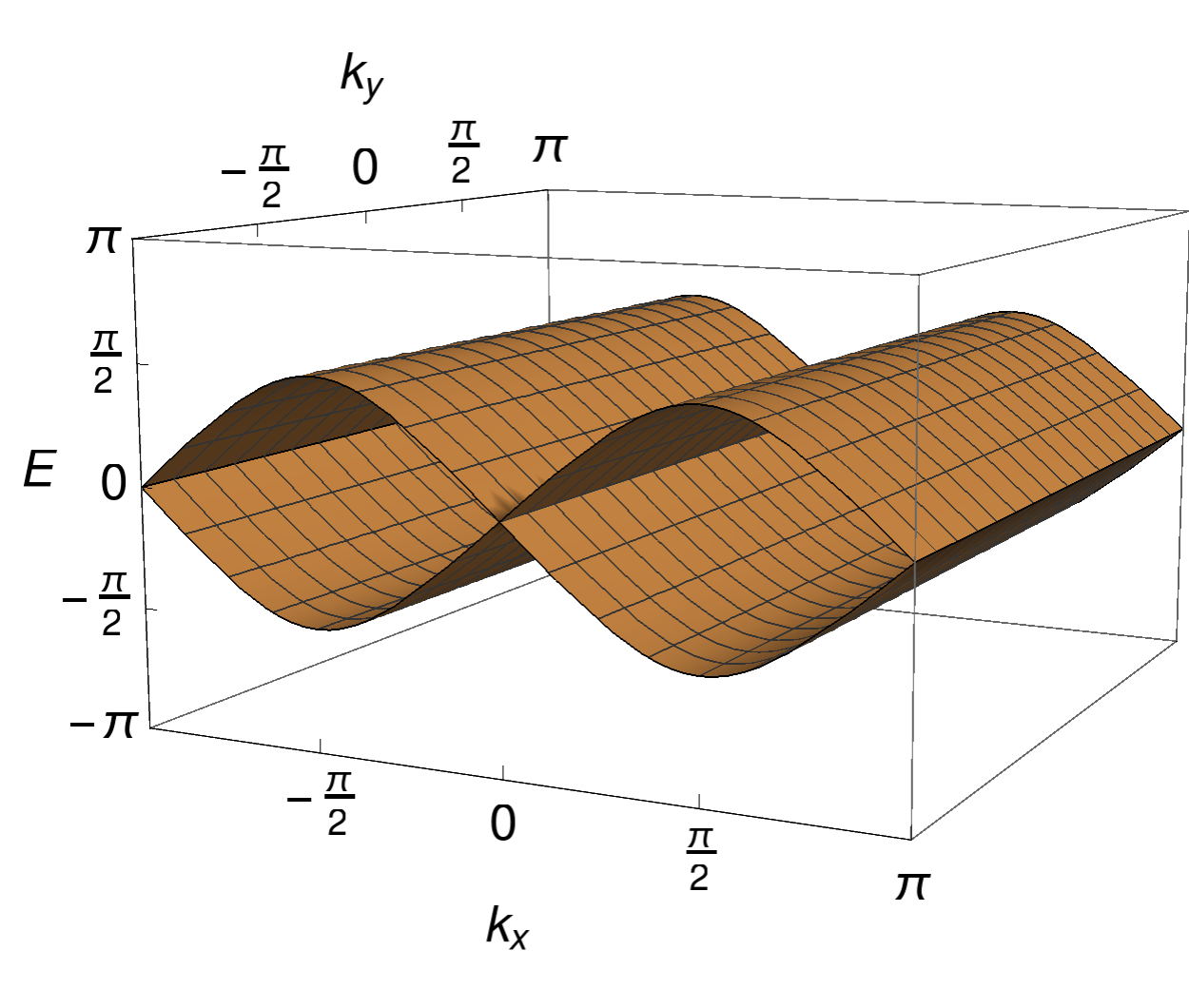}
		}			
		\sidesubfloat[]{
			\includegraphics[width=0.22\linewidth]{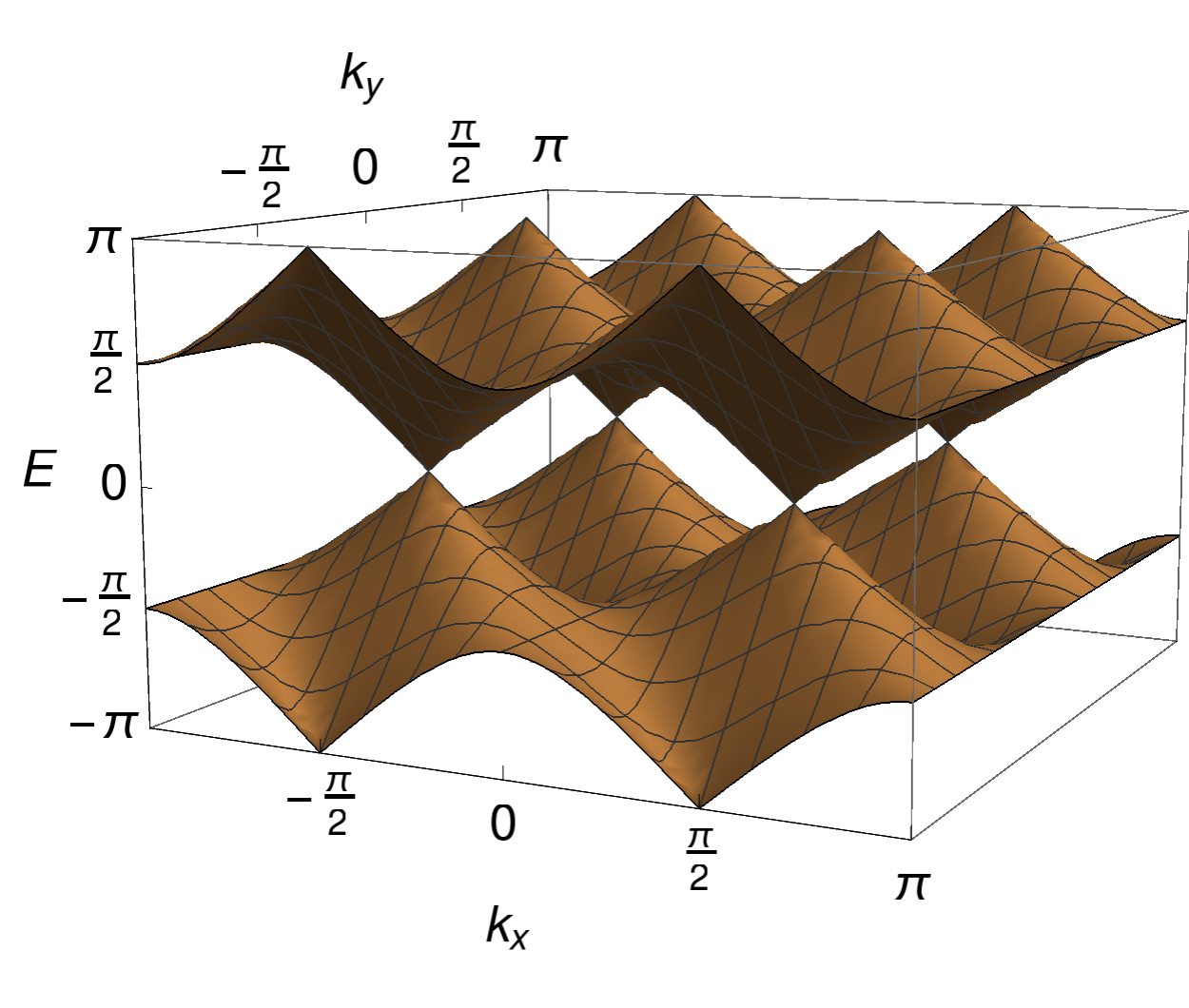}
			\includegraphics[width=0.22\linewidth]{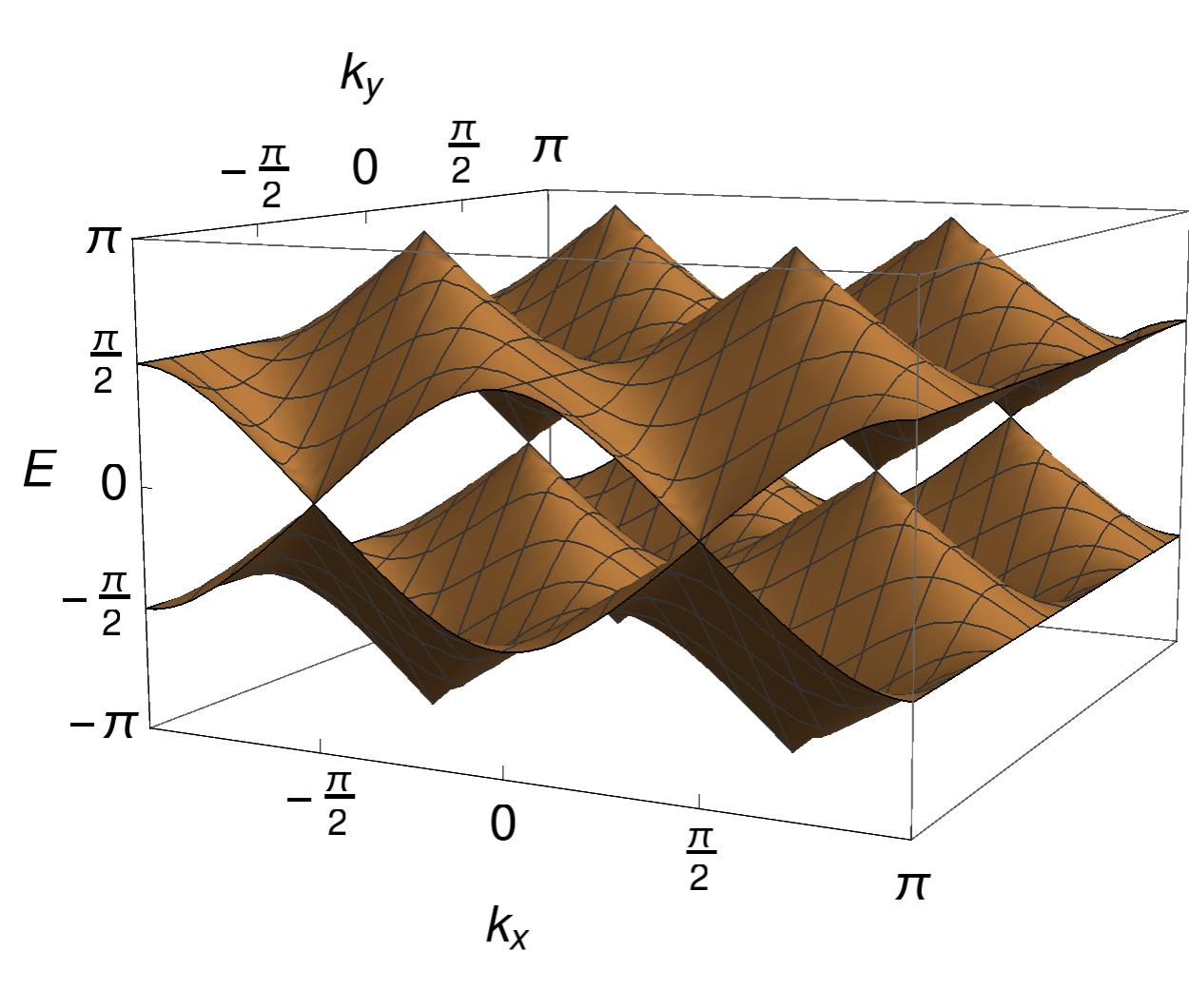}
		}	
		\\[0.0001cm]
		\sidesubfloat[]{
			\includegraphics[width=0.22\linewidth]{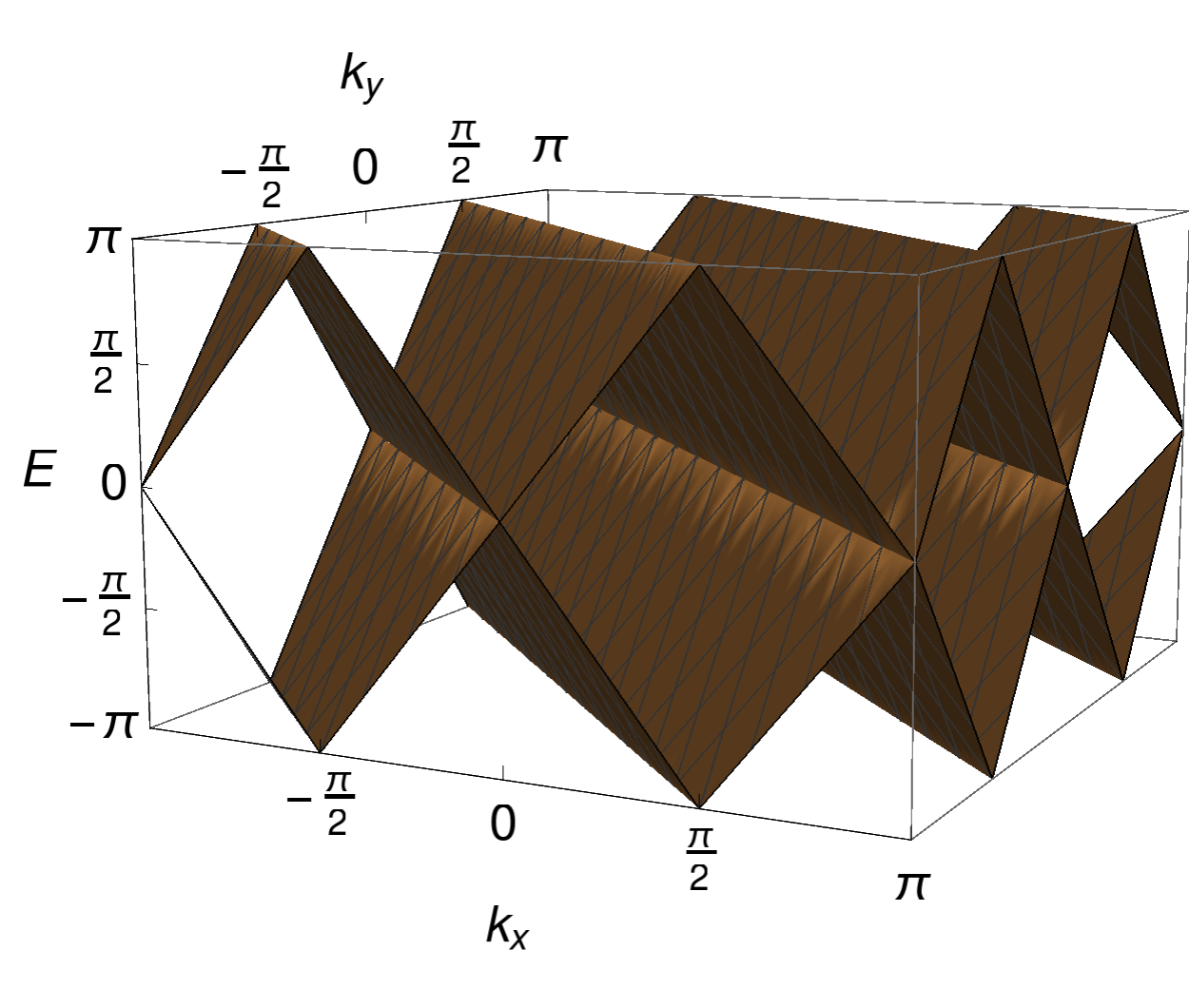}
			\includegraphics[width=0.22\linewidth]{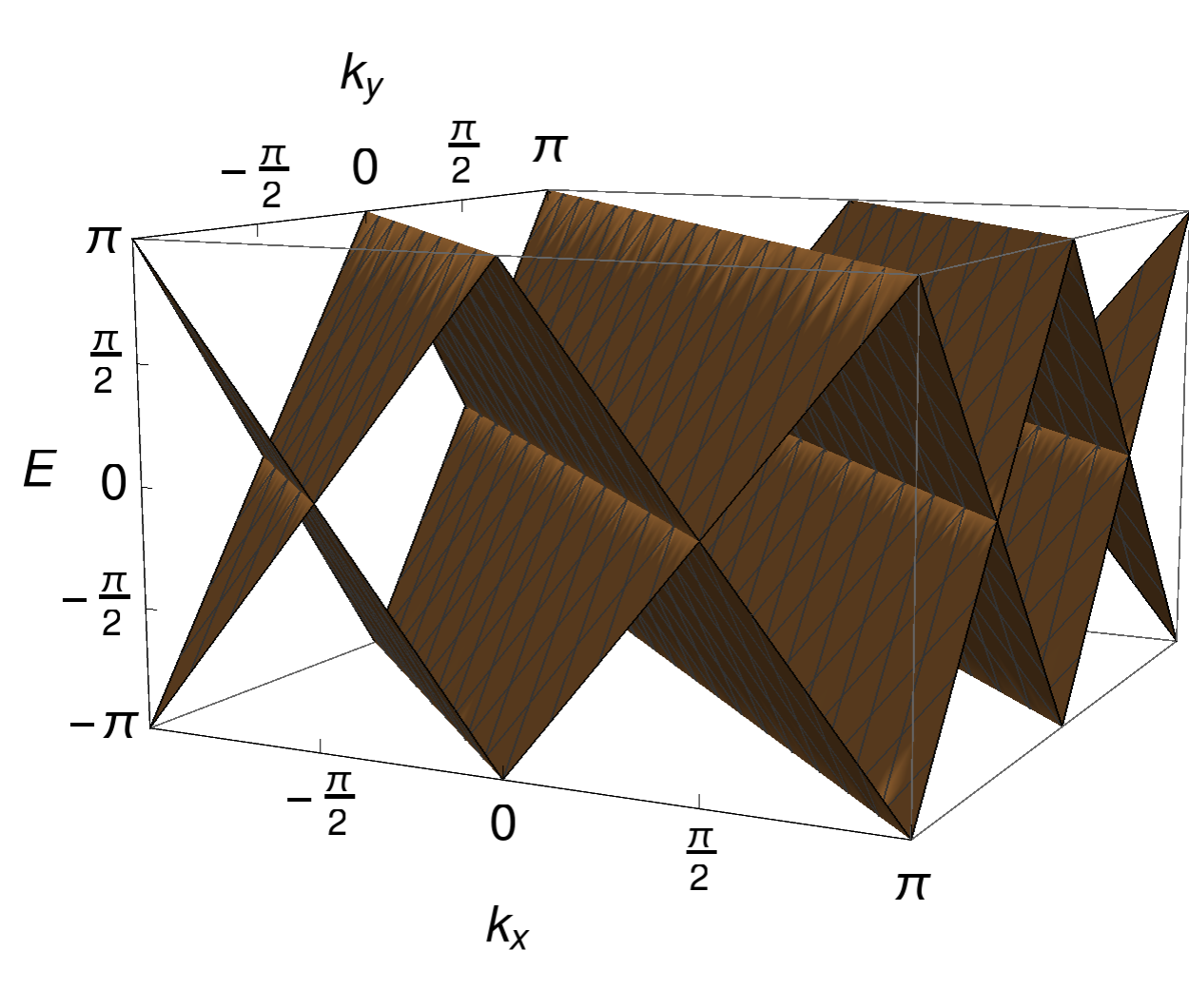}
			\includegraphics[width=0.22\linewidth]{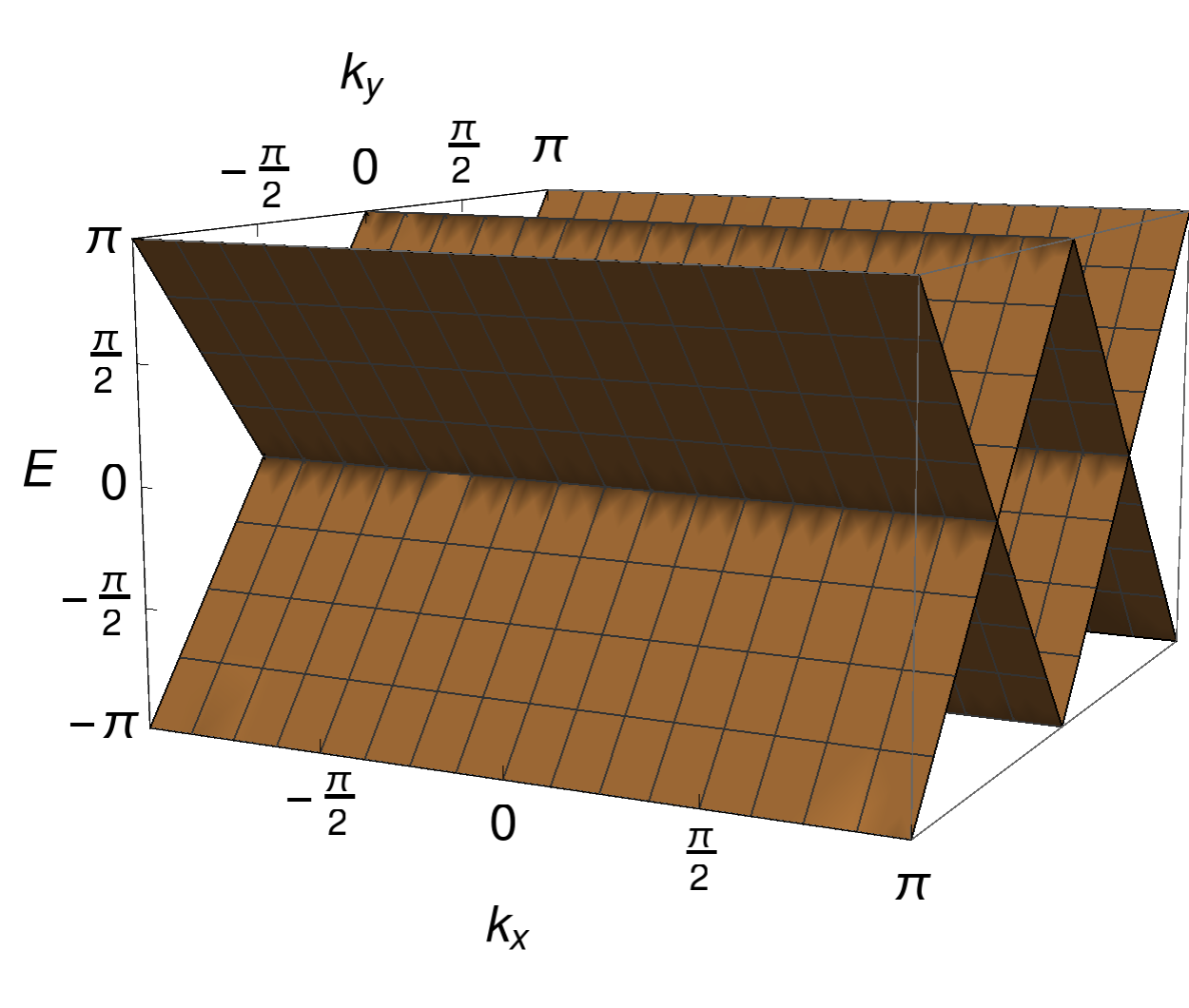}
			\includegraphics[width=0.22\linewidth]{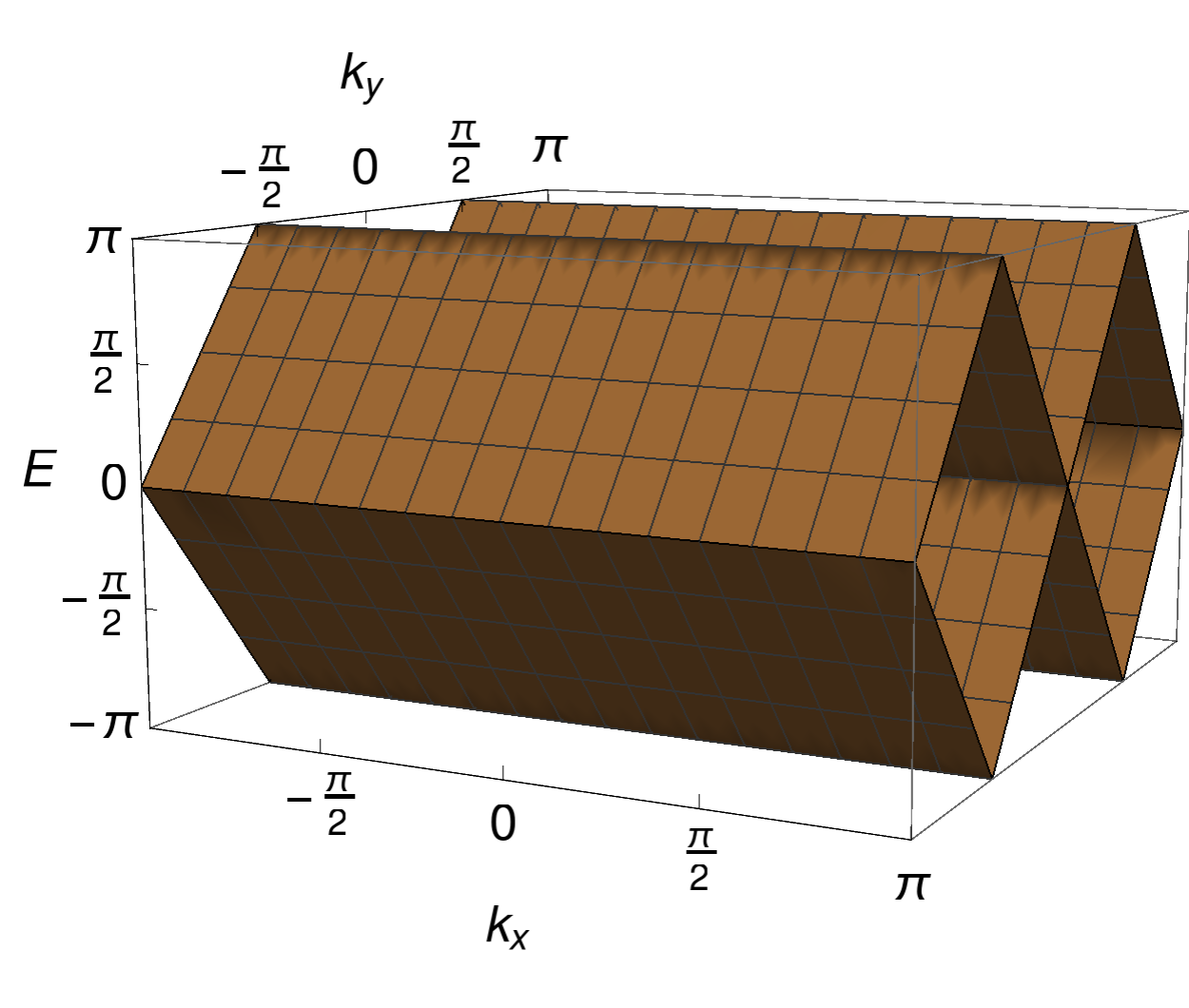}
		}
		\\[0.0001cm]
		\sidesubfloat[]{
			\includegraphics[width=0.22\linewidth]{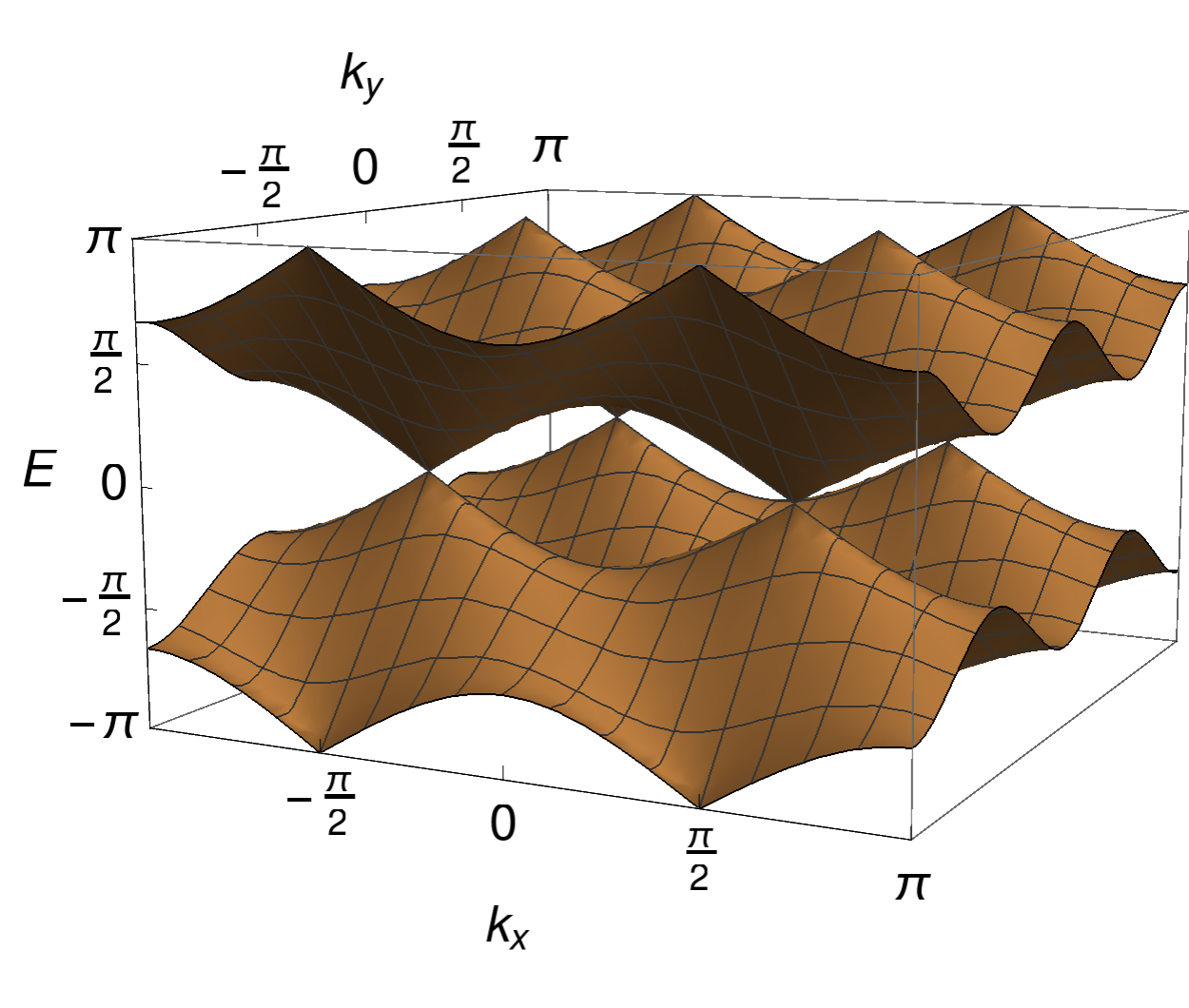}
			\includegraphics[width=0.22\linewidth]{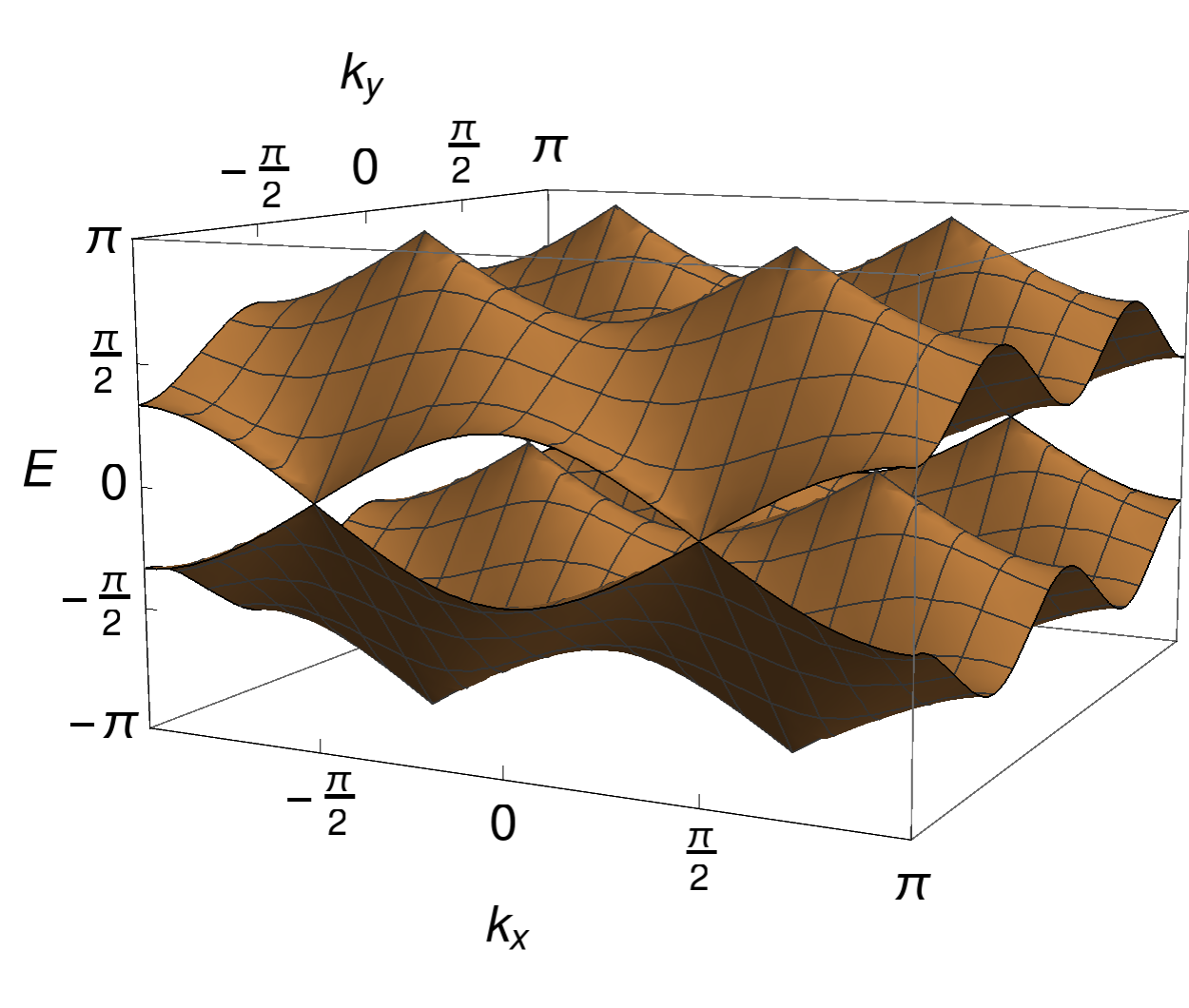}			
			\includegraphics[width=0.22\linewidth]{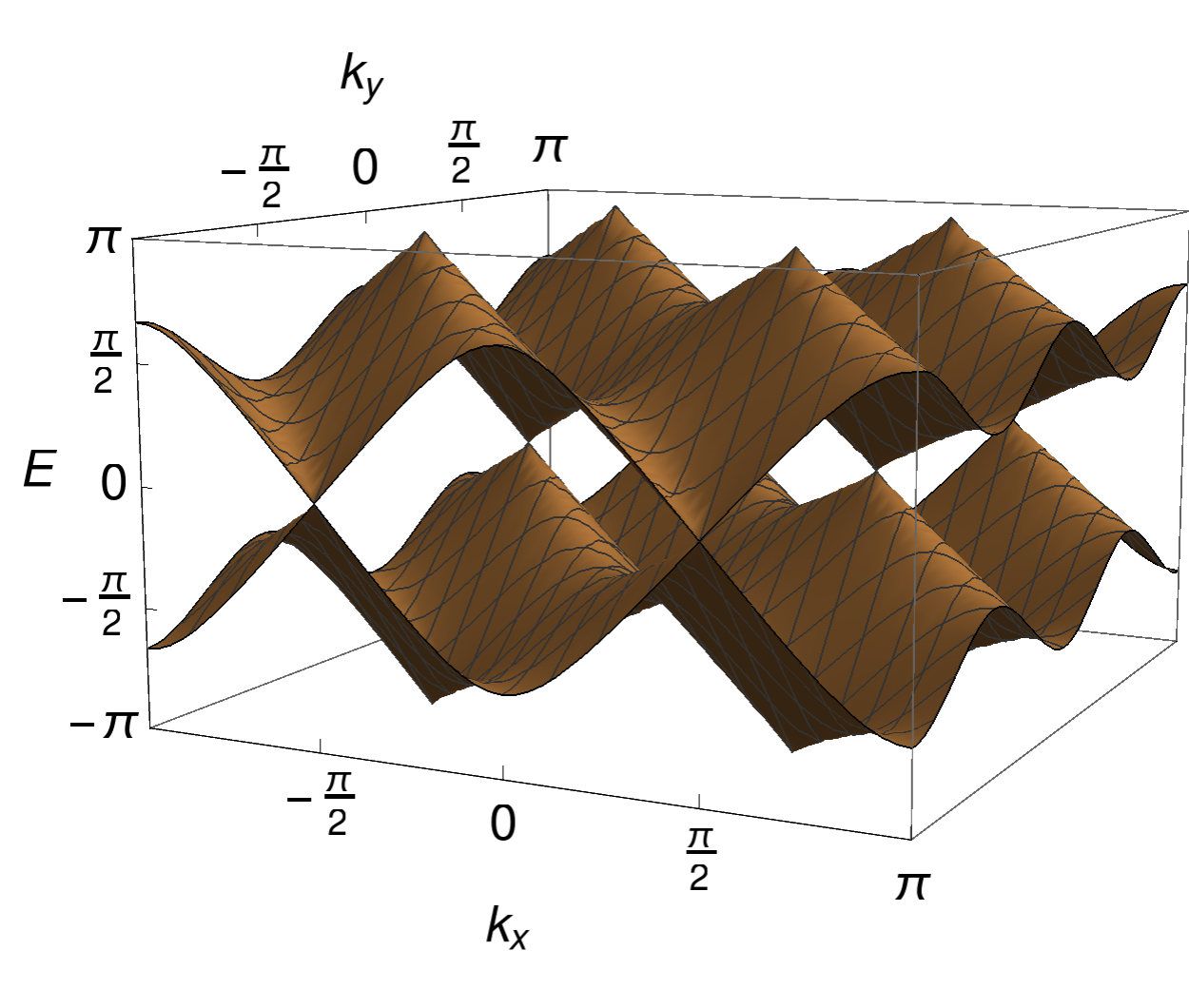}
			\includegraphics[width=0.22\linewidth]{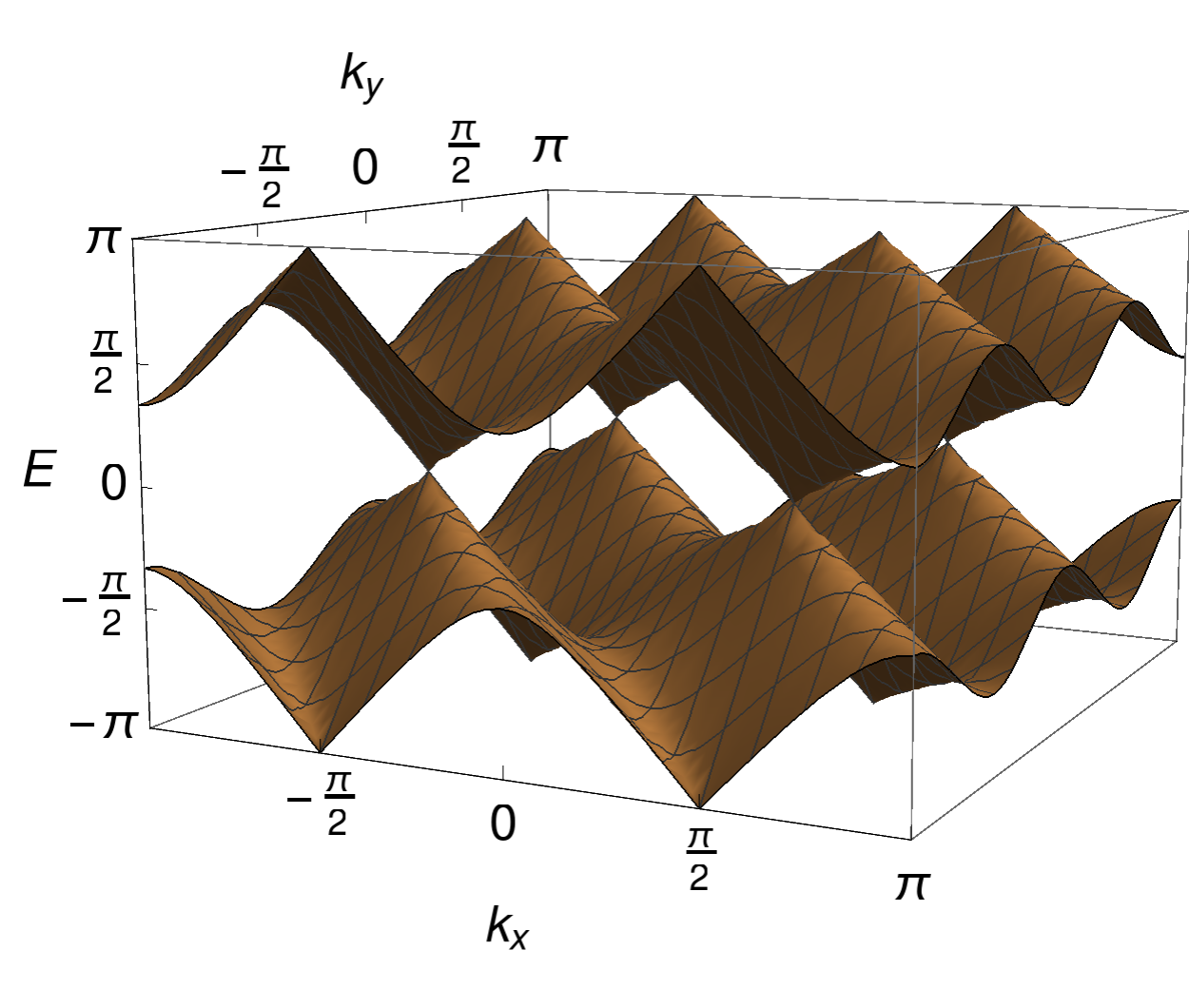}
		}
		\\[0.0001cm]
		\sidesubfloat[]{
			\includegraphics[width=0.22\linewidth]{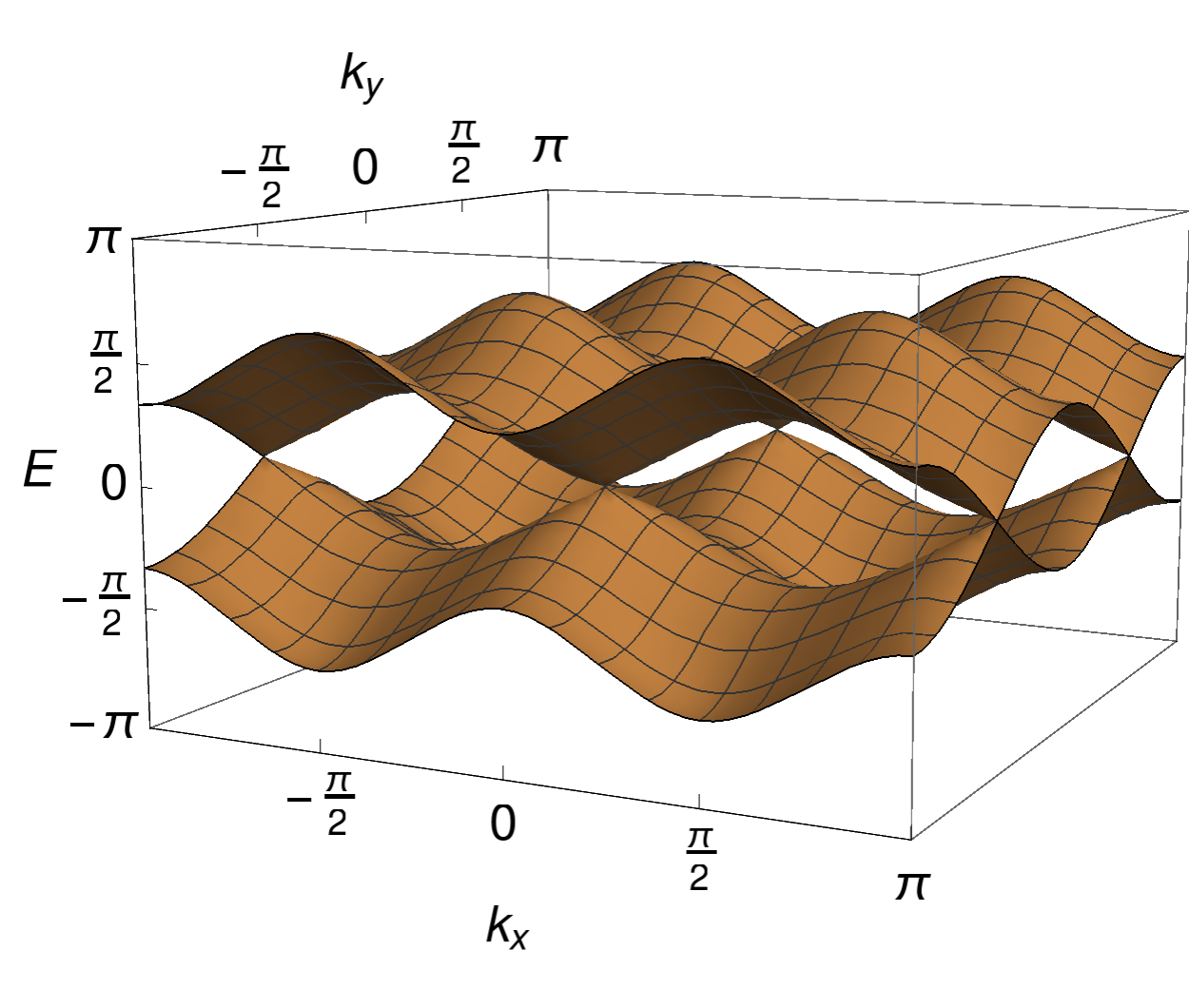}
			\includegraphics[width=0.22\linewidth]{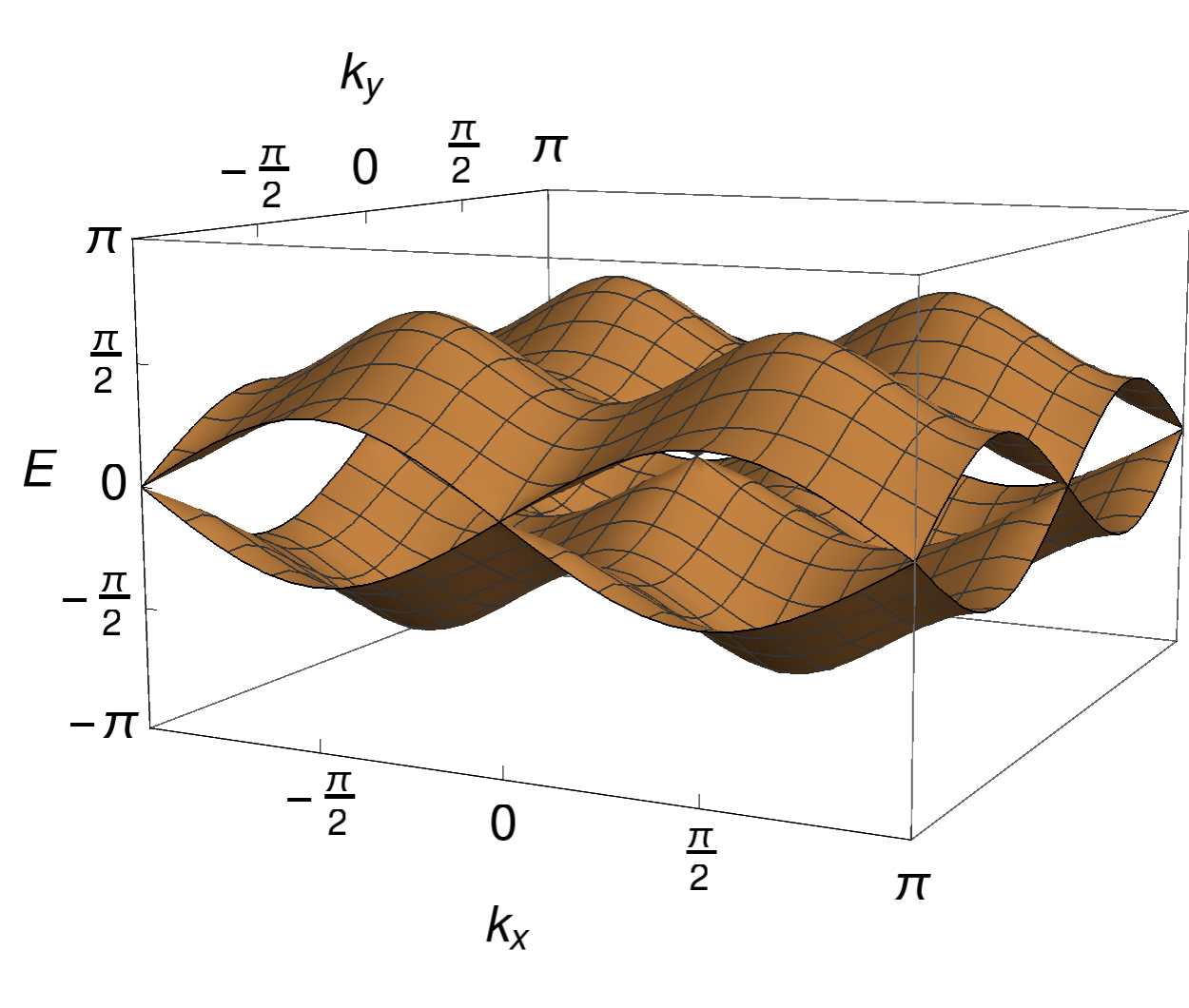}
			\includegraphics[width=0.22\linewidth]{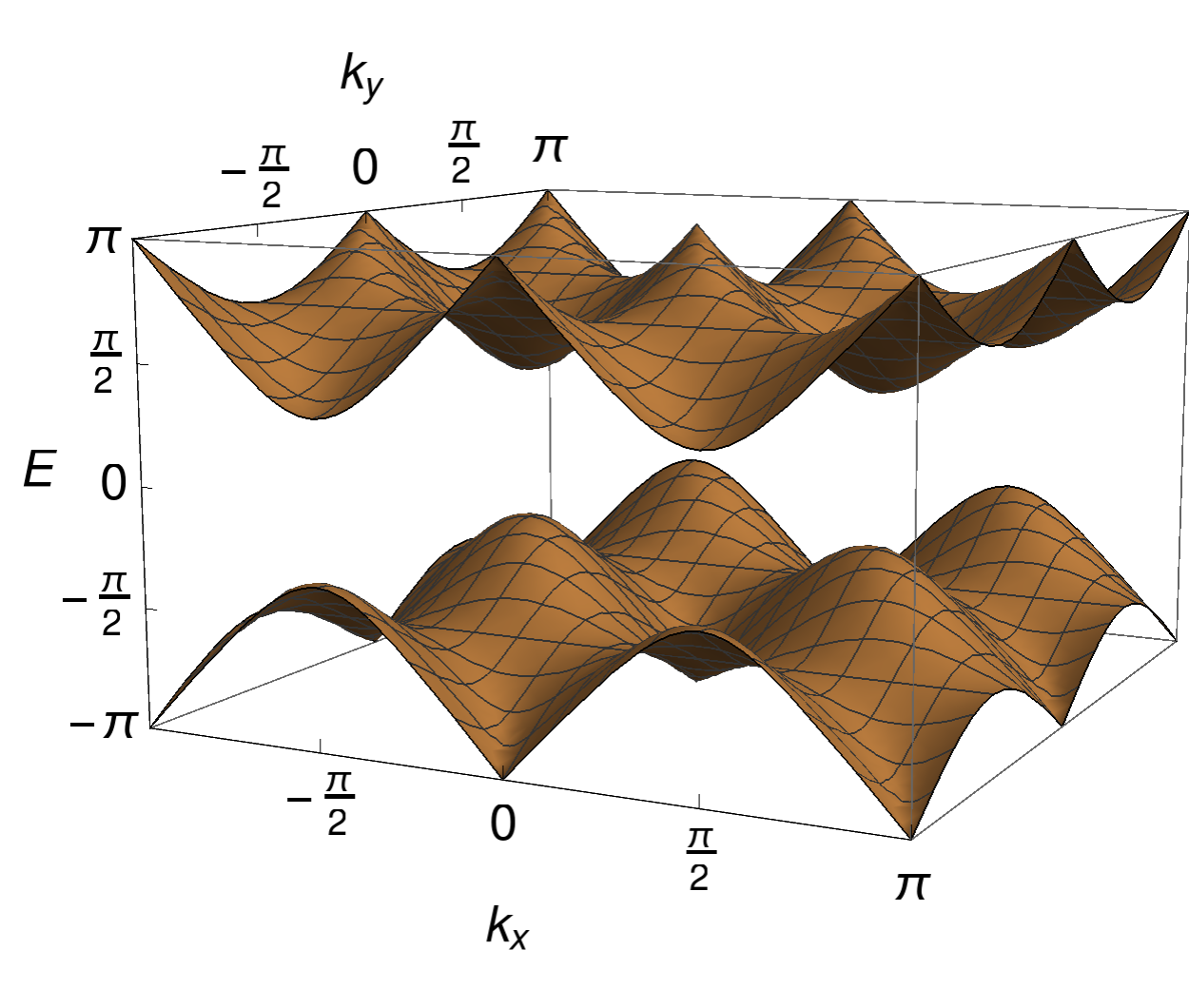}			
			\includegraphics[width=0.22\linewidth]{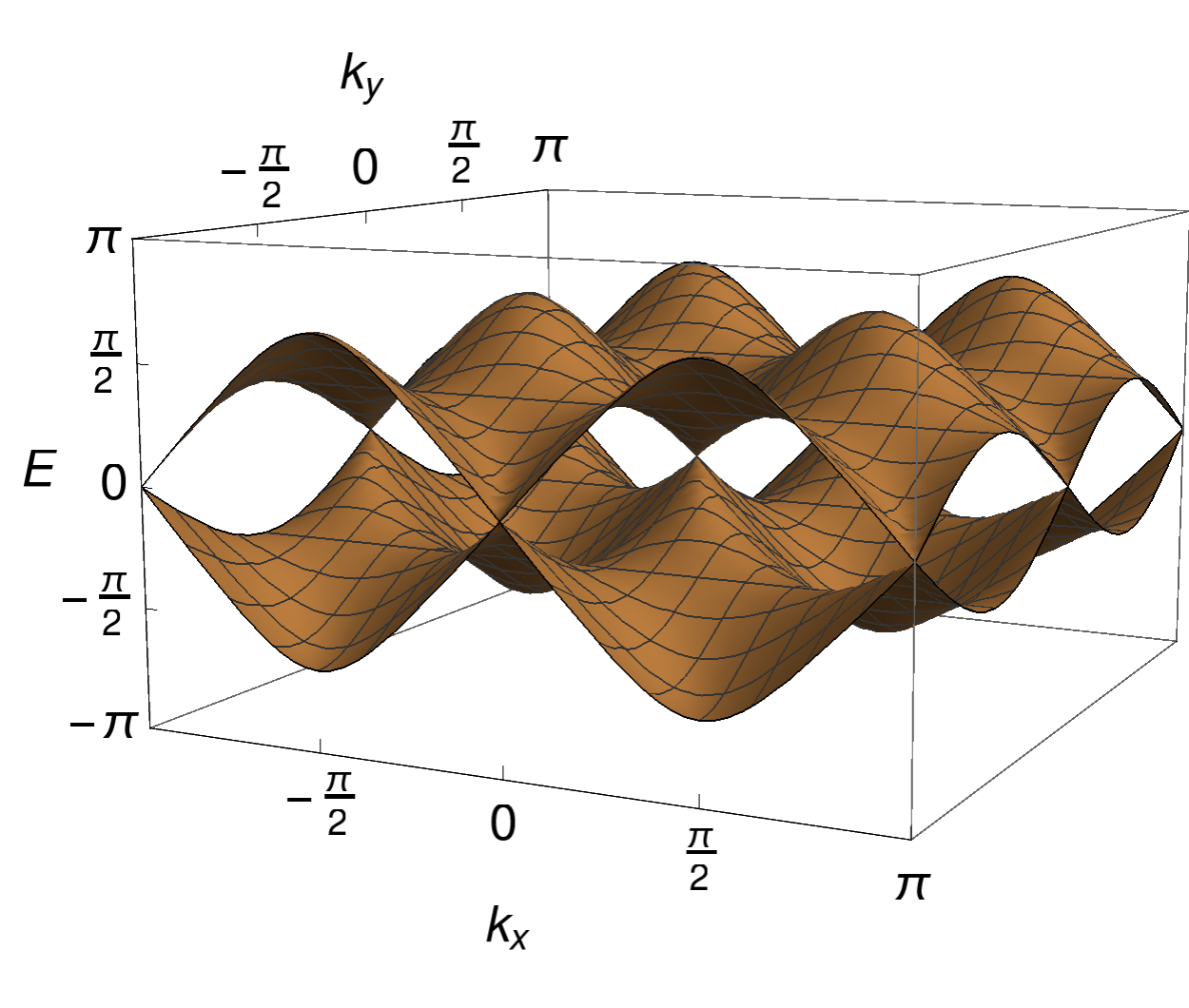}
		}						
	\end{tabular}}				
	\caption{Quantum walk with only PHS (two-dimensional): Energy as a function of rotation angle and momenta for a) $\alpha=\pm (2c+1)\pi/T$ and $\beta=\pm (2c+1)\pi/2T$, b) $\alpha=\pm 2c\pi/T$ and $\beta=\pm (2c+1)\pi/2T$, c) $\alpha=\pm 2c\pi/T$ and $\beta=\pm c\pi/T$, d) $\alpha=\pm 2c\pi/T$ and $\beta=(6c \pm 1)\pi/3T$ and e) $\alpha=\beta=(6c \pm 1)\pi/3T$. In (a), the energy bands are independent of $k_y$. If we fix $k_y$, the observable boundary state is Fermi arc while for variation of $k_y$, we notice boundary states in form lines which could be interpreted as flat band boundary states. In (b), (d) and (e), we explore different ways that Fermi arc boundary states can be formed. It should be noted that for these cases, only Fermi arc boundary states are available for the energy bands. In (c), we observe two type of boundary states: Dirac cone boundary states (type two) if we fix one of the momenta and let the other one varies, Characteristical flat band boundary states of both of the momenta are traversing the first Brillouin zone. Two right panels in (c) present bands of energy independent of $k_x$.} \label{Fig9}
\end{figure*}

Generally, in each step, by tuning the rotation angles alongside of step number, we can make the energy bands independent of one of the momenta (see Fig. \ref{Fig9}) or linear function of them. In such cases, we can simulate three types of the boundary states; Fermi arcs (see (a) in Fig. \ref{Fig9}) and Dirac cone boundary states (see (c) Fig. \ref{Fig9}) are observed if we fix one of the momenta and let the other one traverses the first Brillouin zone. In contrast, if we let both of the momenta varies through the first Brillouin zone, the resultant boundary state has the geometry of a line which is characteristically a type of flat band boundary state. If energy bands are nonlinear functions of both of the momenta, the energy bands only nonlinearly close their gap which indicates that boundary states are only Fermi arcs (see (b), (d) and (e) in Fig. \ref{Fig9}). It should be noted that although we have plotted Fig. \ref{Fig9} for two-dimensional quantum walk with only PHS, the same could be done for the protocol given in Eq. \eqref{protocol33}. 

\begin{figure*}[htb]
	\centering
	{\begin{tabular}[b]{cc}%
		\sidesubfloat[]{
			\includegraphics[width=0.23\linewidth]{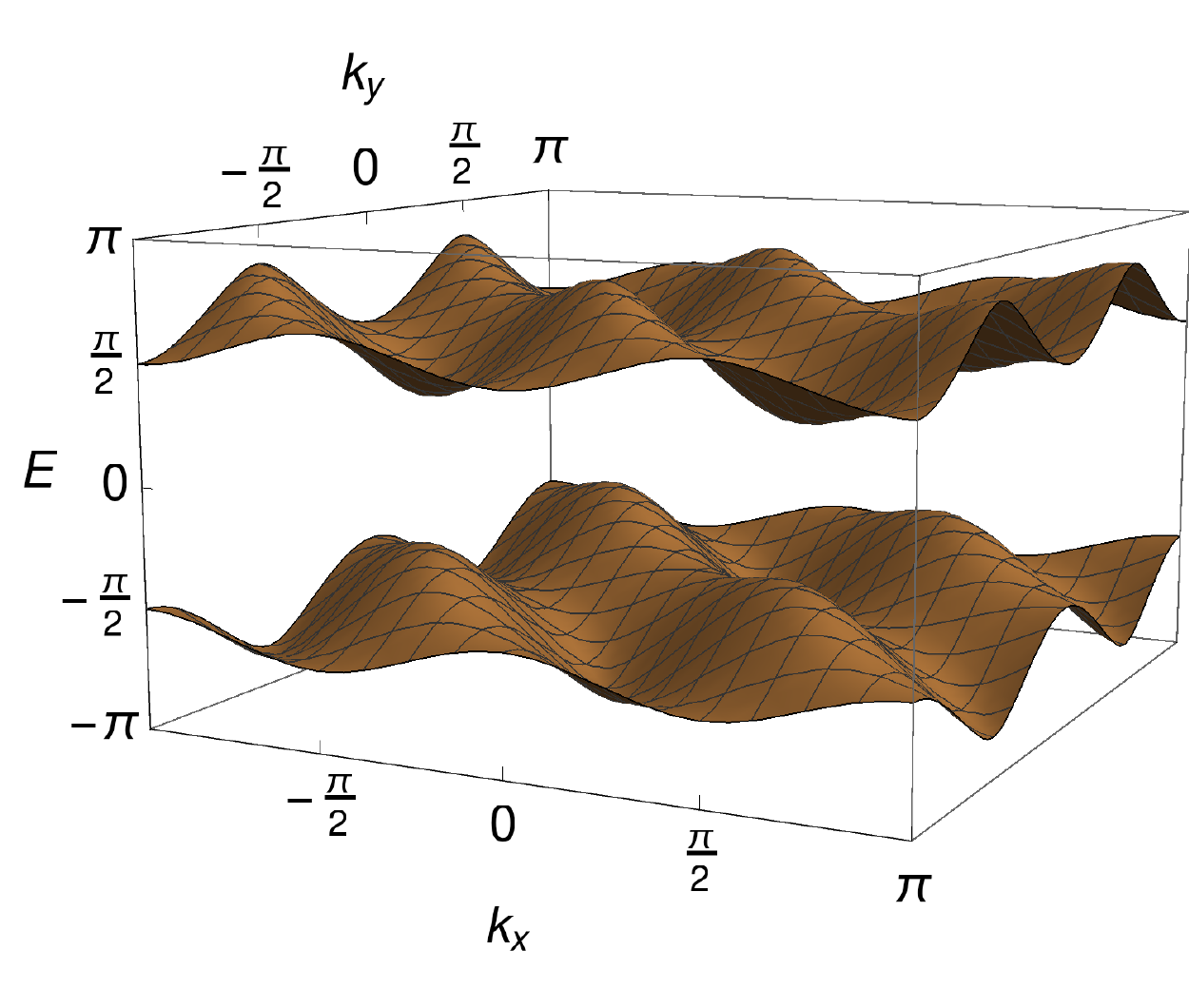}
			\includegraphics[width=0.23\linewidth]{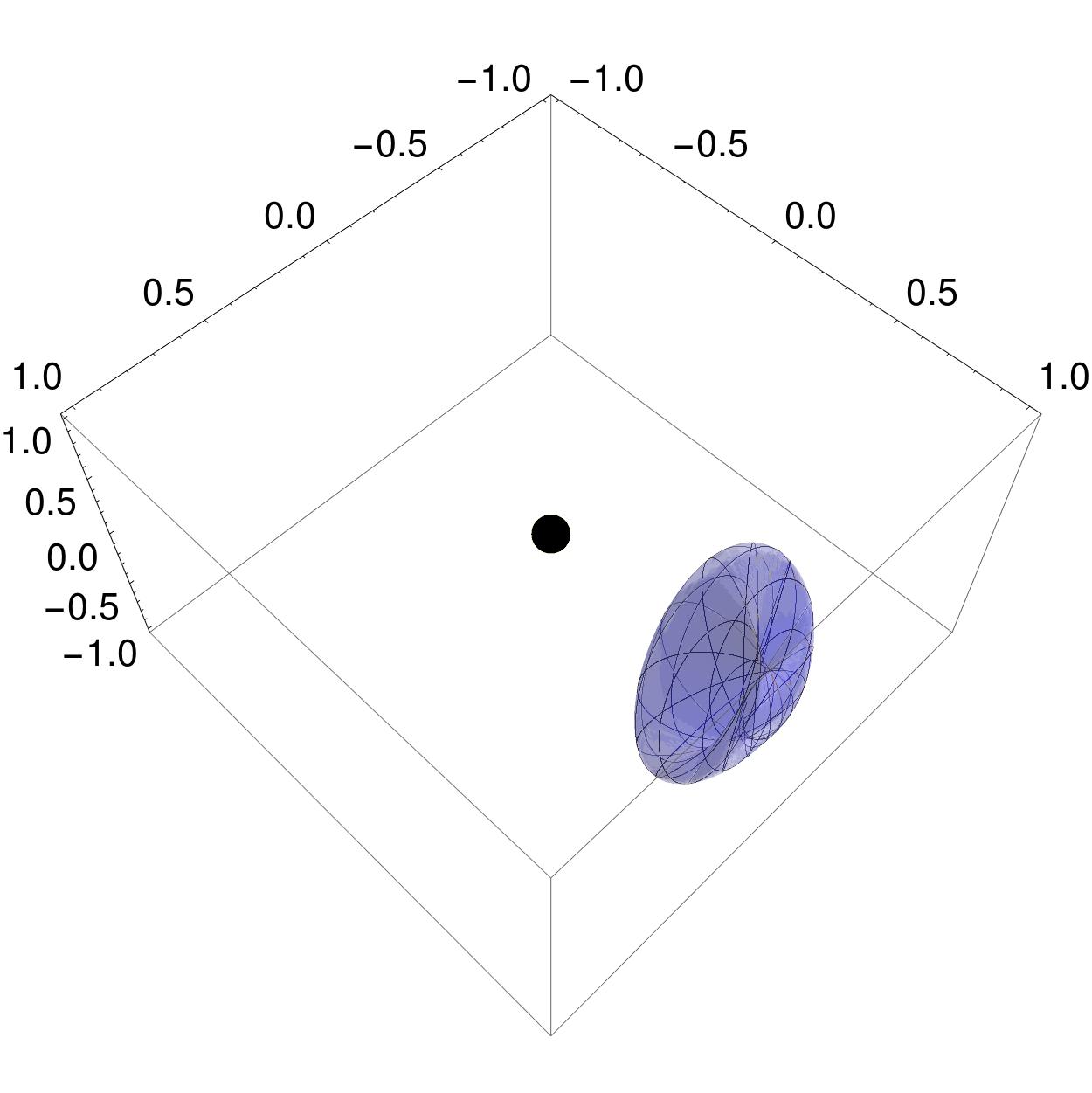}
		}			
		\sidesubfloat[]{
			\includegraphics[width=0.23\linewidth]{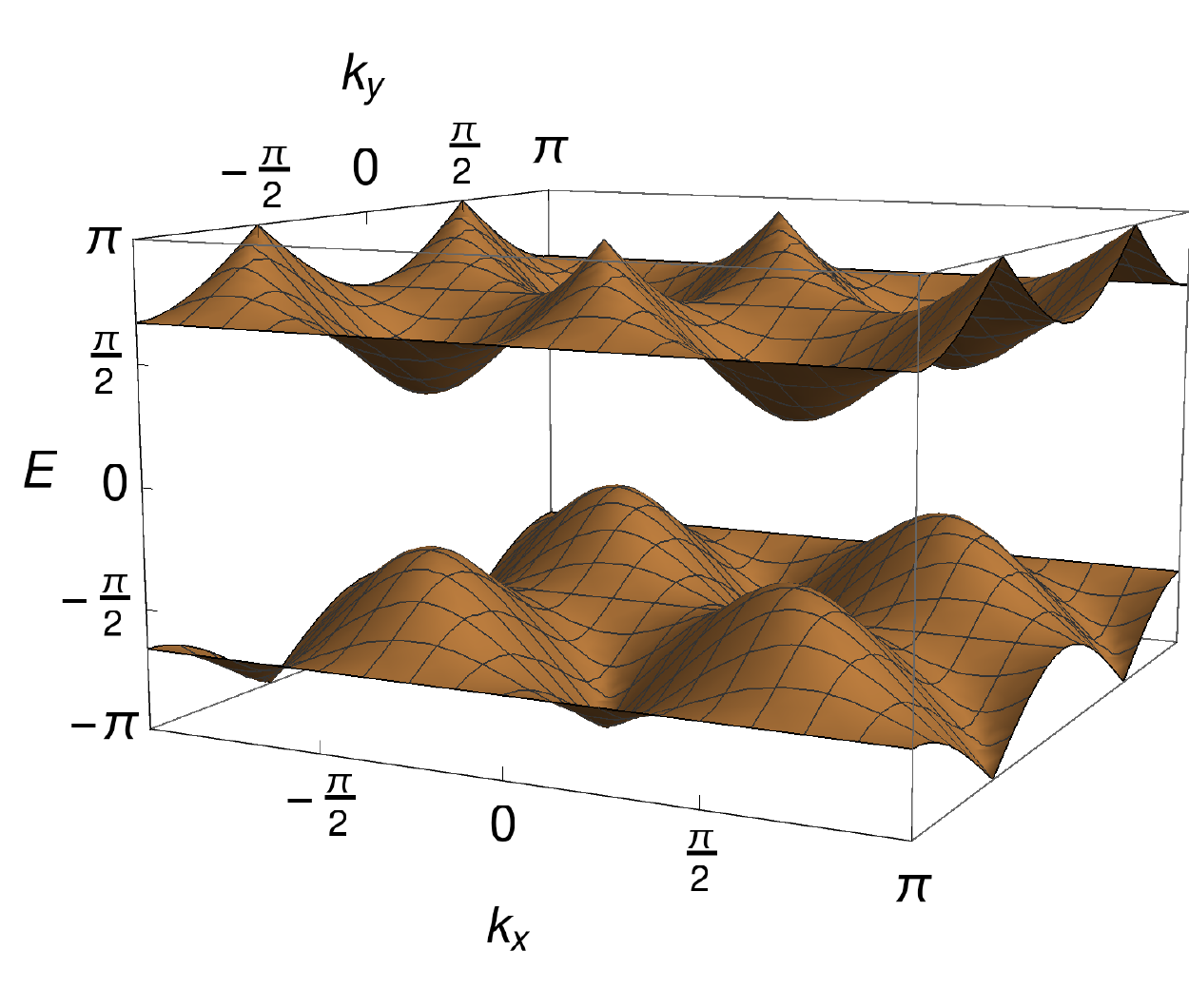}
			\includegraphics[width=0.23\linewidth]{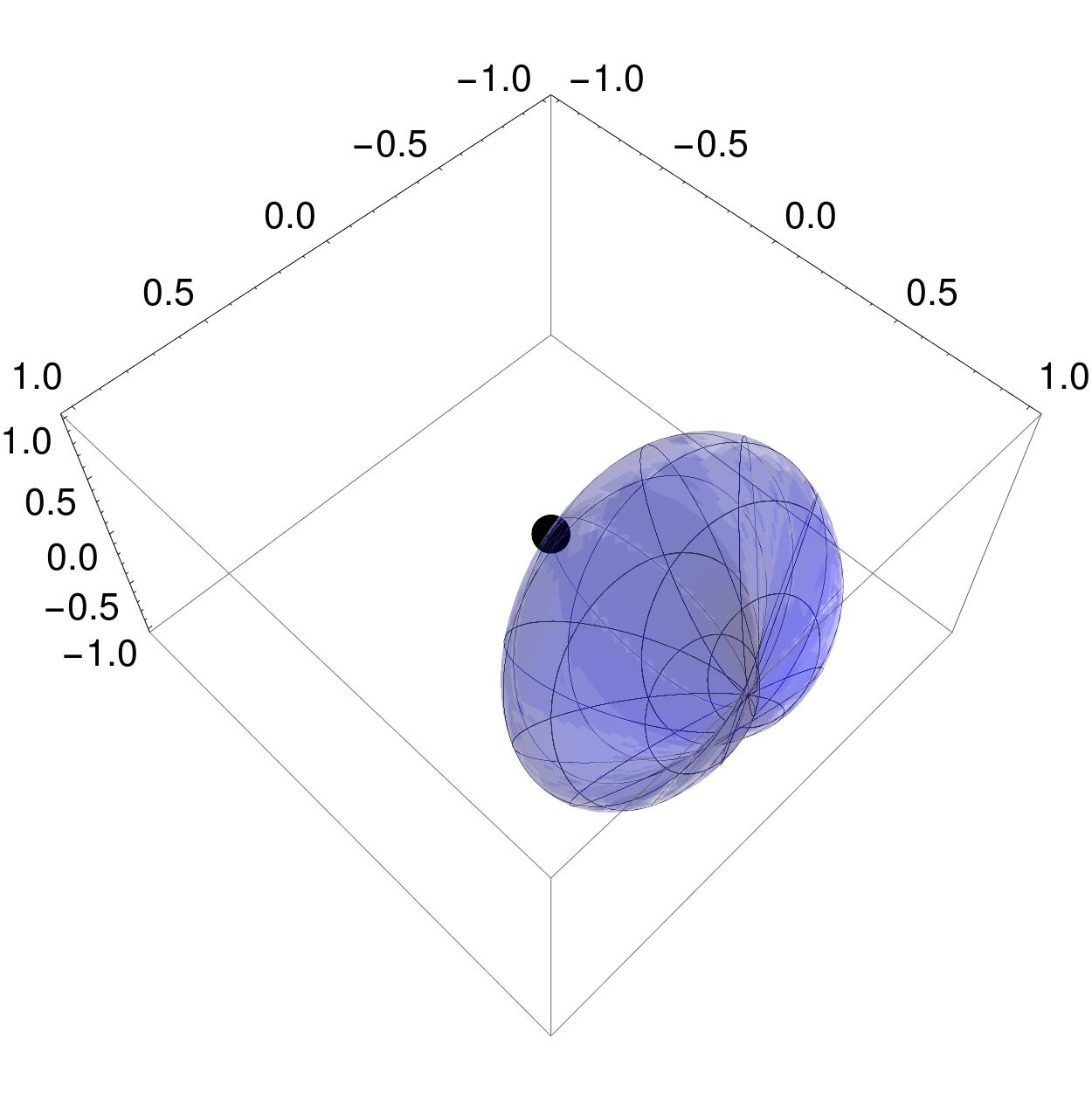}
		}	
		\\[0.0001cm]
		\sidesubfloat[]{
			\includegraphics[width=0.22\linewidth]{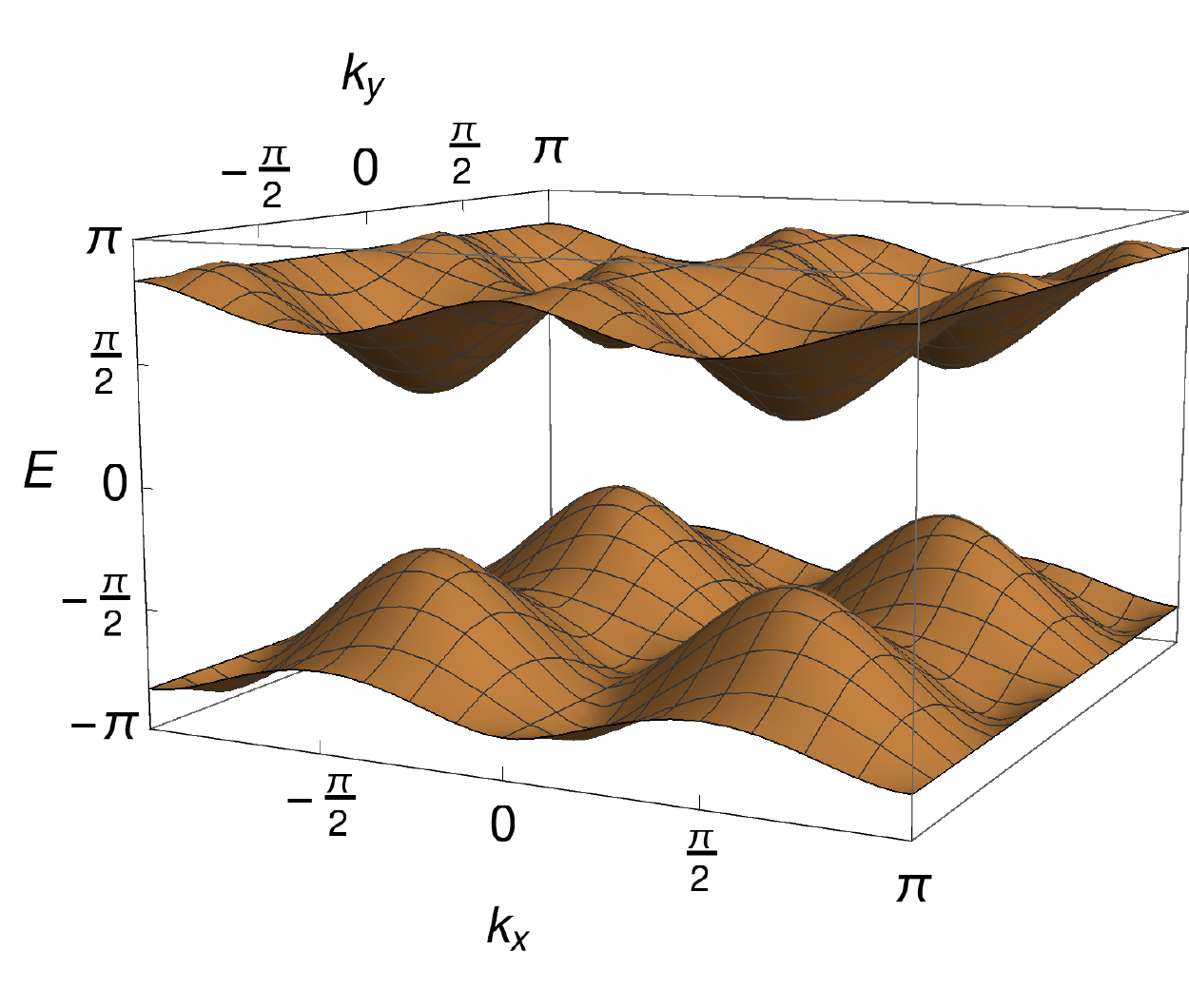}
			\includegraphics[width=0.22\linewidth]{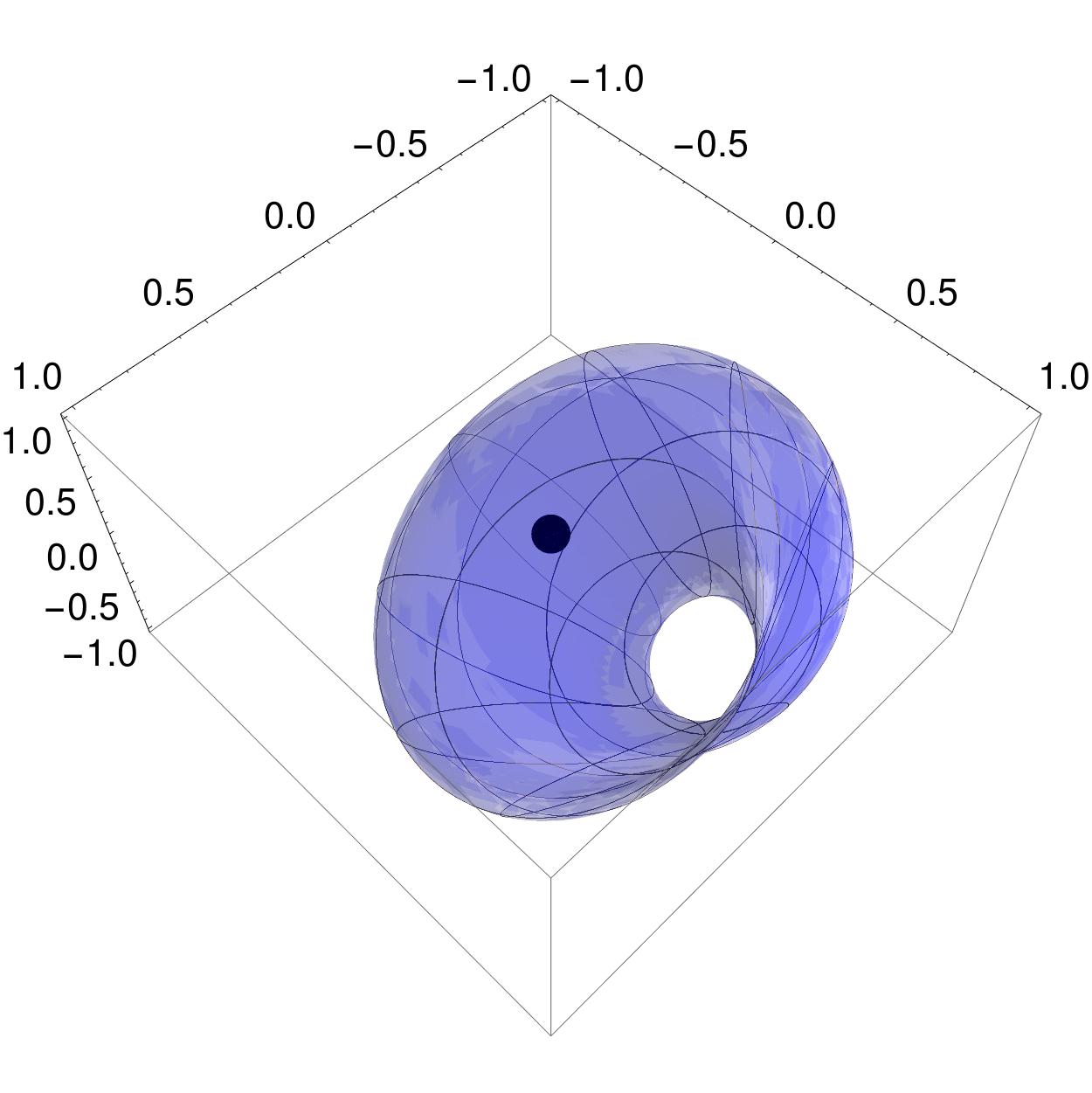}
		}			
		\sidesubfloat[]{
			\includegraphics[width=0.22\linewidth]{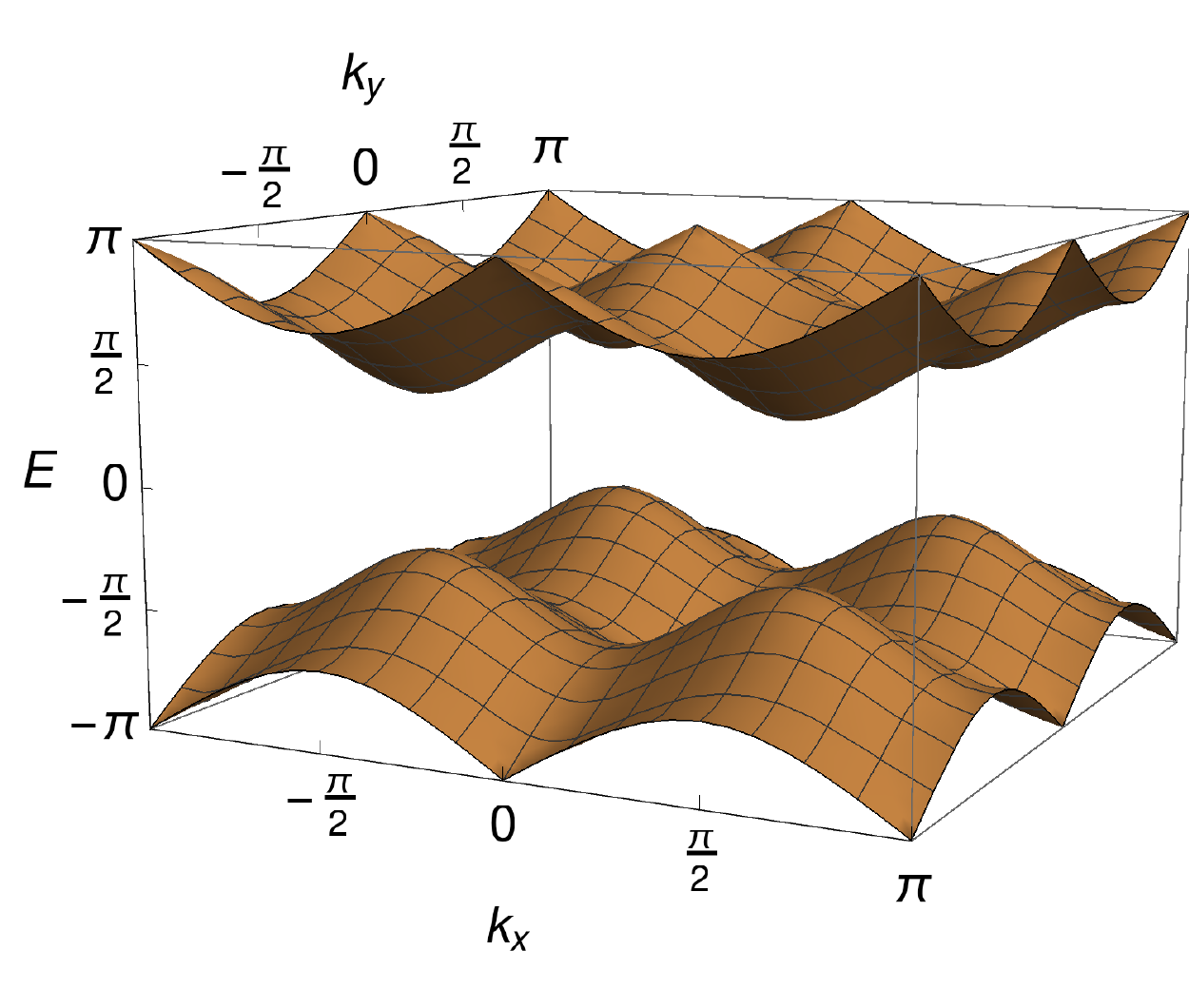}
			\includegraphics[width=0.22\linewidth]{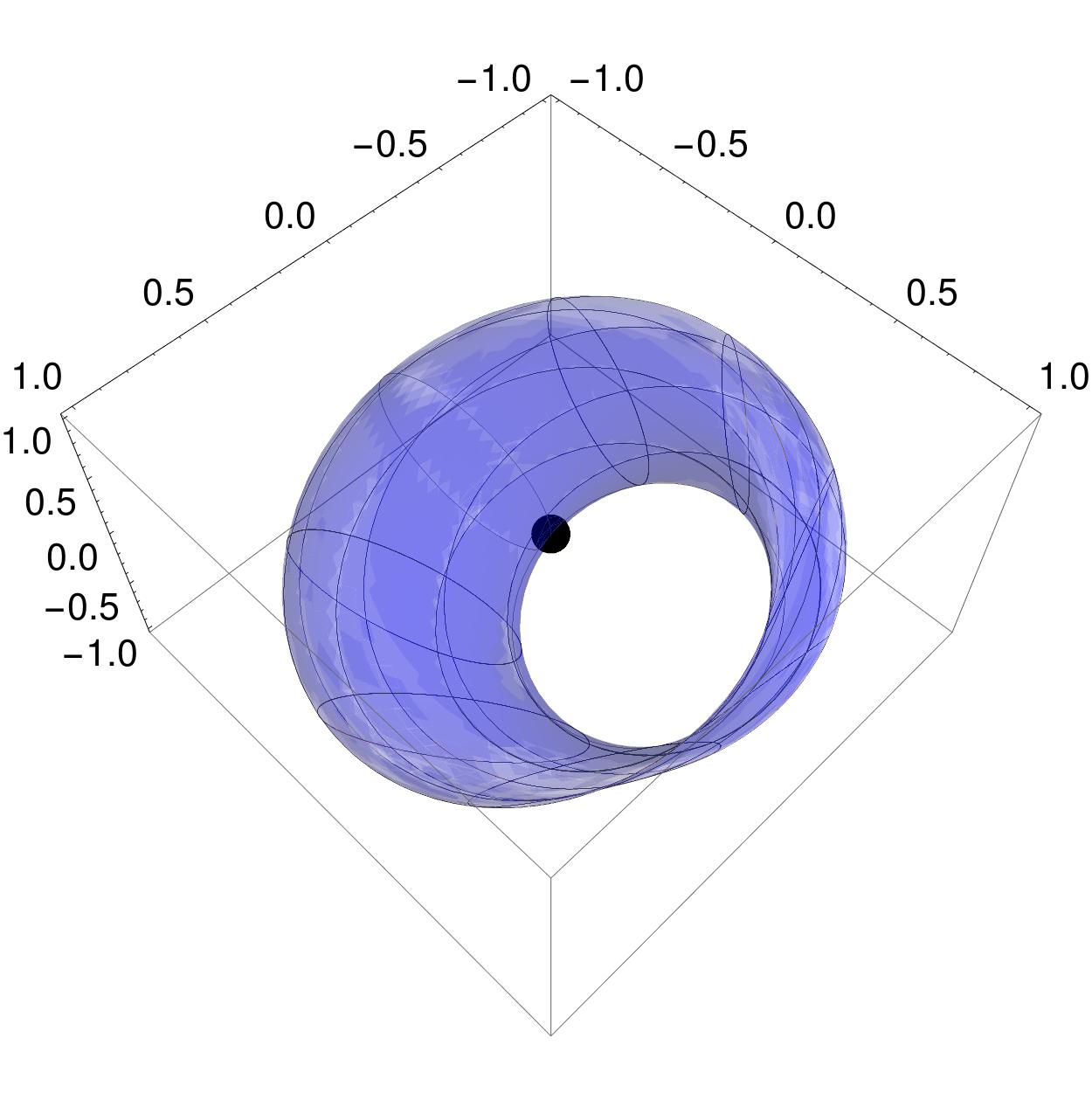}
		}
		\\[0.0001cm]
		\sidesubfloat[]{
			\includegraphics[width=0.22\linewidth]{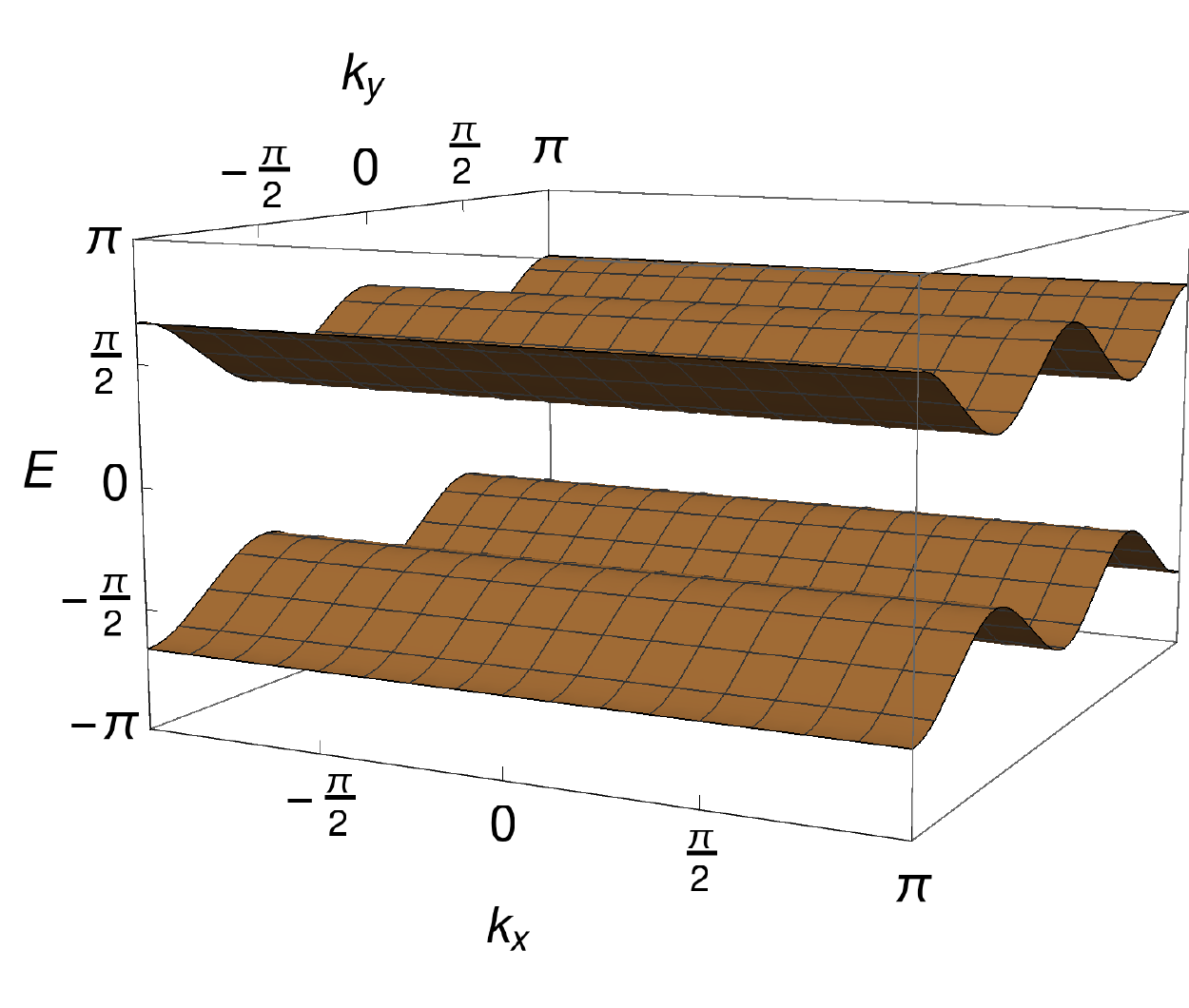}
			\includegraphics[width=0.22\linewidth]{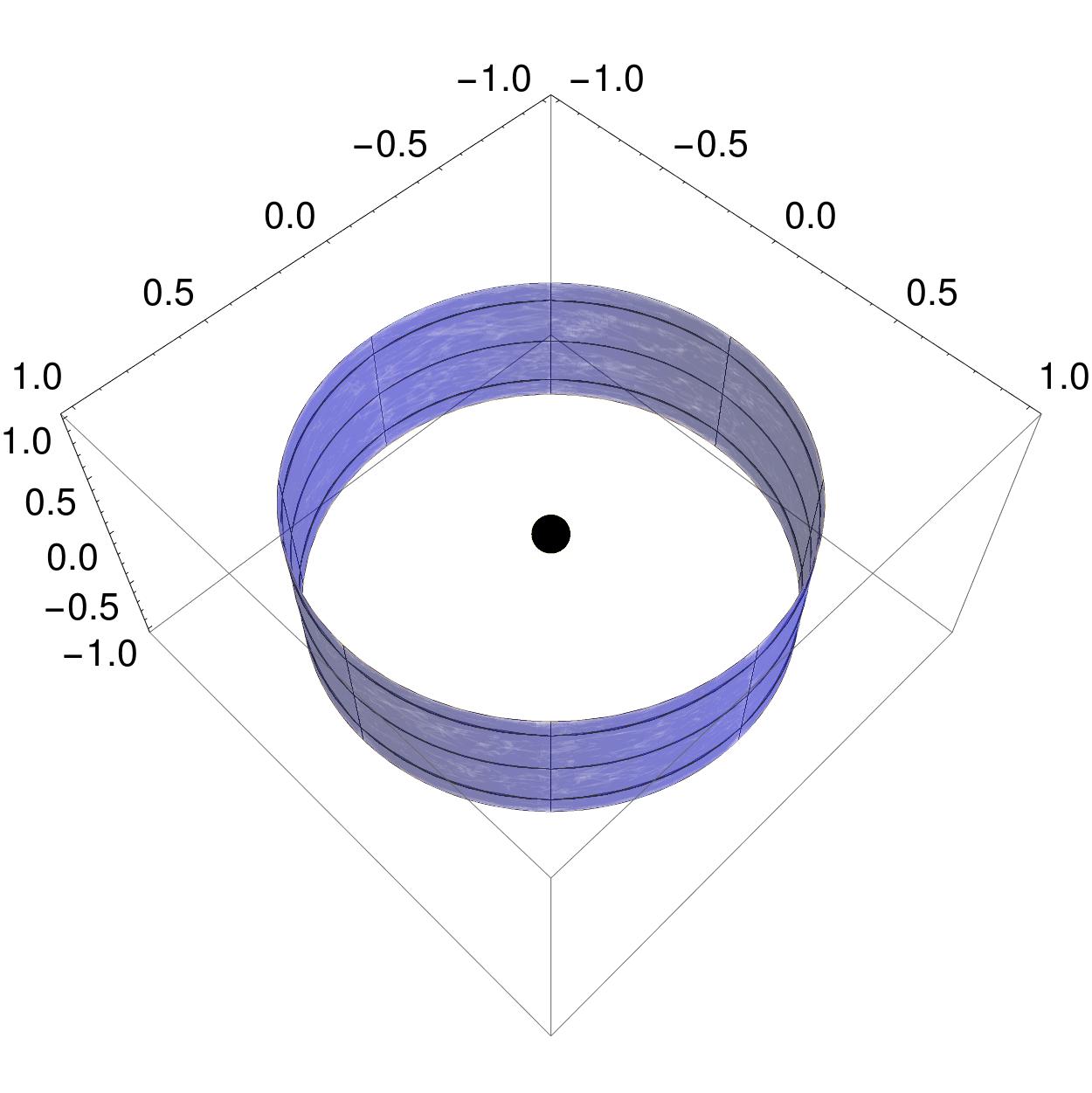}
		}
		\sidesubfloat[]{
			\includegraphics[width=0.22\linewidth]{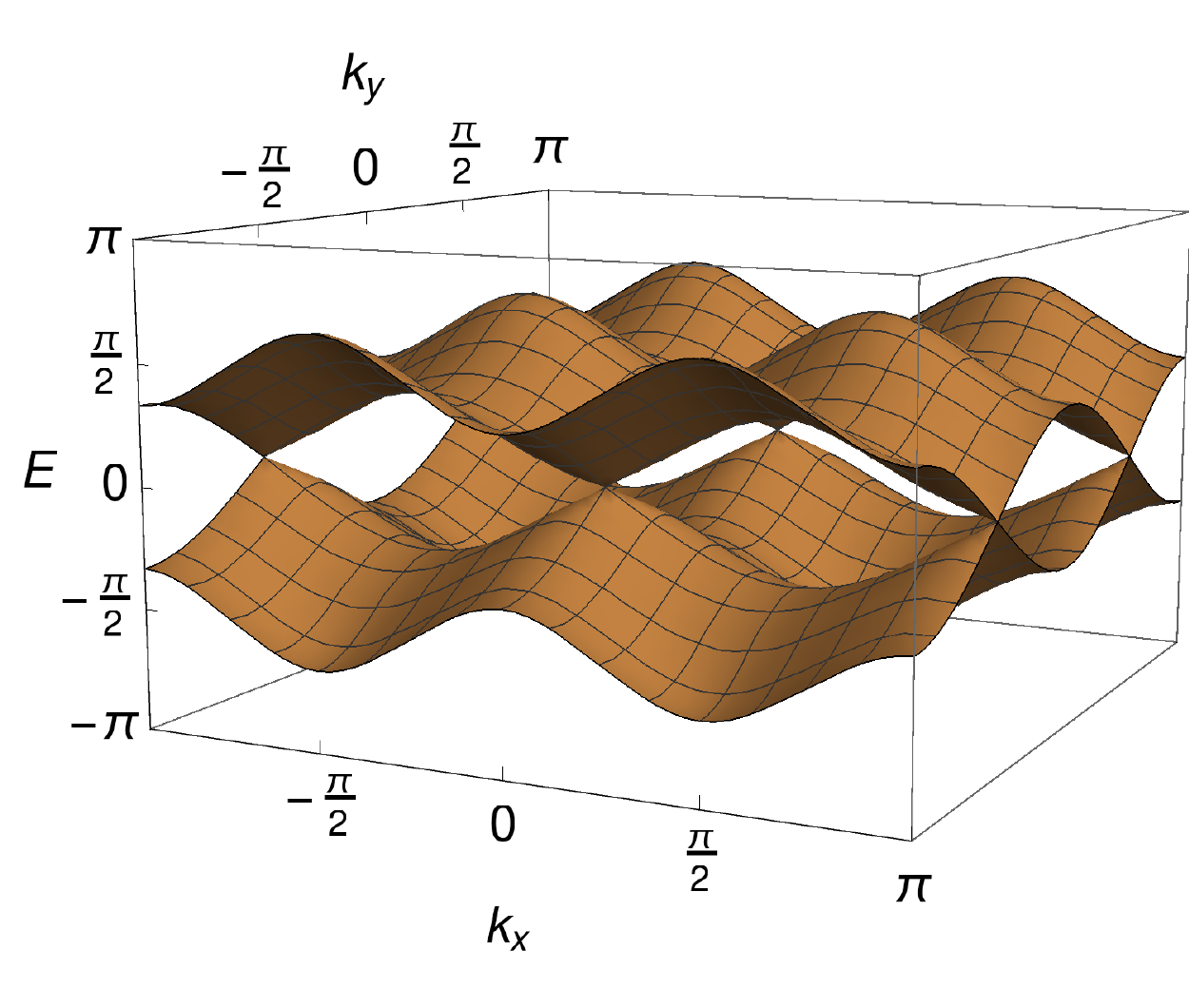}
			\includegraphics[width=0.22\linewidth]{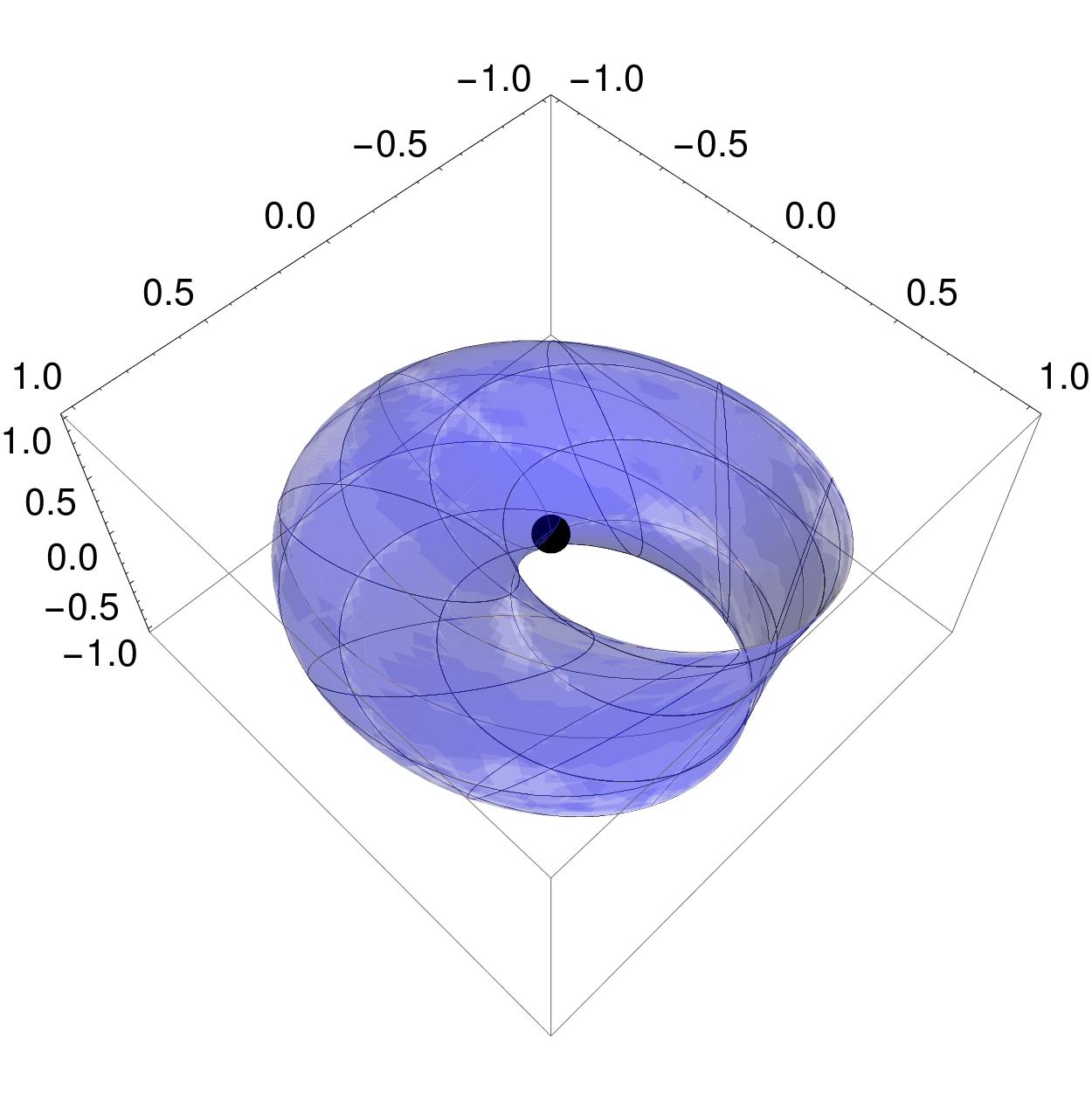}
		}											
	\end{tabular}}				
	\caption{Quantum walk with only PHS (two-dimensional): Modification of energy (left panels) and $\boldsymbol d$ (right panels) with $T=2$ and $\alpha=\pi/3$ for a) $\beta=\pi/12$, b) $\beta=\pi/6$, c) $\beta=\pi/4$, d) $\beta=\pi/3$, e) $\beta=\pi/2$ and f) $\beta=2\pi/3$. The $\boldsymbol d$ surfaces are plotted by variation of the momenta through the first Brillouin zone. In (a) and (e), energy bands of the two topological phases are presented. Since $\boldsymbol d$ do not completely surround the origin, Chern number is $0$ for these phases. In (c), $\boldsymbol d$ surround the origin which indicates a nontrivial topological phase with $+1$ Chern number. In (b), (d) and (f), energy bands close their gaps, therefore we have boundary states. This could be also noted since their corresponding $\boldsymbol d$ passes origin. Chronologically, from (a) to (f), we observe how energy bands change from gapped to gapless and topological phases transition from one to another one as they meet boundary states (phase transition point).} \label{Fig10}
\end{figure*}

\begin{figure*}[htb]
	\centering
	{\begin{tabular}[b]{cc}%
			\sidesubfloat[]{
				\includegraphics[width=0.23\linewidth]{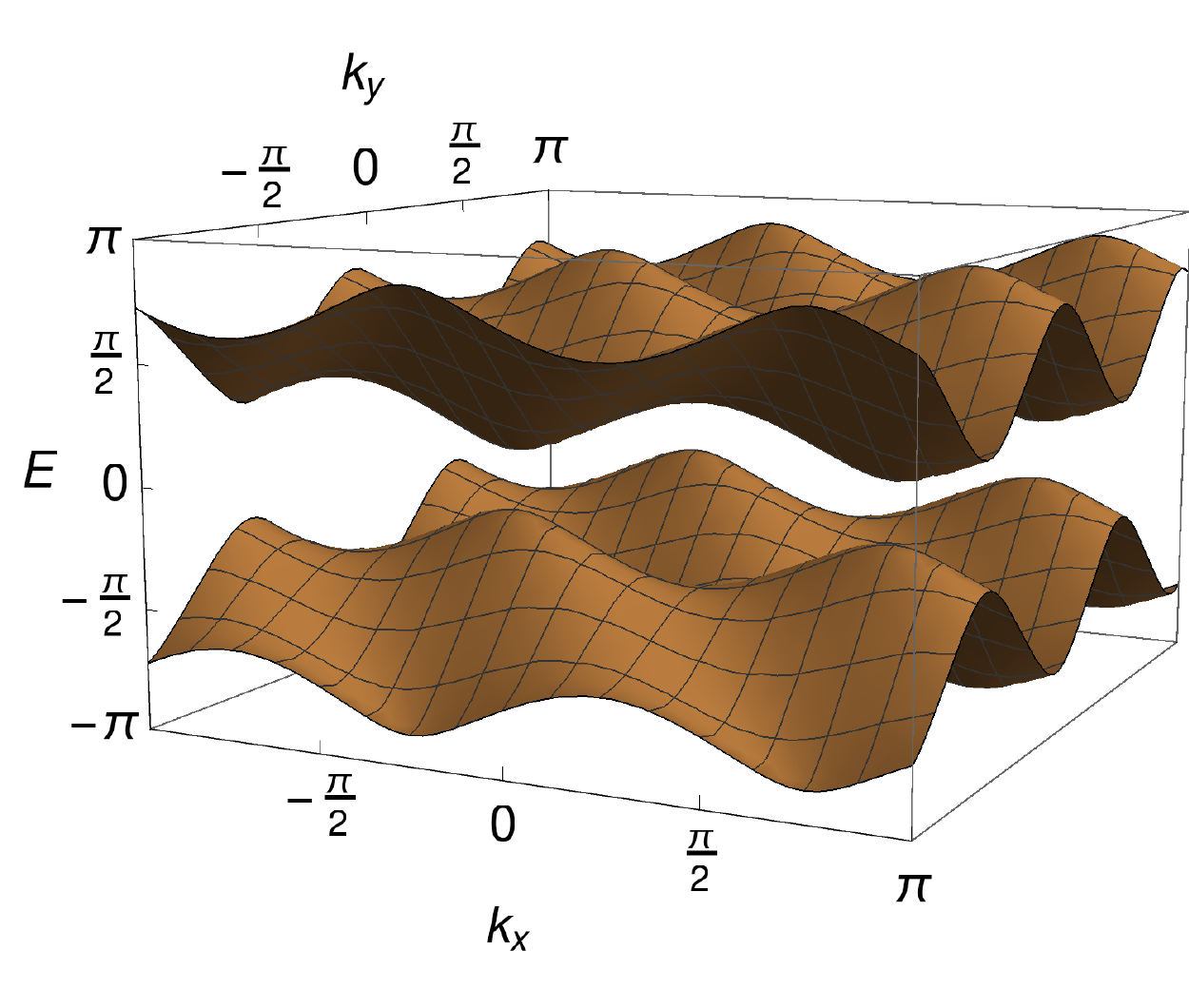}
				\includegraphics[width=0.23\linewidth]{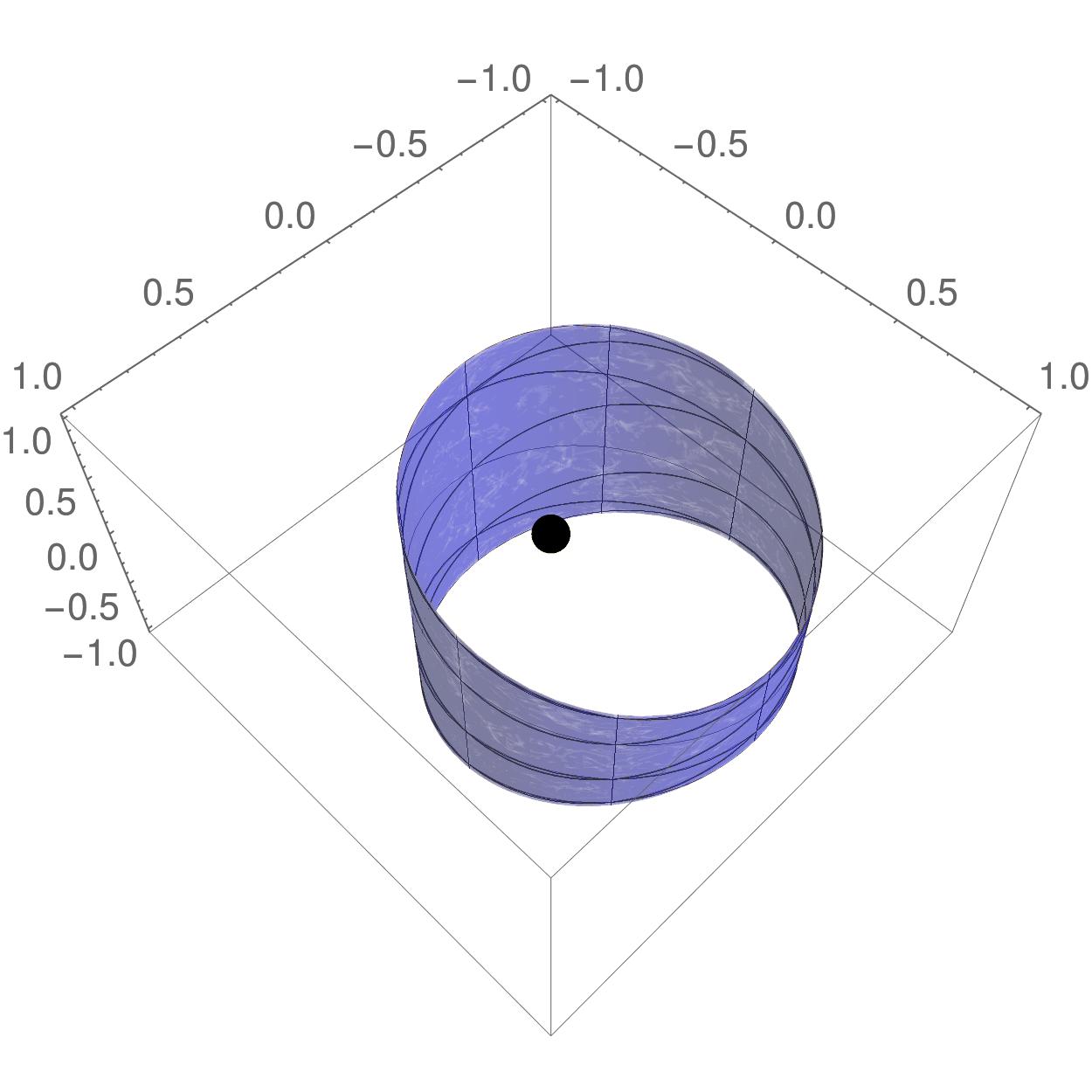}
			}			
			\sidesubfloat[]{
				\includegraphics[width=0.23\linewidth]{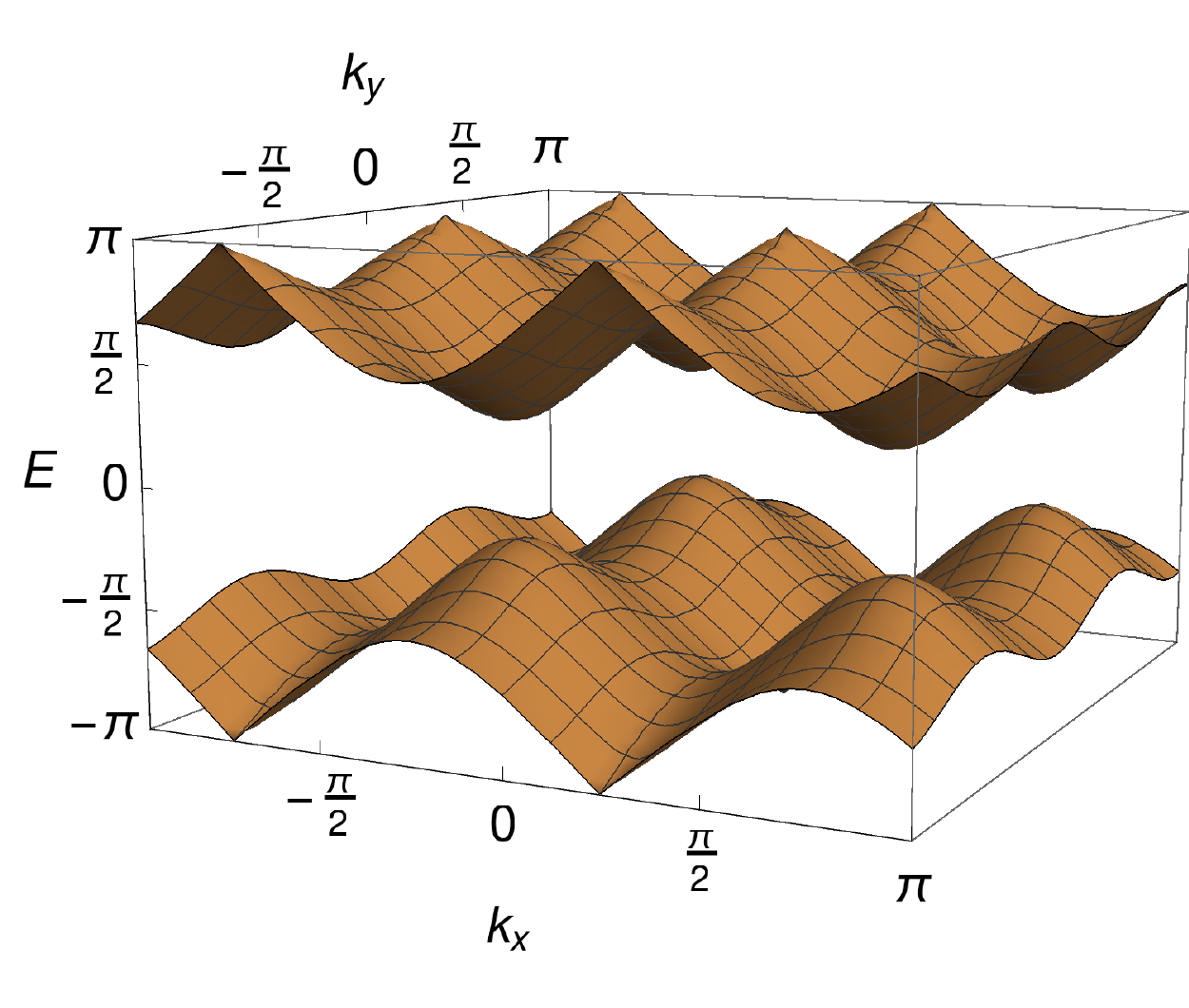}
				\includegraphics[width=0.23\linewidth]{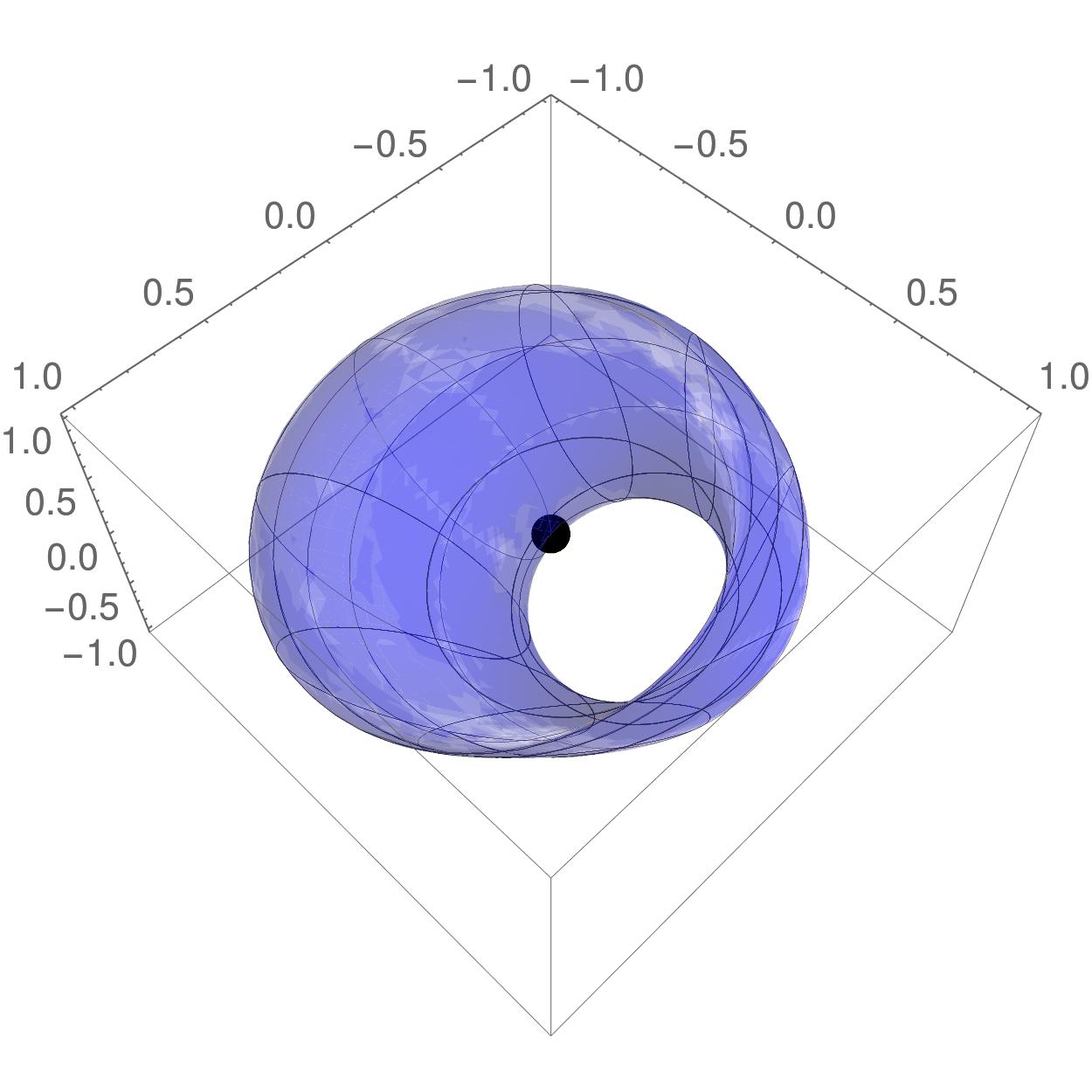}
			}	
			\\[0.0001cm]
			\sidesubfloat[]{
				\includegraphics[width=0.22\linewidth]{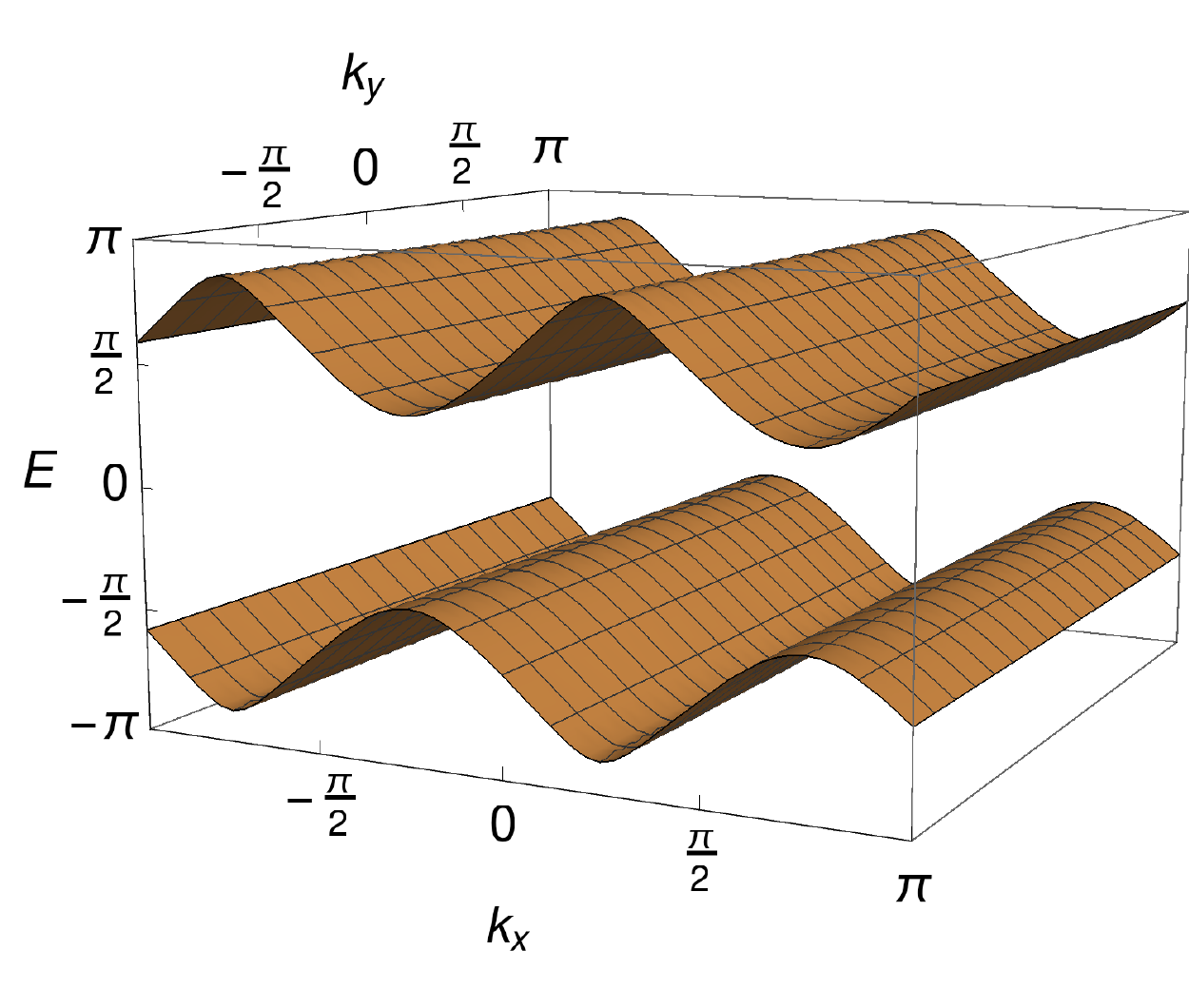}
				\includegraphics[width=0.22\linewidth]{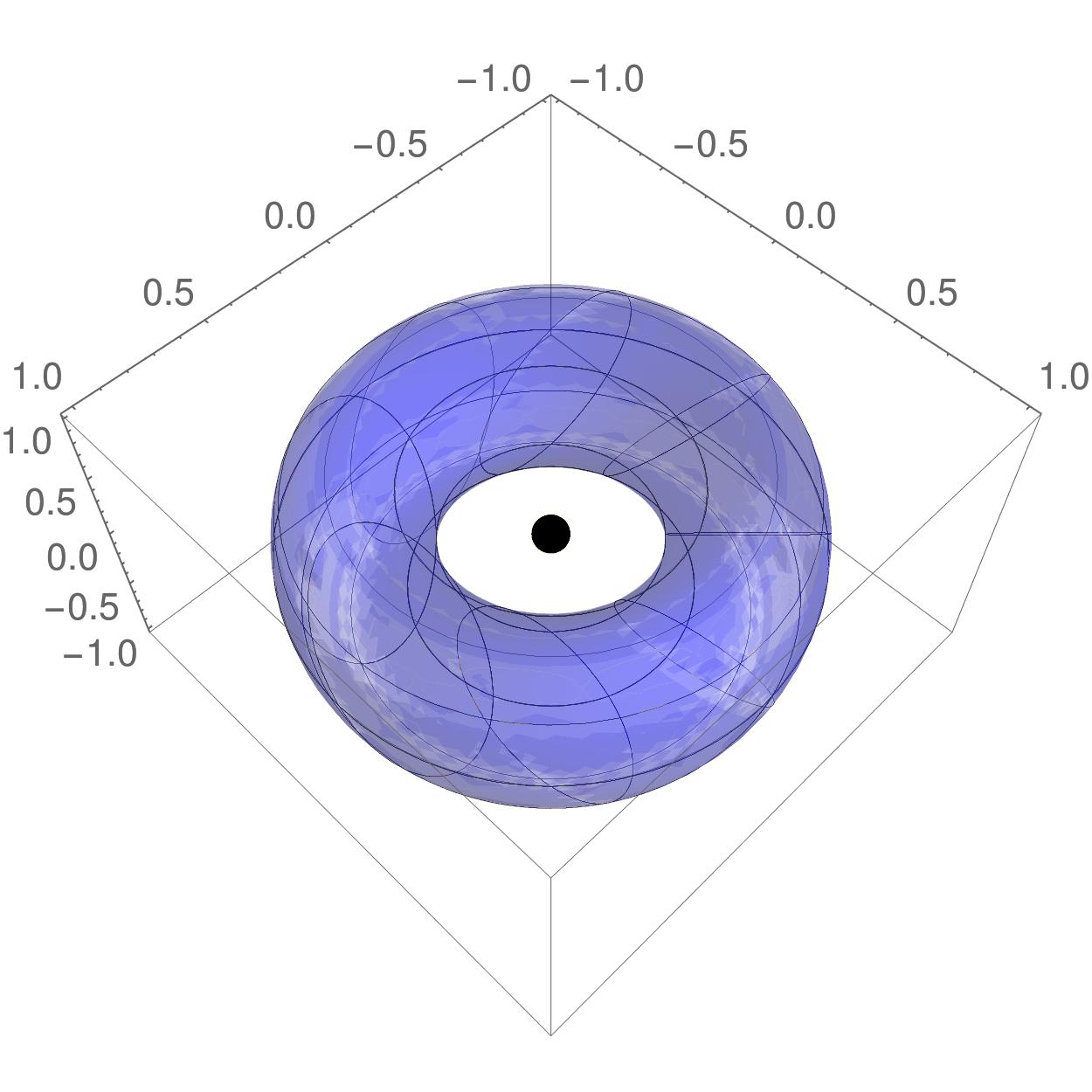}
			}			
			\sidesubfloat[]{
				\includegraphics[width=0.22\linewidth]{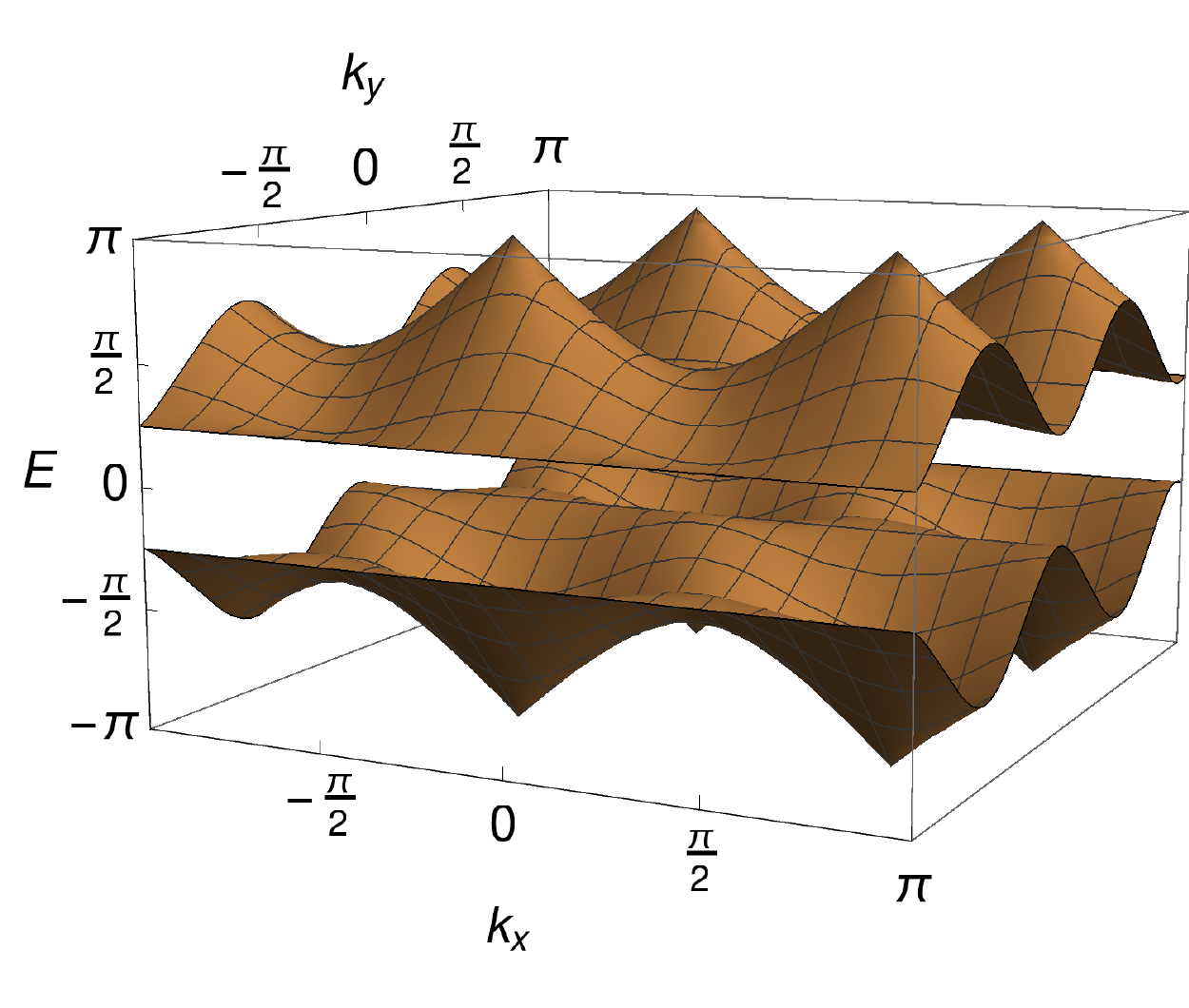}
				\includegraphics[width=0.22\linewidth]{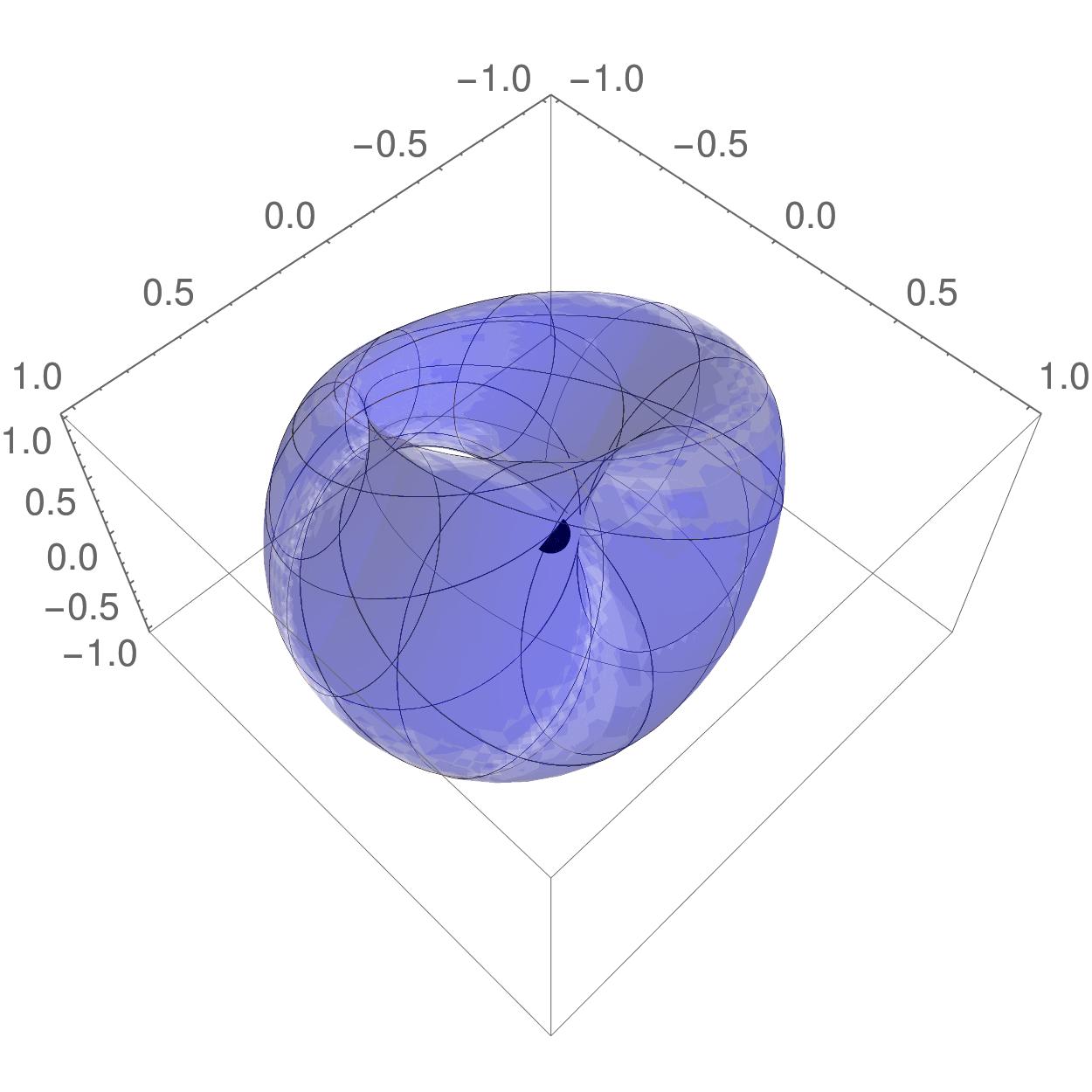}
			}
			\\[0.0001cm]
			\sidesubfloat[]{
				\includegraphics[width=0.22\linewidth]{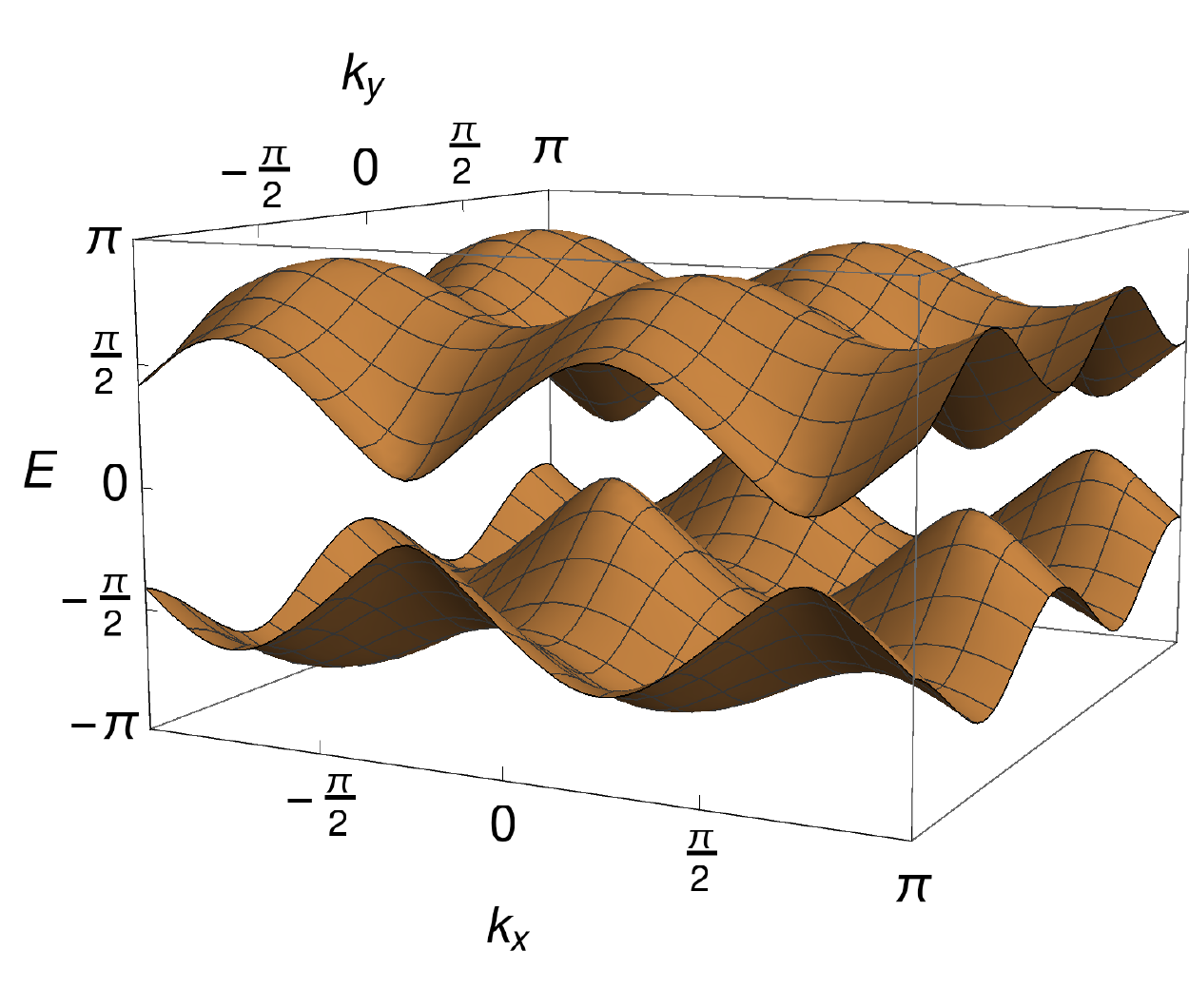}
				\includegraphics[width=0.22\linewidth]{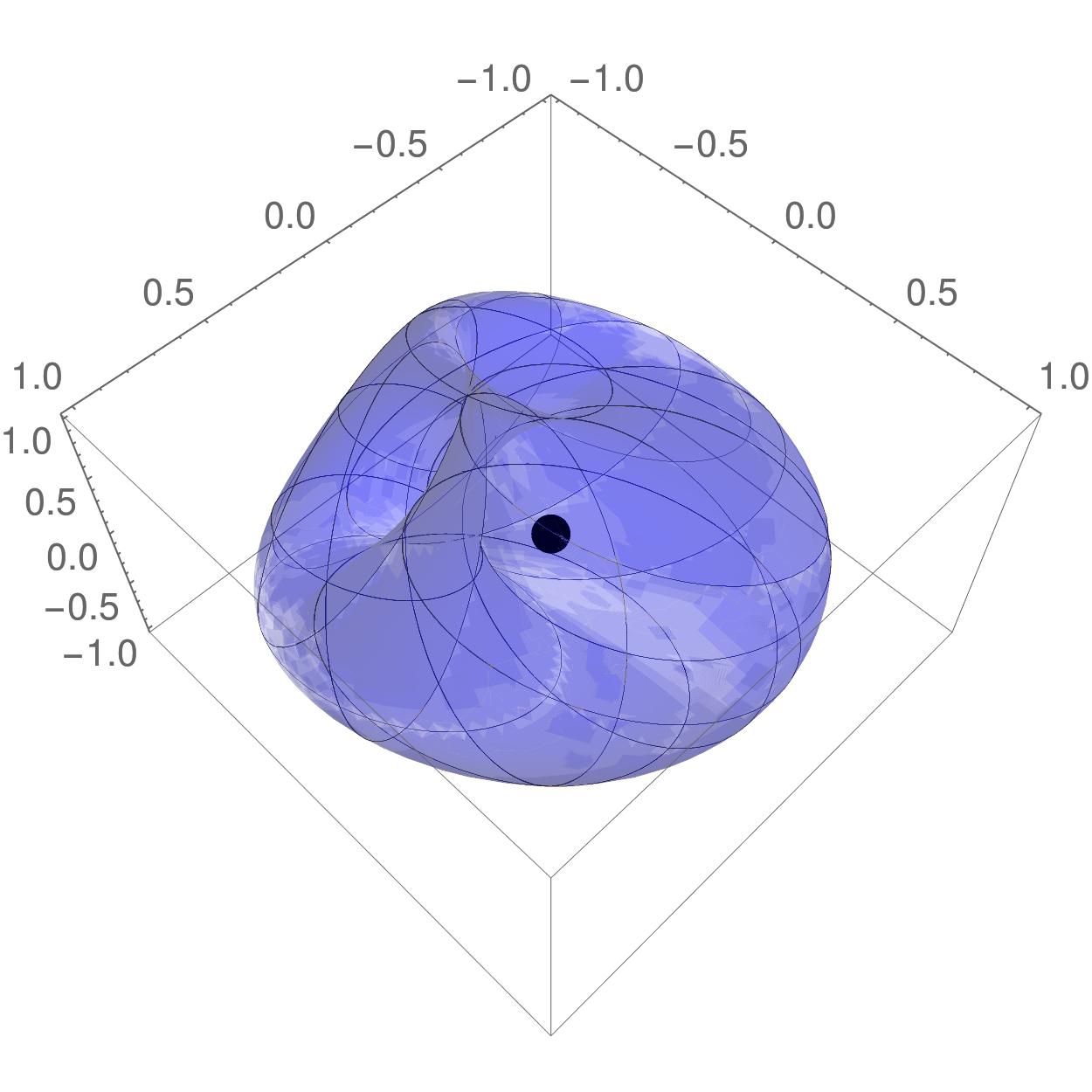}
			}										
	\end{tabular}}				
	\caption{Quantum walk with without three symmetries (two-dimensional): Modification of energy (left panels) and $\boldsymbol d$ (right panels) with $T=3$, $\alpha=\pi/3$ and $\gamma=\pi/4$ for a) $\beta=0$, b) $\beta=\pi/4$, c) $\beta=\pi/3$, d) $\beta=3\pi/4$ and e) $\beta=\pi/2$. The $\boldsymbol d$ surfaces are plotted as the momenta traverse the first Brillouin zone. In (a) and (c), we have two trivial topological phases with Chern number $0$ since $\boldsymbol d$ do not completely surround the origin. In (e), we have a nontrivial topological phase with $+1$ Chern number. In (b) and (d), energy bands close their gaps and $\boldsymbol d$ passes origin, so we have two boundary states. Chronologically, from (a) to (e), we observe two trivial phases next to each other followed by a nontrivial topological phase. Therefore, the phase structure is different from those observed in PHS walk.} \label{Fig11}
\end{figure*}

For both of the two-dimensional protocols studied here, the topological invariant is an integer valued number known as Chern number. Basically, we should vary the momenta in $\boldsymbol n$ through first Brillouin zone. Instead of $\boldsymbol n$, one can do the same for $\boldsymbol d$ and then if the resulted surface surrounds the origin, then we have a nontrivial topological phase with topological invariant being $\pm 1$. If it does not cover the origin completely, then we a have a trivial topological phase with zero Chern number. If the surface passes the origin, we have a boundary state. Based on these concepts, we have plotted Figs. \ref{Fig10} and \ref{Fig11} for the protocol with only PHS \eqref{protocol3} and protocol without any of three symmetries \eqref{protocol33}, respectively. 

For the protocol with only PHS (Fig. \ref{Fig10}), scanning through limited range of the rotation angles confirms the presence of the trivial and nontrivial topological phases separated by boundary states. In neighboring of a nontrivial topological phases, there are two trivial phases. Therefore, the phase transitions are between trivial and nontrivial topological phases. In case of the protocol without the three symmetries (Fig. \ref{Fig11}), as we modify the rotation angle in limited range, we first come across two trivial phases separated by a boundary state and then another boundary state and a nontrivial topological phase. Therefore, in this case, we have phase transition between two trivial phase and transition between trivial and nontrivial phase. It should be noted that phase structure step-dependently changes for both of the protocols. 

In Ref. \cite{Panahiyan2020}, we investigated two protocols of simple- and split-step for two-dimensional quantum walk with 

\begin{eqnarray}
\widehat{U} & = & \widehat{S}_{\uparrow \downarrow} \widehat{C}_{\beta}, \label{SD2}
\\
\widehat{U} & = & \widehat{S}_{\uparrow \downarrow}(y) \widehat{C}_{\alpha} \widehat{S}_{\uparrow \downarrow}(x) \widehat{C}_{\beta}.  \label{SPD2}
\end{eqnarray}

Both of these protocols have PHS, TRS and CHS with $\widehat{\mathcal{P}}^2=+1$, $\widehat{\mathcal{T}}^2=+1$ and $\widehat{\Gamma}^2=+1$. In two-dimensional case, the protocols with such symmetries simulate trivial phases with topological invariants equals to zero. on the other hand, the simple-step protocol \eqref{SD2} has Dirac cone boundary states without any Fermi arc boundary states. Modification to split-step protocol \eqref{SPD2} results into emergences of the Fermi arc boundary states alongside of Dirac cone boundary states. 

In case of the protocol with only PHS \eqref{protocol3}, since $\widehat{\mathcal{P}}^2=+1$, the protocol simulates D family of the topological phases with topological invariant being an integer, $\mathbb{Z}$. The absence of the TRS enables us to employ the doubling procedure to construct the following protocol 

\begin{eqnarray}
\widehat{U} = \begin{pmatrix}
\widehat{U}_{c} & 0 \\
0 & \widehat{U}_{c}^t,
\end{pmatrix} \label{DPHS22}
\end{eqnarray} 
in which $\widehat{U}_{c}$ is given in Eq. \eqref{protocol3} and this protocol has PHS, TRS and CHS with $\widehat{\mathcal{P}}^2=+1$, $\widehat{\mathcal{T}}^2=-1$ and $\widehat{\Gamma}^2=+1$. Therefore, DIII familiy of topological phases coud be simulated by this protocol with topological invariant being a quantity, $\mathbb{Z}_2$.  

The protocol without the three symmetries \eqref{protocol33} can simulate A family of the topological phases in which the topological invariant is an integer, $\mathbb{Z}$. Since the TRS is absent, we use the doubling procedure and find the following protocol

\begin{eqnarray}
\widehat{U} = \begin{pmatrix}
\widehat{U}_{d} & 0 \\
0 & 1 \end{pmatrix} e^{-i \tau_{y} \sigma_{y} \phi/2} \begin{pmatrix}
1 & 0 \\
0 & \widehat{U}_{d}^t 
\end{pmatrix} \label{TRS2D}
\end{eqnarray}
in which $\tau_{y}$ is a Pauli matrix associating to the two additional flavors $A$ and $B$, $\widehat{U}_{d}$ is given in \eqref{protocol33}. This protocol \eqref{TRS2D} possess only TRS with $\widehat{\mathcal{T}}^2=-1$. Therefore, it simulates AII family of the topological phases with topological invariant of $\mathbb{Z}_2$. Additionally, we can also use the doubling procedure to find another new protocol as

\begin{eqnarray}
\widehat{U} = \begin{pmatrix}
\widehat{U}_{d} & 0 \\
0 & \widehat{U}_{d}^{\ast} 
\end{pmatrix} \label{PHS2D}
\end{eqnarray}
where $\ast$ corresponds to complex conjugate. Such a protocol \eqref{PHS2D} has only PHS with $\widehat{\mathcal{P}}^2=-1$ which indicates that it can simulate C family of topological phases with integer topological invariants, $\mathbb{Z}$. 

In summary, we observe that with given protocols in here, we can simulate full classes of topological phases in two dimensions (see table \ref{table}). The presence of step-dependent coins in the protocols of the  quantum walks provided us with different classes of boundary states and controllability over simulation of topological phases, boundary states and their populations.


\section{Three-dimensional quantum walk}  \label{two dimension}

In this section, we investigate simulation of topological phenomena in three-dimensional position space with six protocols including one simple-step and four split-step ones. 

\subsection{Simple-step quantum walk}

The three-dimensional simple-step quantum walk has the following protocol
\begin{eqnarray}
\widehat{U} & = & \widehat{S}_{\uparrow \downarrow}(z)\widehat{S}_{\uparrow \downarrow}(y)\widehat{S}_{\uparrow \downarrow}(x) \widehat{C}_{y}(\beta)  \label{protocol4},
\end{eqnarray}
in which the shift operators are given by 

\begin{eqnarray}
\widehat{S}_{\uparrow \downarrow} (x) = && \ketm{\uparrow} \bram{\uparrow} \otimes \sum_{x,y,z} \ketm{x+1,y,z} \bram{x,y,z} + \notag
\\
 && \ketm{\downarrow} \bram{\downarrow} \otimes \sum_{x,y,z} \ketm{x-1,y,z} \bram{x,y,z}\, , \label{shift7}
\end{eqnarray}
\begin{eqnarray}
\widehat{S}_{\uparrow \downarrow} (y) =  && \ketm{\uparrow} \bram{\uparrow} \otimes \sum_{x,y,z} \ketm{x,y+1,z} \bram{x,y,z} + \notag
\\
 && \ketm{\downarrow} \bram{\downarrow} \otimes \sum_{x,y,z} \ketm{x,y-1,z} \bram{x,y,z}\, , \label{shift8}
\end{eqnarray}
\begin{eqnarray}
\widehat{S}_{\uparrow \downarrow} (z) =  && \ketm{\uparrow} \bram{\uparrow} \otimes \sum_{x,y,z} \ketm{x,y,z+1} \bram{x,y,z} + \notag
\\
 && \ketm{\downarrow} \bram{\downarrow} \otimes \sum_{x,y,z} \ketm{x,y,z-1} \bram{x,y,z}\, , \label{shift9}
\end{eqnarray}
and we can rewrite them as $\widehat{S}_{\uparrow \downarrow} (x)= e^{ik_{x} \sigma_{z}}$, $\widehat{S}_{\uparrow \downarrow} (y)= e^{ik_{y} \sigma_{z}}$ and $\widehat{S}_{\uparrow \downarrow} (z)= e^{ik_{z} \sigma_{z}}$ where $k_{x}$, $k_{y}$ and $k_{z}$ span the first Brillioun zone. A single step of the walk includes rotation of internal states by $\widehat{C}_{y}(\beta)$ and displacement of its walker subsequently in $x$, $y$ and $z$ position spaces.  

Based on the simple-step protocol, we find the energy as 

\begin{eqnarray}
E&=& \pm\cos^{-1} (\kappa_{\beta} \cos(k_{x}+k_{y}+k_{z})). \label{energy4}
\end{eqnarray}
where the presence of $\pm$ indicates existence of two bands of energy. The energy bands traverse $[-\pi,\pi]$ and they close up their gap at $E=0$ and $\pm \pi$ for 

\begin{eqnarray}
	\beta_{E=\pi}&=&\frac{\pm 2\cos ^{-1} [-\sec (k_{x}+k_{y}+k_{z})]+4\pi c}{T}, 
\end{eqnarray}

\begin{eqnarray}
	\beta_{E=0}&=&\frac{\pm 2\cos ^{-1} [\sec (k_{x}+k_{y}+k_{z})]+4\pi c}{T}. 
\end{eqnarray}
where $c$ is an integer. The energy bands close up their gap linearly which indicates that the type of boundary states is Dirac cone. It should be noted that for 

\begin{eqnarray}
	\beta_{E=cte=\pm \pi/2}&=&\frac { 4 \pi  c \pm \pi } {T},
\end{eqnarray}
the energy bands become independent of momenta and flat bands are formed for energy bands.

Next, we find $\boldsymbol n(k_{x},k_{y},k_{z})$ and components of group velocity as 

\begin{equation}
n_{x}  =  \frac{\kappa_{\beta} \sin (k_{x}+k_{y}+k_{z})}{\sin (E)}, \notag
\end{equation}
\begin{equation}
n_{y}  =  \frac{\lambda_{\beta} \cos (k_{x}+k_{y}+k_{z})}{\sin (E)}, \notag
\end{equation}
\begin{equation}
n_{z}  = - \frac{- \kappa_{\beta} \sin (k_{x}+k_{y}+k_{z})}{\sin (E)}. 
\end{equation}
\begin{equation}
V_{k_{x}}= V_{k_{y}}=V_{k_{y}}= \pm n_{z}. \label{groupv4}
\end{equation} 

Both $\boldsymbol n(k_{x},k_{y},k_{z})$ and components of group velocity become ill-defined as the bands of energy close their gap. On the other hand, the components of group velocity, hence group velocity span $[-1,1]$.  

\subsection{Split-step quantum walk}  \label{split}

Next, we modify the simple-step protocol to following split-step one for our three-dimensional quantum walk 

\begin{eqnarray}
\widehat{U} & = &\widehat{S}_{\uparrow \downarrow}(z) \widehat{C}_{y}(\gamma) \widehat{S}_{\uparrow \downarrow}(y) \widehat{C}_{y}(\alpha) \widehat{S}_{\uparrow \downarrow}(x) \widehat{C}_{y} (\beta)  \label{protocol5},
\end{eqnarray}
which indicates that a step of the quantum walk is done through the rotation of internal states of the walker by $\widehat{C}_{y} (\beta)$, displacement of its position in $x$ space by $\widehat{S}_{\uparrow \downarrow}(x)$, additional rotation of internal states by $\widehat{C}_{y}(\alpha)$ followed by a displacement of its position in $y$ space with $\widehat{S}_{\uparrow \downarrow}(y)$, another rotation of internal states by $\widehat{C}_{y}(\gamma)$ and finally a displacement of its position in $z$ space with $\widehat{S}_{\uparrow \downarrow}(z)$.  The coin operators and shift operators are given in Eqs. \eqref{coin1}, \eqref{coin2}, \eqref{shift7}-\eqref{shift9} and coin operator $\widehat{C}_{y}(\gamma)$ is likewise $\widehat{C}_{y} (\beta)$.

It is a matter of calculation to rewrite shift operatores using the Discrete Fourier Transformation and find the two bands of as 

\begin{eqnarray}
E= \pm\cos^{-1}(\rho),  \label{energy5}
\end{eqnarray}
where $\rho=\cos(k_{x}+k_{y}+k_{z})\kappa_{\gamma}\kappa_{\alpha} \kappa_{\beta}-\cos(k_{x}-k_{y}-k_{z})\kappa_{\gamma}\lambda_{\alpha}\lambda_{\beta}-\cos(k_{x}-k_{y}+k_{z})\lambda_{\gamma}\lambda_{\alpha}\kappa_{\beta}-\cos(k_{x}+k_{y}-k_{z})\lambda_{\gamma}\kappa_{\alpha} \lambda_{\beta}$. The energy spans $[-\pi,\pi]$ and the two bands of energy become gapless only at $E=0$ and $E=\pm \pi$. For now, we highlight following cases:

I) If $\beta=\pm (2c+1) \pi/T$ with $\alpha=\pm (2c+1) \pi/T$ and $\gamma=\pm 2c \pi/T$ or vice versa, the energy bands become linear dependent of momenta  which indicates that boundary states are Dirac cone type.

II) Similarly, for $\beta=\pm 2c\pi/T$ with $\alpha=\pm (2c+1) \pi/T$ and $\gamma=\pm 2c \pi/T$, the energy bands will be again linear functions of momenta and the energy gap is closed linearly. 

III) If $\beta=\alpha= \gamma=\pm (2c+1) \pi/T$ or $\beta=\alpha= \gamma=\pm 2c \pi/T$, the energy bands become flat with $E=\pm \pi/2$.

Next, it is a matter of calculation to find $\boldsymbol d$ (hence $\boldsymbol n$) and components of group velocity as

\begin{widetext}
\begin{equation}
d_{x} = \lambda_{\beta} [\kappa_{\alpha}\kappa_{\gamma} \sin (k_x+k_y+k_z)-\lambda_{\alpha} \lambda_{\gamma} \sin (k_x-k_y+k_z)] - \kappa_{\beta} [\lambda_{\alpha} \kappa_{\gamma} \sin (k_x-k_y-k_z)+\kappa_{\alpha} \lambda_{\gamma} \sin (k_x+k_y-k_z)]                \notag
\end{equation}
\begin{equation}
d_{y}=   \lambda_{\beta} [\lambda_{\alpha}\lambda_{\gamma} \cos (k_x-k_y+k_z)-\kappa_{\alpha} \kappa_{\gamma} \cos (k_x+k_y+k_z)]-\kappa_{\beta} [\lambda_{\alpha} \kappa_{\gamma} \cos (k_x-k_y-k_z)+\kappa_{\alpha}  \lambda_{\gamma} \cos (k_x+k_y-k_z)], \notag
\end{equation}
\begin{equation}
d_{z}= \lambda_{\gamma} [\lambda_{\alpha} \kappa_{\beta} \sin (k_x-k_y+k_z)-\kappa_{\alpha} \lambda_{\beta} \sin (k_x+k_y-k_z)]- \kappa_{\gamma} [\lambda_{\alpha} \lambda_{\beta} \sin (k_x-k_y-k_z)+\kappa_{\alpha} \kappa_{\beta} \sin (k_x+k_y+k_z)],
\end{equation}

\begin{eqnarray}
V(k_{i})=& & \pm (1- \rho^2)^{-\frac{1}{2}} \bigg(	
A_{i}^{1}\kappa_{\gamma}\lambda_{\alpha} \lambda_{\beta}\sin (k_{x}-k_{y}-k_{z})+
A_{i}^{2}\kappa_{\alpha} \lambda_{\gamma} \lambda_{\beta}\sin(k_{x}+k_{y}-k_{z})+                                         \notag
\\& &
A_{i}^{3}\kappa_{\beta}\lambda_{\gamma} \lambda_{\alpha}\sin (k_{x}-k_{y}+k_{z})+ A_{i}^{4}\kappa_{\alpha}\kappa_{\gamma} \kappa_{\beta} \sin (k_{x}+k_{y}+k_{z}) \bigg), \label{groupv5}
\end{eqnarray} 

\end{widetext}
where $k_{i}$ indicates $k_{x},k_{y}$ and $k_{z}$, $A_{x}^{j}=(+1,+1,+1,-1)$, $A_{y}^{j}=(-1,-1,-1,+1)$ and $A_{z}^{j}=(+1,+1,-1,+1)$. Evidently, the $\boldsymbol n$ and group velocity are ill-defined at the gapless points of the energy bands. Additionally, group velocity traverse $[-1,1]$.

\subsection{Split-step quantum walk with only PHS}  

In this section, we employ a protocol for three dimensional walk that has only PHS. To do so, first, we consider the protocl to be real and the rotation axis is in $xy$ plane. This is given by 

\begin{equation}
\widehat{U} =\widehat{S}_{\uparrow \downarrow}(z) \widehat{C}_{y}(\zeta)\widehat{S}_{\uparrow \downarrow}(y) \widehat{C}_{y}(\gamma) \widehat{S}_{\uparrow \downarrow}(x) \widehat{C}_{y}(\alpha)\widehat{S}_{\uparrow \downarrow} \widehat{C}_{y} (\beta)  \label{protocol6},
\end{equation}
where a single step of the walk includes rotation of internal states with $\widehat{C}_{y} (\beta) $, movement of the walker in $xzy$ positions space with $\widehat{S}_{\uparrow \downarrow}$, another rotation of internal states by $\widehat{C}_{y}(\alpha)$, displacement of the walker in $x$ position space by $\widehat{S}_{\uparrow \downarrow}(x)$ which is followed by a rotation of internal states by $ \widehat{C}_{y}(\gamma)$, change the $y$ position of the walker by $\widehat{S}_{\uparrow \downarrow}(y)$ and subsecquntly, rotation of internal states and movement of the walker in $z$ position space by $\widehat{C}_{y}(\zeta)$ and $\widehat{S}_{\uparrow \downarrow}(z)$, respectively. The presence of $\widehat{S}_{\uparrow \downarrow}$ breaks off CHS and the absence of CHS indicates that TRS is absent too. $\widehat{S}_{\uparrow \downarrow}$ is given by

\begin{eqnarray}
\widehat{S}_{\uparrow \downarrow} (z) =  && \ketm{\uparrow} \bram{\uparrow} \otimes \sum_{x,y,z} \ketm{x+1,y+1,z+1} \bram{x,y,z} + \notag
\\
&& \ketm{\downarrow} \bram{\downarrow} \otimes \sum_{x,y,z} \ketm{x-1,y-1,z-1} \bram{x,y,z}\,. \label{shift10}
\end{eqnarray}

The momenta and rotation angles span the first Brillouin zone. It is a matter of calculation to find two bands of energy for such protocol as

\begin{eqnarray}
E= \pm\cos^{-1}(\rho),  \label{energy6}
\end{eqnarray}
where 

\begin{eqnarray}
\rho= &&\kappa _{\alpha } \kappa _{\beta }[ \kappa _{\gamma } \kappa _{\zeta } \cos \left(2 k_x+2 k_y+2 k_z\right)- \lambda_{\gamma } \lambda _{\zeta } \cos \left(2 k_x+2k_z\right)]                           \notag
\\ &&
-\kappa _{\alpha }\lambda _{\beta }[\lambda _{\gamma } \kappa _{\zeta } \cos \left(2 k_x\right)
+\kappa_{\gamma } \lambda _{\zeta } \cos \left(2 k_x+2 k_y\right)]                         \notag
\\ && 
-\lambda _{\alpha } \kappa _{\beta }[ \lambda _{\gamma }\kappa _{\zeta } \cos \left(2 k_y+2 k_z\right)+ \kappa _{\gamma } \lambda _{\zeta } \cos \left(2 k_z\right)]                       \notag
\\ &&
-\lambda _{\alpha } \lambda _{\beta }[ \kappa _{\gamma } \kappa _{\zeta } - \lambda _{\gamma } \lambda _{\zeta } \cos \left(2 k_y\right)]   .
\end{eqnarray}

The two bands of energy close their gap at $E=0$ and $\pm \pi$. Generally, this closing gap would be nonlinearly, hence the boundary state are Fermi arc. In special cases though, the energy bands can close their gaps differently such as:

I) For $\beta=\pm (2c+1) \pi/T$, $\alpha=\gamma=\zeta=\pm 2c\pi/T$, the energy bands become linear function of the momenta. Therefore, the energy bands close their gap linearly and Dirac cone boundary states are formed.

II) If $\gamma=\zeta=\pm 2c\pi/T$, the energy bands reduce to $E=\pm\cos^{-1}(\pm \lambda _{\alpha } \lambda _{\beta })$. This indicates that gapless energy bands are independent of momenta and boundary states are indeed flat bands. It should be noted that in such case, we have a network of flat bands which includes flat bands boundary states. 

III) In case of $\alpha=\zeta=\pm 2c\pi/T$ and $\beta=\gamma=\pm (2c+1) \pi/T$, the energy bands are independent momenta $k_y$ and $k_z$ while it is linear function of $k_x$. 

Based on these cases, we can see that three types of the boundary states are observable for three-dimensional quantum walk with only PHS. Next, we calculate $\boldsymbol d$ (hence $\boldsymbol n$) and components of group velocity as

\begin{widetext}
\begin{eqnarray}
d_{x} = & & -\kappa _{\alpha } \kappa _{\beta } \lambda _{\gamma } \kappa _{\zeta } \sin \left(2 k_x\right)-\kappa _{\alpha } \kappa _{\beta } \kappa _{\gamma } \lambda _{\zeta } \sin \left(2 k_x+2 k_y\right)+\kappa _{\alpha }
\lambda _{\beta } \kappa _{\gamma } \kappa _{\zeta } \sin \left(2 k_x+2 k_y+2 k_z\right)-   \notag
\\ & &                                             
\kappa _{\alpha } \lambda _{\beta } \lambda _{\gamma } \lambda _{\zeta } \sin \left(2 k_x+2 k_z\right)+\lambda _{\alpha }
\kappa _{\beta } \lambda _{\gamma } \lambda _{\zeta } \sin \left(2 k_y\right)-\lambda _{\alpha } \lambda _{\beta } \lambda _{\gamma } \kappa _{\zeta } \sin \left(2 k_y+2 k_z\right)-\lambda _{\alpha } \lambda
_{\beta } \kappa _{\gamma } \lambda _{\zeta } \sin \left(2 k_z\right),                      \notag
\end{eqnarray}
\begin{eqnarray}
d_{y}= & &  \lambda _{\alpha } \kappa _{\beta } \kappa _{\gamma } \kappa _{\zeta }+\kappa _{\alpha } \kappa _{\beta } \lambda _{\gamma } \kappa _{\zeta } \cos \left(2 k_x\right)+\kappa _{\alpha } \kappa _{\beta } \kappa
_{\gamma } \lambda _{\zeta } \cos \left(2 k_x+2 k_y\right)+\kappa _{\alpha } \lambda _{\beta } \kappa _{\gamma } \kappa _{\zeta } \cos \left(2 k_x+2 k_y+2 k_z\right)-                    \notag
\\ & &   
\kappa _{\alpha } \lambda _{\beta } \lambda
_{\gamma } \lambda _{\zeta } \cos \left(2 k_x+2 k_z\right)-\lambda _{\alpha } \kappa _{\beta } \lambda _{\gamma } \lambda _{\zeta } \cos \left(2 k_y\right)-\lambda _{\alpha } \lambda _{\beta } \lambda _{\gamma }
\kappa _{\zeta } \cos \left(2 k_y+2 k_z\right)-\lambda _{\alpha } \lambda _{\beta } \kappa _{\gamma } \lambda _{\zeta } \cos \left(2 k_z\right), \notag
\end{eqnarray}
\begin{eqnarray}
d_{z}= & & \kappa _{\alpha } \kappa _{\beta } \lambda _{\gamma } \lambda _{\zeta } \sin \left(2 k_x+2k_z\right)+\lambda _{\alpha } \lambda _{\beta } \lambda _{\gamma } \lambda _{\zeta } \sin \left(2 k_y\right)+\lambda _{\alpha } \kappa _{\beta } \lambda _{\gamma } \kappa _{\zeta } \sin \left(2 k_y+2k_z\right)+\lambda _{\alpha } \kappa _{\beta } \kappa _{\gamma } \lambda _{\zeta } \sin \left(2 k_z\right)                                         \notag
\\ & & 
-\kappa _{\alpha } \lambda _{\beta } \lambda _{\gamma } \kappa _{\zeta } \sin \left(2 k_x\right)-\kappa _{\alpha } \lambda _{\beta } \kappa _{\gamma } \lambda _{\zeta } \sin \left(2 k_x+2 k_y\right)-\kappa _{\alpha} \kappa _{\beta } \kappa _{\gamma } \kappa _{\zeta } \left(\sin \left(2 k_x+2 k_y+2 k_z\right)\right),
\end{eqnarray}
	
\begin{equation}
V(k_{x})=\pm\frac{2\kappa _{\alpha }[  \lambda _{\beta } \lambda _{\gamma } \kappa _{\zeta } \sin \left(2 k_x\right)+ \lambda _{\beta } \kappa _{\gamma } \lambda _{\zeta } \sin \left(2 k_x+2 k_y\right)- \kappa _{\beta } \kappa _{\gamma } \kappa _{\zeta } \sin \left(2 k_x+2 k_y+2 k_z\right)+ \kappa _{\beta } \lambda _{\gamma } \lambda _{\zeta } \sin \left(2 k_x+2 k_z\right)]}{\sqrt{1-\rho^2}}, \notag
\end{equation}
\begin{equation}
V(k_{y})=\pm\frac{2[\kappa _{\alpha } \lambda _{\beta } \kappa _{\gamma } \lambda _{\zeta } \sin \left(2 k_x+2 k_y\right)-\kappa _{\alpha } \kappa _{\beta } \kappa _{\gamma } \kappa _{\zeta } \sin \left(2 k_x+2 k_y+2 k_z\right)-\lambda _{\alpha } \lambda _{\beta } \lambda _{\gamma } \lambda _{\zeta } \sin \left(2 k_y\right)+\lambda _{\alpha } \kappa _{\beta } \lambda _{\gamma } \kappa _{\zeta } \sin \left(2 k_y+2 k_z\right)]}{\sqrt{1-\rho^2}}, \notag
\end{equation}
\begin{equation}
V(k_{z})=\pm\frac{2\kappa _{\beta }[ \kappa _{\alpha } \lambda _{\gamma } \lambda _{\zeta } \sin \left(2 k_x+2 k_z\right)+\lambda _{\alpha } \lambda _{\gamma } \kappa _{\zeta } \sin \left(2 k_y+2 k_z\right)+ \lambda _{\alpha } \kappa _{\gamma } \lambda _{\zeta } \sin \left(2 k_z\right)- \kappa _{\alpha }\kappa _{\gamma } \kappa _{\zeta } \sin \left(2 k_x+2 k_y+2 k_z\right)]}{\sqrt{1-\rho^2}}, \label{groupv6}
\end{equation} 	 	 	
\end{widetext}
which similar to previous cases, the $\boldsymbol n$ and group velocity are ill-defined when energy bands close their gap.

\subsection{Split-step quantum walk with only CHS}  

Here, we focus on three-dimensional quantum walk where its protocol has only CHS. To break PHS, we need at least one rotation matrices to rotate the internal states outside of $xy$ plane. Therefore, we use the following protocol 

\begin{eqnarray}
\widehat{U} & = &\widehat{S}_{\uparrow \downarrow}(z) \widehat{C}_{\nu}(\gamma) \widehat{S}_{\uparrow \downarrow}(y) \widehat{C}_{\nu}(\alpha) \widehat{S}_{\uparrow \downarrow}(x) \widehat{C}_{\nu} (\beta)  \label{protocol7},
\end{eqnarray}
where the rotation matrices rotate the internal states with respect to $\nu=\frac{1}{\sqrt{2}}(0,1,1)$. A single step of the walk is completed when internal states are rotated by $\widehat{C}_{\nu} (\beta)$, walker is displaced in $x$ position space by $\widehat{S}_{\uparrow \downarrow}(x)$, another rotation of internal states by $\widehat{C}_{\nu}(\alpha)$, movement of the walker in $y$ position space by $\widehat{S}_{\uparrow \downarrow}(y)$, final rotation of internal states through $\widehat{C}_{\nu}(\gamma)$ and displacement of the walker in $z$ position space by $\widehat{S}_{\uparrow \downarrow}(z)$. 

Using the protocol of the quantum walk \eqref{protocol7} and rewriting shift operators by Discrete Fourier Transformation, we can find the energy 

\begin{eqnarray}
E= \pm\cos^{-1}(\rho),  \label{energy7}
\end{eqnarray}
where $\pm$ indicates two bands of energy with 

\begin{widetext}
\begin{eqnarray}
\rho= &&\left[2\lambda _{\alpha } \kappa _{\beta } \kappa _{\gamma }+2\kappa _{\alpha } \lambda _{\beta } \kappa _{\gamma }+2\kappa _{\alpha } \kappa _{\beta } \lambda _{\gamma}-\lambda _{\alpha } \lambda _{\beta } \lambda _{\gamma }\right] \frac{\sin \left(k_x+k_y+k_z\right)}{2\sqrt{2}}
-\frac{1}{2}[ \lambda _{\alpha } \lambda _{\beta } \kappa _{\gamma } \cos \left(k_x-k_y-k_z\right)+                \notag
\\ &&
\kappa _{\alpha } \lambda _{\beta } \lambda _{\gamma } \cos \left(k_x+k_y-k_z\right)+ \lambda _{\alpha } \kappa _{\beta } \lambda _{\gamma } \cos \left(k_x-k_y+k_z\right)]+
[2\kappa _{\alpha } \kappa _{\beta } \kappa _{\gamma }- \lambda _{\alpha } \lambda _{\beta } \kappa _{\gamma }-\lambda _{\alpha } \kappa _{\beta } \lambda _{\gamma }-                \notag
\\ &&
 \kappa _{\alpha } \lambda _{\beta}\lambda_{\gamma}]\frac{\cos\left(k_x+k_y+k_z\right)}{2}
+ \left[\sin \left(k_x-k_y-k_z\right)-\sin \left(k_x+k_y-k_z\right)-\sin \left(k_x-k_y+k_z\right)\right] \frac{\lambda _{\alpha } \lambda _{\beta } \lambda _{\gamma }}{2 \sqrt{2}}
\end{eqnarray}
\end{widetext}

The energy bands traverse $[-\pi,\pi]$ and close their gaps at $E=0$ and $\pm \pi$. We notice that for $\alpha=\beta=\gamma=\pm 2c \pi/T$, the energy bands become linear functions of momenta which indicates that Dirac cone boundary states are observable. Otherwise, the energy bands close their gap nonlinearly, hence Fermi arc boundary states.  

Due to the size of the calculated $\boldsymbol d$ and components of group velocity, we present them in the appendix in Eqs. \eqref{n3CHS} and \eqref{groupv7}. The obtained show that $\boldsymbol d$ and group velocity become ill-defined as energy bands close their gap.

\subsection{Split-step quantum walk without PHS, CHS and TRS}

Finally, we focus on a three-dimensional quantum walk which its protocol is without PHS, CHS and TRS. We find such protocol as

\begin{equation}
\widehat{U} =\widehat{S}_{\uparrow \downarrow}(z) \widehat{C}_{y}(\zeta) \widehat{S}_{\uparrow \downarrow}(y) \widehat{C}_{y}(\gamma)  \widehat{S}_{\uparrow \downarrow}(x) \widehat{C}_{\nu}(\alpha) \widehat{S}_{\uparrow \downarrow} \widehat{C}_{y} (\beta)  \label{protocol8},
\end{equation}
in which rotation matrix $\widehat{C}_{\nu}(\alpha)$ breaks the PHS while $\widehat{S}_{\uparrow \downarrow}$ does the the same for CHS. Each step of the walk consists of rotation of internal states by $\widehat{C}_{y} (\beta)$, displacement of the walker in $xzy$ position space by $\widehat{S}_{\uparrow \downarrow}$, rotation of internal states and movement of walker in $x$ position space by $ \widehat{C}_{\nu}(\alpha)$ and $\widehat{S}_{\uparrow \downarrow}(x)$, additional rotation of internal states by $\widehat{C}_{y}(\gamma)$ and displacement of the walker in $y$ position space by $\widehat{S}_{\uparrow \downarrow}(y)$ and final rotation of internal states by $\widehat{C}_{y}(\zeta)$ followed by movement in $z$ position space by $\widehat{S}_{\uparrow \downarrow}(z)$. We find the energy of this protocol as 

\begin{eqnarray}
E= \pm\cos^{-1}(\rho),  \label{energy8}
\end{eqnarray}
where we have two bands of energy in which 
 
\begin{widetext}
\begin{eqnarray}
\rho= &&\kappa _{\alpha } \kappa _{\beta } \kappa _{\gamma } \kappa _{\zeta } \cos \left(2 k_x+2 k_y+2 k_z\right)+\frac{\lambda _{\alpha } \kappa_{\beta } \kappa _{\gamma } \kappa _{\zeta } \sin \left(2 k_x+2 k_y+2 k_z\right)}{\sqrt{2}}+[\lambda _{\zeta } \cos \left(2 k_y\right)-\kappa _{\zeta } \sin \left(2 k_x\right)] \frac{\lambda_{\alpha } \lambda _{\beta } \lambda _{\gamma }}{\sqrt{2}}-                     \notag
\\ &&
\frac{\lambda _{\alpha } \lambda _{\beta } \kappa _{\gamma } \kappa _{\zeta }}{\sqrt{2}}-\kappa _{\alpha } \lambda _{\beta } \lambda _{\gamma } \kappa _{\zeta } \cos \left(2 k_x\right)-\kappa _{\alpha } \lambda _{\beta } \kappa _{\gamma } \lambda _{\zeta } \cos \left(2 k_x+2 k_y\right)-\kappa _{\alpha } \kappa _{\beta } \lambda _{\gamma } \lambda _{\zeta } \cos \left(2 k_x+2k_z\right)+                                  \notag
\\ &&
\frac{\lambda _{\alpha }}{\sqrt{2}}[\lambda _{\beta } \kappa _{\gamma } \lambda _{\zeta } \sin \left(2 k_x+2 k_y\right)+ \kappa _{\beta } \lambda _{\gamma } \lambda _{\zeta } \sin \left(2 k_x+2k_z\right)+ \kappa _{\beta } \lambda _{\gamma } \kappa _{\zeta } \cos \left(2 k_y+2 k_z\right)+ \kappa _{\beta } \kappa _{\gamma } \lambda _{\zeta } \cos\left(2 k_z\right)]
\end{eqnarray}
\end{widetext}

The value of energy is limited to first Brillouin zone similar to momenta and rotation angles, and we can highlight the following issues:

I) If $\alpha=\beta=\gamma=\zeta=\pm 2c\pi/T$, the energy will be a linear function of the momenta which signals to presence of Dirac cone boundary states.

II) For $\alpha=\beta=\pm (2c+1) \pi/T$ and $\gamma=\zeta=\pm 2c\pi/T$, energy becomes independent of momenta and flat bands with $E=\pm \pi/4$ are formed. 

In case of $\alpha=\beta=\gamma=\pm (2c+1) \pi/T$ and $\zeta=\pm 2c\pi/T$, the enegry bands will be independent of momenta $k_y$ and $k_z$ and linear functions of $k_x$.

Finally, similar to previous section, due to sizes of $\boldsymbol d$ and components of group velocity, we give them in the appendix as Eqs. \eqref{nTRS} and \eqref{groupv8} which have characteristic behavior of being ill-defined at boundary states.

\subsection{Discussion on results of three-dimensional quantum walks}

For simulation of some groups of topological phases such as insulators and superconductors in three dimensions, the minimal number of the energy bands that the Hamiltonian should posses is four \cite{Schnyder2008,Ryu2010,Chiu,Zhang2018}. This indicates existence of four internal degrees of the freedom for the system. In contrast, they are other types of phases such as Weyl semimetallic phases which can be described by Hamiltonians with only two bands of energy. The protocols that we have presented for three-dimensional quantum walks all have only two bands of energy due to two internal degrees of freedom. There are three motivations and reasons for presenting such protocols: 

I) It was shown that using the method of the dimensional reduction, it is possible to relate different families of the topological phases from one dimension to another one \cite{Ryu2010}. This is done via compactification of one or more spatial dimensions similar to Kaluza-Klein approach. 

II) The three-dimensional quantum walks which admit Hamiltonian with two bands of energy could be used to simulate Weyl semimetallic phases \cite{Weng,Okugawa,Burkov2016,Yokomizo,Armitage,Rafi}. Naively speaking, the Hamiltonians with two bands of energy have only three anticommuting terms given by Pauli matrices which vanish at discrete points in first Brillioun zone. This will lead to formation of the Weyl semimetallic phases \cite{Li2019}. 

III) Finally, by taking the advantages of the doubling procedure \cite{Kitagawa}, we can build protocols which their corresponding Hamiltonina has four bands of energy using the protocols that provide two-bands Hamiltonian.  

In this paper, we only focus on the last motivation. Although we discuss all the protocols that have been given in this paper, the main focus us would be on the protocols that are built based on doubling procedure. 

Due to step-dependent coins in the stucture of the three-dimensional protocols, the resultant simulated topological phenamena and their properties are step-dependent as well. This introduces dynamicality as a feature to simulated topological phases and boundary states. This dynamicality can be used as a mean to control the number of topological phases and boundary states, their populations and even their occurrences. Therefore, the step-number is a mean to engineer or control the simulation of topological phenomena on a very high level. 

In case of simple-step protocol \eqref{protocol4}, the energy bands close their gap only linearly. This indicates that the only present boundary states are Dirac cone type. This protocol has three symmetries of PHS, TRS and CHS with $\widehat{\mathcal{P}}^2=+1$, $\widehat{\mathcal{T}}^2=+1$ and $\widehat{\Gamma}^2=+1$. If we neglect being two-bands Hamiltonian, simple-step protocol in three dimensions simulate BDI family of topological phases with $0$ topological invariants (absence of topological phases). In contrast, remembering the definition for Weyl semimetallic phases, this protocol can be used to simulate these type of phases.  

The split-step protocol given in Eq. \eqref{protocol5} has three symmetries of PHS, TRS and CHS with $\widehat{\mathcal{P}}^2=+1$, $\widehat{\mathcal{T}}^2=+1$ and $\widehat{\Gamma}^2=+1$ and similar to simple step-protocol, no topological phase can be simulated by it in three-dimensional case. The boundary states observed for this protocol are two types of the Dirac cone and Fermi arcs. Under certain circumstances, we might be able to achieve simulation of the Weyl semimetallic phases with this protocol. 

The split-step protocol with only PHS \eqref{protocol6} (with $\widehat{\mathcal{P}}^2=+1$) is the first protocol for three-dimensional quantum walk that can includes three types of boundary states of Dirac cone, Fermi arc and flat bands. The flat bands egde states are independent of the momenta. This is reminiscence of what we observed in one-dimensional case where the flat bands are also independent of momentum \cite{Panahiyan2020}. This protocol, similar to the other previous protocols in three-dimensional walks, simulate no topological phases. Instead, we can use the doubling procedure to build the following protocol   

\begin{eqnarray}
\widehat{U} = \begin{pmatrix}
\widehat{U}_{e} & 0 \\
0 & \widehat{U}_{e}^t 
\end{pmatrix}, 
\end{eqnarray}
in which $\widehat{U}_{e}$ is given by Eq. \eqref{protocol6}. This protocol produces four-bands Hamiltonian which meets the minimal condition to simulate topological phases that are reported as insulators and superconductors. The protocol has PHS, TRS and CHS with $\widehat{\mathcal{P}}^2=+1$, $\widehat{\mathcal{T}}^2=-1$ and $\widehat{\Gamma}^2=+1$ and it can simulate DIII family of the topological phases with topological invariant being $\mathbb{Z}$.

Next, for the protocol with only CHS \eqref{protocol7}, the resultant boundary states are of two types of Dirac cone and Fermi arc. According to present symmetries in the system and dimensionality of the quantum walk (neglecting Hamiltonian being two-bands), this protocol can simulate AIII family of the topological phases which has integer=valued topological invariant, $\mathbb{Z}$. The absence of the TRS enables us to implement doubling procedure and find the following protocol    

\begin{eqnarray}
\widehat{U} = \begin{pmatrix}
\widehat{U}_{f} & 0 \\
0 & \widehat{U}_{f}^t 
\end{pmatrix}, 
\end{eqnarray} 
in which $\widehat{U}_{f}$ is the one in Eq. \eqref{protocol7}. The symmetries of this protocol are PHS, TRS and CHS with $\widehat{\mathcal{P}}^2=-1$, $\widehat{\mathcal{T}}^2=-1$ and $\widehat{\Gamma}^2=+1$ and since its corresponding Hamiltonian has four bands of energy, the topological phases simulated by it are CII family with topological invariants of $\mathbb{Z}_2$. 

In case of the protocol without any of the symmetries \eqref{protocol8}, the boundary states that can be simulated by it are of two types of the Fermi arc and Dirac cone boundary states. Neglecting the absence of the four-bands Hamiltonian, this protocol can simulate AIII family of the topological phases with topological invariants of $\mathbb{Z}$. Additionally, we can build two protocols using the doubling procedure which would have four-bands Hamiltonian. The first protocol is given by

\begin{eqnarray}
\widehat{U} = \begin{pmatrix}
\widehat{U}_{g} & 0 \\
0 & \widehat{U}_{g}^{\ast} 
\end{pmatrix},
\end{eqnarray}  
where $\widehat{U}_{g}$ is given by Eq. \eqref{protocol8}. This protocol has only PHS with $\widehat{\mathcal{P}}^2=-1$, therefore, it can not simulate any type of the topological phases in three dimensions. The other protocol will be 

\begin{eqnarray}
\widehat{U} = \begin{pmatrix}
\widehat{U}_{g} & 0 \\
0 & 1 \end{pmatrix} e^{-i \tau_{y} \sigma_{y} \phi/2} \begin{pmatrix}
1 & 0 \\
0 & \widehat{U}_{g}^t 
\end{pmatrix}
\end{eqnarray}
which has only TRS with $\widehat{\mathcal{T}}^2=-1$. This indicates that AII family of the topological phases can be simulated by this protocol which has topological invariant of $\mathbb{Z}_{2}$. 

As a final remark, we point it out that for protocols devised by doubling procedure, we can decompose them into followings 

\begin{equation}
\widehat{U}' = d_{0} \tau_{0} \otimes \sigma_{0} - I d_{1}  \tau_{0} \otimes \sigma_{1} -I d_{2} \tau_{3} \otimes \sigma_{2} - I \tau_{0} \otimes \sigma_{3},
\end{equation}
if the protocols are produced by $\widehat{U}' = \begin{pmatrix} \widehat{U} & 0 \\
0 & \widehat{U}^t \end{pmatrix} $. In contrast, if the protocol is given by $\widehat{U} = \begin{pmatrix}\widehat{U} & 0 \\0 & \widehat{U}^{\ast}\end{pmatrix}$, its decomposition would be 

\begin{equation}
\widehat{U}' = d_{0} \tau_{0} \otimes \sigma_{0} - I d_{1}  \tau_{3} \otimes \sigma_{1} -I d_{2} \tau_{0} \otimes \sigma_{2} - I \tau_{3} \otimes \sigma_{3},
\end{equation}
in which $\sigma_{i}$ and $\tau_{i}$ are Pauli matrices for two internal degrees of freedom  $ \ketm{0}$ and $\ketm{1}$, and two flavors of $A$ and $B$. Such decompositions would be highly used in calculation of Chern and winding numbers for even and odd number of spatial dimensions in the quantum walks, respectively.

\begin{table*}
	\caption {Topological phases realized by different protocols of quantum walk for one (1D), two (2D) and three dimensions (3D). Presence of certain symmetry is depicted by $\pm 1$ while the absence is given by $0$. $+1$ and $-1$ specify whether the symmetry squares to $+1$ and $-1$ (e. g.  $\widehat{\mathcal{P}}^2=+1$ or $-1$). The family of the topological phases are determined by the dimensionality, available symmetries and corresponding topological invariant which could be an integer, $\mathbb{Z}$ , or a binary $\mathbb{Z}_{2}$.} \label{table}	
	\begin{tabular}{|c|c|c|c|c|c|c|c|c|}
		\hline
		&
		& 
		\multicolumn{3}{|c|}{} & 
		& 
		\\
		\multicolumn{1}{|c|}{Dimension} &
		\multicolumn{1}{|c|}{Protocol} & 
		\multicolumn{3}{|c|}{Symmetry} & 
		\multicolumn{1}{|c|}{Topological} & 
		\multicolumn{1}{|c|}{Topological} 			 
		\\			  
		&			  
		\multicolumn{1}{c|}{} & 
		\multicolumn{1}{c|}{ $\widehat{\mathcal{P}}$} & \multicolumn{1}{c|}{$\widehat{\mathcal{T}}$} & 
		\multicolumn{1}{c|}{$\widehat{\Gamma}$} & 
		\multicolumn{1}{c|}{Invariant} & 
		\multicolumn{1}{c|}{Family} 								
		\\ \hline
		\parbox[t]{7mm}{\multirow{6}{*}{\rotatebox[origin=c]{90}{1D}}}   
		& $\widehat{U} = \widehat{S}_{\uparrow \downarrow}(x) \widehat{C}_{y}(\beta)$  & +1  & +1  & +1 & $\mathbb{Z}$ & BDI \\ \cline{2-7}			 
		& $\widehat{U} =\widehat{S}_{\uparrow}(x) \widehat{C}_{y}(\alpha) \widehat{S}_{\downarrow}(x) \widehat{C}_{y}(\beta)$  &  +1 & +1  & +1  & $\mathbb{Z}$ & BDI \\	\cline{2-7} 
		& $\widehat{U}_{a} = \widehat{S}_{\uparrow}(x) \widehat{C}_{y}(\alpha) \widehat{S}_{\downarrow}(x) \widehat{C}_{y}(\beta)\widehat{S}_{\uparrow \downarrow}(x)$  & +1  & 0  & 0 & $\mathbb{Z}_{2}$ & D \\ 	\cline{2-7} 
		&  $\widehat{U} = \begin{pmatrix}
		\widehat{U}_{a} & 0 \\
		0 & \widehat{U}_{a}^t 
		\end{pmatrix} $ & +1  &  -1 & +1  & $\mathbb{Z}_{2}$ & DIII  \\ \cline{2-7}
		& $\widehat{U}_{b}= \widehat{S}_{\uparrow}(x) \widehat{C}_{\nu}(\alpha) \widehat{S}_{\downarrow}(x) \widehat{C}_{\nu}(\beta) $  &  0 &  0 & +1  & $\mathbb{Z}$ & AIII \\ \cline{2-7} 
		&  $\widehat{U} = \begin{pmatrix}
		\widehat{U}_{b} & 0 \\
		0 & \widehat{U}_{b}^t 
		\end{pmatrix} $  & -1  &  -1 & +1  & $\mathbb{Z}$ & CII 
		\\ \hline
		\parbox[t]{7mm}{\multirow{6}{*}{\rotatebox[origin=c]{90}{2D}}}   
		& $\widehat{U} = \widehat{S}_{\uparrow \downarrow}(y)\widehat{S}_{\uparrow \downarrow}(x) \widehat{C}_{y}(\beta)$  & +1  &  +1 & +1  &  0 & BDI \\ \cline{2-6} 
		& $\widehat{U} = \widehat{S}_{\uparrow \downarrow}(y) \widehat{C}_{y}(\alpha) \widehat{S}_{\uparrow \downarrow}(x) \widehat{C}_{y} (\beta)$  & +1  & +1  &  +1 & 0 & BDI  \\ \cline{2-7} 
		& $\widehat{U}_{c} = \widehat{S}_{\uparrow \downarrow}(x) \widehat{C}_{y}(\beta) \widehat{S}_{\uparrow \downarrow}(y) \widehat{C}_{y}(\alpha) \widehat{S}_{\uparrow \downarrow} \widehat{C}_{y}(\beta)$  & +1  & 0 &  0 & $\mathbb{Z}$ &  D \\ \cline{2-7}
		& $\widehat{U} = \begin{pmatrix}
		\widehat{U}_{c} & 0 \\
		0 & \widehat{U}_{c}^t 
		\end{pmatrix} $  & +1  &  -1 &  +1 & $\mathbb{Z}_{2}$ & DIII \\ \cline{2-7} 
		& $\widehat{U}_{d} = \widehat{S}_{\uparrow \downarrow}(y) \widehat{C}_{y}(\gamma)  \widehat{S}_{\uparrow \downarrow}(x) \widehat{C}_{\nu}(\alpha) \widehat{S}_{\uparrow \downarrow} \widehat{C}_{y} (\beta)$  & 0  &  0 &  0 & $\mathbb{Z}$ & A \\ \cline{2-7}			 
		& $\widehat{U} = \begin{pmatrix}
		\widehat{U}_{d} & 0 \\
		0 & 1 \end{pmatrix}$ $e^{-i \tau_{y} \sigma_{y} \phi/2}$ $\begin{pmatrix}
		1 & 0 \\
		0 & \widehat{U}_{d}^t 
		\end{pmatrix}$  & 0  &  -1 &  0 & $\mathbb{Z}_{2}$ & AII \\ \cline{2-7}			 
		& $\widehat{U} = \begin{pmatrix}
		\widehat{U}_{d} & 0 \\
		0 & \widehat{U}_{d}^{\ast} 
		\end{pmatrix} $  & -1  &  0 & 0  & $\mathbb{Z}$ & C \\ \hline					
		\parbox[t]{7mm}{\multirow{8}{*}{\rotatebox[origin=c]{90}{3D}}}   
		& $\widehat{U} = \widehat{S}_{\uparrow \downarrow}(z)\widehat{S}_{\uparrow \downarrow}(y)\widehat{S}_{\uparrow \downarrow}(x) \widehat{C}_{y}(\beta)$  & +1  & +1  & +1  & 0 & BDI \\ \cline{2-7} 
		& $\widehat{U} =\widehat{S}_{\uparrow \downarrow}(z) \widehat{C}_{y}(\gamma) \widehat{S}_{\uparrow \downarrow}(y) \widehat{C}_{y}(\alpha) \widehat{S}_{\uparrow \downarrow}(x) \widehat{C}_{y} (\beta)$  & +1  & +1  & +1  & 0 & BDI \\ \cline{2-7} 
		& $\widehat{U}_{e} =\widehat{S}_{\uparrow \downarrow}(z) \widehat{C}_{y}(\zeta)\widehat{S}_{\uparrow \downarrow}(y) \widehat{C}_{y}(\gamma) \widehat{S}_{\uparrow \downarrow}(x) \widehat{C}_{y}(\alpha)\widehat{S}_{\uparrow \downarrow} \widehat{C}_{y} (\beta)$  & +1  & 0  & 0  & 0 & D \\ \cline{2-7}
		& $\widehat{U} = \begin{pmatrix}
		\widehat{U}_{e} & 0 \\
		0 & \widehat{U}_{e}^t 
		\end{pmatrix}$  & +1  & -1  & +1  & $\mathbb{Z}$ & DIII \\ \cline{2-7} 
		& $\widehat{U}_{f} =\widehat{S}_{\uparrow \downarrow}(z) \widehat{C}_{\nu}(\gamma) \widehat{S}_{\uparrow \downarrow}(y) \widehat{C}_{\nu}(\alpha) \widehat{S}_{\uparrow \downarrow}(x) \widehat{C}_{\nu} (\beta)$  & 0  &  0 &  +1 & $\mathbb{Z}$ & AIII \\ \cline{2-7}			 
		& $\widehat{U} = \begin{pmatrix}
		\widehat{U}_{f} & 0 \\
		0 & \widehat{U}_{f}^t 
		\end{pmatrix} $  & -1  &  -1 & +1  & $\mathbb{Z}_{2}$ & CII \\  \cline{2-7}
		& $\widehat{U}_{g} =\widehat{S}_{\uparrow \downarrow}(z) \widehat{C}_{y}(\zeta) \widehat{S}_{\uparrow \downarrow}(y) \widehat{C}_{y}(\gamma)  \widehat{S}_{\uparrow \downarrow}(x) \widehat{C}_{\nu}(\alpha) \widehat{S}_{\uparrow \downarrow} \widehat{C}_{y} (\beta)$   & 0  &  0 &  0 & 0 & A  \\ \cline{2-7}			 
		& $\widehat{U} = \begin{pmatrix}
		\widehat{U}_{g} & 0 \\
		0 & 1 \end{pmatrix}$ $e^{-i \tau_{y} \sigma_{y} \phi/2}$ $\begin{pmatrix}
		1 & 0 \\
		0 & \widehat{U}_{g}^t 
		\end{pmatrix}$    & 0  & -1  &  0 &  $\mathbb{Z}_{2}$ & AII \\ \cline{2-7}			 
		& $\widehat{U} = \begin{pmatrix}
		\widehat{U}_{g} & 0 \\
		0 & \widehat{U}_{g}^{\ast} 
		\end{pmatrix} $  & -1  & 0  &  0 & 0 & C  \\ \hline				
	\end{tabular}
\end{table*}  
	
\section{Conclusion}  \label{Conclusion}

In this paper, we studied the simulation of topological phenomena observed in condensed matter using one-, two- and three-dimensional quantum walks. To enrich the simulations further, we used step-dependent coins in the protocols of the quantum walks. The result of such consideration was step-dependency of the simulated topological phenomena and their properties which introduced  dynamicality as a feature to simulated topological phases and boundary states. This dynamicality enabled us to use the step-number of the quantum walk as a mean to control and engineer the number of topological phases and boundary states, their populations, their types and even their occurrences.
It should be noted that by adjusting the step-number to $1$ in considered quantum walks in this paper, the obtained results would reduce to quantum walks with step-independent coins.  

In previous study \cite{Panahiyan2020}, we investigated a set of protocols for one- and two-dimensional quantum walks which could simulate limited number of topological phases. Here, we included additional protocols for one- and two-dimensional quantum walks and completed the table of protocols necessary for simulation of all type of topological phases in one- and two-dimensional cases. We also highlighted the conditions for emergences of different boundary states, the possibility of their coexistence for a single step of the walk and possible phase structures for different protocols.  

In addition, We took the first steps to construct the protocols for quantum walks that can be used to simulate topological phenomena in three dimensions. Our strategy mainly relied on using the doubling procedure to build the protocols that have four-bands Hamiltonians correspondingly. This enabled us to simulate DIII, CII and AII families of the topological phases in three dimensions using quantum walks. If we neglect the condition of Hamiltonian with minimal four bands, we were also able to simulate AIII family of topological phases with quantum walks. In addition, we extracted the conditions for presence of different types of boundary states and we confirmed that it is possible to simulate Dirac cone, Fermi and flat bands boundary states with these quantum walks. 

The presented results of this paper can used in two directions; first of all, it would be interesting to see if one can use the method of the dimensional reduction to connect a certain family of a topological phases in higher dimension to another family in lower dimensions. In addition, the three-dimensional protocols with two-bands of energy can be investigated for the possibility of the simulation of Weyl semimetallic phases. Furthermore, although we constructed several four-bands protocols by using the doubling procedure, it is interesting to see if it is possible to have similar protocols that are not made through such method and can simulate different family of the topological phases in three dimensions. Finally, it may be possible to use the protocols provided int his paper to investigate spectral magnetization ratchets (absence or presence of spectral magnetization) \cite{Mallick}. 

\section{Appendix}

For three-dimensional quantum walk with only PHS \eqref{protocol7}, it is a matter of calculation to find $\boldsymbol d$ and components of group velocity as

\begin{widetext}
\begin{eqnarray}
d_{x} = & &[\lambda _{\alpha } \lambda _{\beta } \lambda _{\gamma } -2 \kappa _{\alpha } \kappa _{\beta } \lambda _{\gamma }]\frac{ \sin \left(k_x+k_y-k_z\right)}{2 \sqrt{2}}-[2\lambda _{\alpha } \kappa _{\beta } \kappa _{\gamma }+\lambda _{\alpha } \lambda_{\beta } \lambda _{\gamma }] \frac{ \sin \left(k_x-k_y-k_z\right)}{2 \sqrt{2}}+                                      \notag
\\ &&
[2\kappa _{\alpha } \lambda _{\beta } \kappa _{\gamma }-\lambda _{\alpha } \lambda _{\beta } \lambda_{\gamma }]\frac{ \sin \left(k_x+k_y+k_z\right)}{2 \sqrt{2}}-\frac{\lambda _{\alpha } \lambda _{\beta } \lambda _{\gamma } \sin \left(k_x-k_y+k_z\right)}{2\sqrt{2}}+                                                            \notag
\\ &&
[ \kappa _{\alpha } \lambda _{\beta }\lambda _{\gamma } + \lambda _{\alpha } \kappa _{\beta } \lambda _{\gamma }] \frac{\cos \left(k_x+k_y-k_z\right)}{2}+[ \lambda _{\alpha } \lambda _{\beta } \kappa _{\gamma } - \lambda _{\alpha } \kappa _{\beta } \lambda _{\gamma }]\frac{\cos \left(k_x-k_y-k_z\right)}{2} -                                     \notag
\\ &&
[ \lambda _{\alpha } \lambda _{\beta } \kappa _{\gamma }+\frac{1}{2} \kappa _{\alpha } \lambda _{\beta } \lambda_{\gamma}]\frac{\cos\left(k_x+k_y+k_z\right)}{2}                                              \notag
\end{eqnarray}
\begin{eqnarray}
d_{y}= & &  [\lambda _{\alpha } \kappa _{\beta } \lambda _{\gamma } + \kappa _{\alpha } \lambda _{\beta } \lambda _{\gamma }]\frac{\sin\left(k_x+k_y-k_z\right)}{2}+[\lambda _{\alpha } \lambda _{\beta } \kappa _{\gamma }- \lambda _{\alpha } \kappa _{\beta } \lambda _{\gamma }]\frac{\sin \left(k_x-k_y-k_z\right)}{2}+                                             \notag
\\ &&
[\lambda _{\alpha } \lambda _{\beta} \kappa _{\gamma }+ \kappa _{\alpha } \lambda _{\beta } \lambda_{\gamma}]\frac{\sin\left(k_x+k_y+k_z\right)}{2}+[2\kappa _{\alpha } \lambda _{\beta } \kappa _{\gamma}-\lambda _{\alpha } \lambda _{\beta } \lambda _{\gamma }] \frac{\cos \left(k_x+k_y+k_z\right)}{2\sqrt{2}}+            \notag
\\ &&following
[2\lambda _{\alpha } \kappa _{\beta } \kappa _{\gamma }+\lambda _{\alpha } \lambda _{\beta } \lambda _{\gamma }]\frac{\cos
\left(k_x-k_y-k_z\right)}{2\sqrt{2}}+[2\kappa _{\alpha } \kappa _{\beta } \lambda _{\gamma }-\lambda _{\alpha } \lambda _{\beta } \lambda _{\gamma }]\frac{\cos \left(k_x+k_y-k_z\right)}{2\sqrt{2}}-                                         \notag
\\ &&
\frac{\lambda _{\alpha } \lambda _{\beta } \lambda _{\gamma }\cos \left(k_x-k_y+k_z\right)}{2 \sqrt{2}}                        \notag
\end{eqnarray}
\begin{eqnarray}
d_{z}= & &\frac{1}{2}[\lambda_{\alpha } \kappa _{\beta } \lambda _{\gamma } \sin \left(k_x-k_y+k_z\right) - \lambda _{\alpha } \lambda _{\beta } \kappa _{\gamma } \sin \left(k_x-k_y-k_z\right)- \kappa _{\alpha } \lambda _{\beta } \lambda _{\gamma } \sin \left(k_x+k_y-k_z\right) ]+                                                         \notag
\\ &&
[ \lambda _{\alpha } \lambda _{\beta } \kappa _{\gamma }+ \lambda _{\alpha } \kappa _{\beta } \lambda_{\gamma }+ \kappa _{\alpha } \lambda _{\beta } \lambda _{\gamma }-2\kappa _{\alpha } \kappa _{\beta } \kappa _{\gamma }] \frac{\sin\left(k_x+k_y+k_z\right)}{2}+ [2\lambda _{\alpha } \kappa	_{\beta } \kappa _{\gamma }+2\kappa _{\alpha } \lambda _{\beta } \kappa _{\gamma }+
                        \notag
\\ &&    
2\kappa _{\alpha } \kappa _{\beta } \lambda _{\gamma }-\lambda _{\alpha }	\lambda _{\beta } \lambda _{\gamma }] \frac{\cos \left(k_x+k_y+k_z\right)}{2 \sqrt{2}} + [\cos\left(k_x+k_y-k_z\right)-\cos \left(k_x-k_y-k_z\right)-                         \notag
\\ && 
\cos \left(k_x-k_y+k_z\right)]\frac{\lambda _{\alpha } \lambda _{\beta } \lambda _{\gamma }}{2 \sqrt{2}}  \label{n3CHS}
\end{eqnarray}
	
\begin{eqnarray}
V(k_{i})=& &\pm (1-\rho^2)^{-\frac{1}{2}}\bigg(
\frac{1}{2}[A_{i}^{1}\lambda_{\alpha } \kappa _{\beta } \lambda _{\gamma } \sin \left(k_x-k_y+k_z\right) +A_{i}^{2} \lambda _{\alpha } \lambda _{\beta } \kappa _{\gamma } \sin \left(k_x-k_y-k_z\right)+A_{i}^{3} \kappa _{\alpha } \lambda _{\beta } \lambda _{\gamma } \sin \left(k_x+k_y-k_z\right) ]+                                                         \notag
\\ &&
[ \lambda _{\alpha } \lambda _{\beta } \kappa _{\gamma }+ \lambda _{\alpha } \kappa _{\beta } \lambda_{\gamma }+ \kappa _{\alpha } \lambda _{\beta } \lambda _{\gamma }-2\kappa _{\alpha } \kappa _{\beta } \kappa _{\gamma }] \frac{\sin\left(k_x+k_y+k_z\right)}{2}+ [2\lambda _{\alpha } \kappa	_{\beta } \kappa _{\gamma }+2\kappa _{\alpha } \lambda _{\beta } \kappa _{\gamma }+
2\kappa _{\alpha } \kappa _{\beta } \lambda _{\gamma }-                \label{groupv7}          
\\ && 
\lambda _{\alpha }	\lambda _{\beta } \lambda _{\gamma }] \frac{\cos \left(k_x+k_y+k_z\right)}{2 \sqrt{2}} + [A_{i}^{4}\cos\left(k_x+k_y-k_z\right)+A_{i}^{5}\cos \left(k_x-k_y-k_z\right)+A_{i}^{6}\cos \left(k_x-k_y+k_z\right)]\frac{\lambda _{\alpha } \lambda _{\beta } \lambda _{\gamma }}{2 \sqrt{2}}\bigg),    \notag 
\end{eqnarray} 	
\end{widetext}
where $k_{i}$ indicates $k_{x},k_{y}$ and $k_{z}$, $A_{x}^{j}=(+1,+1,+1,-1,+1,-1)$, $A_{y}^{j}=(-1,-1,+1,-1,-1,+1)$ and $A_{z}^{j}=(+1,-1,-1,+1,-1,-1)$. 

Next, we calculate the $\boldsymbol d$ and components of group velocity for three-dimensional quantum walk with for protocol in Eq. \eqref{protocol8} which leads to

\begin{widetext}
\begin{eqnarray}
d_{x} = & & \kappa _{\alpha } \lambda _{\beta } \kappa _{\gamma } \kappa _{\zeta } \sin \left(2 k_x+2 k_y+2 k_z\right)-\frac{\lambda _{\alpha } \lambda _{\beta } \kappa _{\gamma } \kappa _{\zeta } \cos\left(2 k_x+2 k_y+2 k_z\right)}{\sqrt{2}}-\kappa_{\alpha } \kappa _{\beta } \kappa _{\gamma } \lambda _{\zeta } \sin \left(2 k_x+2k_y\right)+                          \notag
\\ &&
\frac{\lambda _{\alpha } \lambda _{\beta } \lambda _{\gamma }}{\sqrt{2}} [\lambda _{\zeta } \cos \left(2 k_x+2 k_z\right)-\kappa _{\zeta } \sin \left(2 k_y+2 k_z\right)]-\kappa _{\alpha } \lambda _{\beta } \lambda _{\gamma } \lambda _{\zeta } \sin \left(2 k_x+2 k_z\right)-\kappa _{\alpha } \kappa _{\beta } \lambda _{\gamma } \kappa _{\zeta } \sin \left(2 k_x\right)+                           \notag
\\ &&
\frac{\lambda _{\alpha }}{\sqrt{2}}[\kappa _{\beta } \lambda _{\gamma } \kappa _{\zeta } \cos \left(2 k_x\right)+ \kappa _{\beta } \lambda _{\gamma } \lambda _{\zeta }\sin \left(2 k_y\right)-\lambda _{\beta } \kappa _{\gamma } \lambda _{\zeta } \sin \left(2 k_z\right)+\kappa _{\beta } \kappa _{\gamma } \lambda _{\zeta }\cos\left(2k_x+2k_y\right)]                 \notag
\end{eqnarray}
\begin{eqnarray}
d_{y} = & & \frac{\lambda _{\alpha } \lambda _{\beta } \kappa _{\gamma } \kappa _{\zeta } \sin\left(2 k_x+2 k_y+2 k_z\right)}{\sqrt{2}}+\kappa _{\alpha } \lambda _{\beta } \kappa _{\gamma } \kappa _{\zeta } \cos \left(2 k_x+2 k_y+2 k_z\right) + \frac{\lambda _{\alpha } \kappa _{\beta } \kappa _{\gamma } \kappa _{\zeta }}{\sqrt{2}}+\kappa_{\alpha } \kappa _{\beta } \lambda _{\gamma } \kappa _{\zeta } \cos \left(2 k_x\right)-             \notag
\\ &&
\frac{\lambda _{\alpha } \lambda _{\beta } \lambda _{\gamma }}{\sqrt{2}}[\lambda _{\zeta } \sin \left(2 k_x+2 k_z\right)+\kappa _{\zeta } \cos \left(2 k_y+2k_z\right)]+\kappa _{\alpha } \kappa _{\beta } \kappa _{\gamma } \lambda _{\zeta } \cos \left(2 k_x+2 k_y\right)
-\kappa _{\alpha } \lambda _{\beta } \lambda _{\gamma } \lambda_{\zeta } \cos \left(2 k_x+2 k_z\right)+                   \notag
\\ &&
\frac{\lambda _{\alpha }}{\sqrt{2}}[\kappa _{\beta } \kappa _{\gamma } \lambda _{\zeta } \sin \left(2 k_x+2k_y\right)+\kappa _{\beta } \lambda _{\gamma } \kappa _{\zeta } \sin \left(2 k_x\right)-\kappa _{\beta } \lambda _{\gamma } \lambda _{\zeta } \cos \left(2 k_y\right)-\lambda _{\beta } \kappa _{\gamma} \lambda _{\zeta } \cos \left(2 k_z\right)]             \notag
\end{eqnarray}
\begin{eqnarray}
d_{z} = & &\frac{\lambda _{\alpha } \kappa _{\beta } \kappa _{\gamma } \kappa _{\zeta } \cos \left(2 k_x+2 k_y+2	k_z\right)}{\sqrt{2}}-\kappa _{\alpha } \kappa _{\beta } \kappa _{\gamma } \kappa _{\zeta } \sin \left(2 k_x+2 k_y+2k_z\right)+ [\kappa _{\zeta } \cos \left(2k_x\right)+\lambda _{\zeta } \sin \left(2k_y\right)]\frac{\lambda _{\alpha } \lambda _{\beta } \lambda _{\gamma }}{\sqrt{2}}+                   \notag
\\ &&
\kappa _{\alpha } \kappa _{\beta } \lambda _{\gamma } \lambda _{\zeta} \sin \left(2 k_x+2 k_z\right)-\kappa _{\alpha } \lambda _{\beta } \lambda _{\gamma } \kappa _{\zeta } \sin \left(2 k_x\right)-\kappa _{\alpha } \lambda _{\beta } \kappa _{\gamma } \lambda _{\zeta } \sin \left(2 k_x+2 k_y\right)+                     \notag
\\ &&
\frac{\lambda_{\alpha }}{\sqrt{2}}[ \lambda _{\beta } \kappa _{\gamma } \lambda _{\zeta } \cos \left(2 k_x+2 k_y\right)- \kappa _{\beta } \lambda _{\gamma } \lambda _{\zeta } \cos \left(2 k_x+2 k_z\right)+ \kappa _{\beta } \lambda _{\gamma } \kappa _{\zeta } \sin \left(2 k_y+2 k_z\right)+ \kappa _{\beta } \kappa _{\gamma } \lambda _{\zeta } \sin \left(2 k_z\right)]   \label{nTRS}                
\end{eqnarray}
	
\begin{eqnarray}
V(k_{x})= && \pm (1-\rho^2)^{-\frac{1}{2}} \bigg( 2 \kappa _{\alpha } \lambda _{\beta } \lambda _{\gamma } \kappa _{\zeta } \sin \left(2 k_x\right)-\sqrt{2} \lambda _{\alpha } \lambda _{\beta } \lambda _{\gamma } \kappa _{\zeta } \cos \left(2 k_x\right)+2 \kappa
_{\alpha } \lambda _{\beta } \kappa _{\gamma } \lambda _{\zeta } \sin \left(2 k_x+2 k_y\right)-                  \notag
\\ &&
\sqrt{2} \lambda _{\alpha } \lambda _{\beta } \kappa _{\gamma } \lambda _{\zeta } \cos \left(2 k_x+2k_y\right)+\sqrt{2} \lambda _{\alpha } \kappa _{\beta } \kappa _{\gamma } \kappa _{\zeta } \cos \left(2 k_x+2 k_y+2 k_z\right)-2 \kappa _{\alpha } \kappa _{\beta } \kappa _{\gamma } \kappa _{\zeta } \sin \left(2k_x+2 k_y+2 k_z\right)+                 \notag
\\ &&
2 \kappa _{\alpha } \kappa _{\beta } \lambda _{\gamma } \lambda _{\zeta } \sin \left(2 k_x+2 k_z\right)-\sqrt{2} \lambda _{\alpha } \kappa _{\beta } \lambda _{\gamma } \lambda _{\zeta }\cos \left(2 k_x+2 k_z\right) \bigg),                  \notag
\end{eqnarray}
\begin{eqnarray}
V(k_{y})= && \pm (1-\rho^2)^{-\frac{1}{2}} \bigg(2 \kappa _{\alpha } \lambda _{\beta } \kappa _{\gamma } \lambda _{\zeta } \sin \left(2 k_x+2 k_y\right)-\sqrt{2} \lambda _{\alpha } \lambda _{\beta } \kappa _{\gamma } \lambda _{\zeta } \cos \left(2 k_x+2k_y\right)+\sqrt{2} \lambda _{\alpha } \kappa _{\beta } \kappa _{\gamma } \kappa _{\zeta } \cos \left(2 k_x+2 k_y+2 k_z\right)-                                   \notag
\\ &&
2 \kappa _{\alpha } \kappa _{\beta } \kappa _{\gamma } \kappa _{\zeta } \sin \left(2k_x+2 k_y+2 k_z\right)-\sqrt{2} \lambda _{\alpha } \lambda _{\beta } \lambda _{\gamma } \lambda _{\zeta } \sin \left(2 k_y\right)+\sqrt{2} \lambda _{\alpha } \kappa _{\beta } \lambda _{\gamma } \kappa _{\zeta }\sin \left(2 k_y+2 k_z\right) \bigg),              \notag
\end{eqnarray}
\begin{eqnarray}
V(k_{z})= && \pm (1-\rho^2)^{-\frac{1}{2}} \bigg(\sqrt{2} \lambda _{\alpha } \kappa _{\beta } \kappa _{\gamma } \kappa _{\zeta } \cos \left(2 k_x+2 k_y+2 k_z\right)-2 \kappa _{\alpha } \kappa _{\beta } \kappa _{\gamma } \kappa _{\zeta } \sin \left(2 k_x+2 k_y+2k_z\right)+2 \kappa _{\alpha } \kappa _{\beta } \lambda _{\gamma } \lambda _{\zeta } \sin \left(2 k_x+2 k_z\right)- \notag
\\ &&
\sqrt{2} \lambda _{\alpha } \kappa _{\beta } \lambda _{\gamma } \lambda _{\zeta } \cos \left(2k_x+2 k_z\right)+\sqrt{2} \lambda _{\alpha } \kappa _{\beta } \lambda _{\gamma } \kappa _{\zeta } \sin \left(2 k_y+2 k_z\right)+\sqrt{2} \lambda _{\alpha } \kappa _{\beta } \kappa _{\gamma } \lambda _{\zeta }	\sin \left(2 k_z\right)\bigg), \label{groupv8}
\end{eqnarray} 
\end{widetext}

\end{document}